\documentclass[11pt,twoside]{article}
\pdfoutput=1
  \usepackage{./mylay2,jmlr2e}
  \usepackage{comment}
  \usepackage{bbold}
  \usepackage{hyperref}
  \usepackage{bbm}
  \newcommand{\Expct}{\mathbb E}
  \newcommand{\E}{\Expct}
  \newcommand{\dist}{{\cal J}}

  \newcommand{\prob}{{\mathbb P}}
  \newcommand{\tildep}{\tilde{p}}
  \newcommand{\tildeq}{\tilde{q}}
  \newcommand{\ngpe}{B(p,\epsilon; \cP)}
  \newcommand{\ngpte}{B(p,\tilde\epsilon_p; \cP)}
  \newcommand{\ngpom}{B(p,1/m; \cP)}
  \newcommand{\ngpep}{B(p,\epsilon_{{}_p}; \cP)}
  \newcommand{\ngpepp}{B(p,\epsilon'; \cP)}

  \newcommand{\bb}{{\textbf b}}
  \newcommand{\x}{{\textbf x}}
  \newcommand{\y}{{\textbf y}}
  \newcommand{\z}{{\textbf z}}
  
  \newcommand{\X}{\cX}
  \newcommand{\dwc}{\emph{d.w.c.}\xspace}
  
  \newcommand{\epsilonpd}{\epsilon_{p,\delta}}
  \newcommand{\Qpd}{Q_{p,\delta}}
  \newcommand{\Qpm}{Q_{p,m}}
  \newcommand{\Qpprimem}{Q_{p',m}}
  \newcommand{\Qppd}{Q_{p,\delta}}

  \newcommand{\barq}{{\bar{q}}}
  
  \newcommand{\dotF}{ {\dot F}}

\newif\ifmaybe
\maybefalse

\newif\ifvenkat
\venkatfalse

\newif\ifprasad

\newtheorem{Remark*}{Remark}

\ShortHeadings{Data-Derived Consistency}{Santhanam, Anantharam and
  Szpankowski}
\usepackage{lastpage}
\jmlrheading{23}{2022}{1-\pageref{LastPage}}{6/20; Revised 12/21}{12/21}{20-644}{Narayana Santhanam, Venkatachalam Anantharam, and Wojciech Szpankowski}
  
\begin{document}
\title{Data-Derived Weak Universal Consistency}
\author{\name Narayana Santhanam  \email nsanthan@hawaii.edu\\
  \addr Department of Electrical Engineering,\\
  University of Hawaii, Manoa\\
  Honolulu, HI 96822, USA \AND
  \name Venkat Anantharam   \email ananth@eecs.berkeley.edu \\
  \addr Department of Electrical Engineering and Computer Science,\\
  University of California, Berkeley \\
  Berkeley, CA 94720, USA\\
  \AND
  \name Wojciech Szpankowski \email szpan@purdue.edu\\
  \addr Department of Computer Science,\\
  Purdue University\\
  W. Lafayette, IN 47907, USA}

\editor{Nicholas Vayatis}
\maketitle 
  \begin{abstract}%
    Many current applications in data science need rich model classes to
    adequately represent the statistics that may be driving the observations.
    Such rich model classes may be too complex to admit uniformly
    consistent estimators.
    In such cases, it is conventional to settle for
    estimators with guarantees on convergence rate where the
    performance can be bounded in a model-dependent way,
    i.e. pointwise consistent estimators.  But this viewpoint has the
    practical drawback that estimator performance is a function of the
    unknown model within the model class that is being estimated.
    Even if an estimator is consistent, how well
    it is doing at any given time may not be clear, no matter what the
    sample size of the observations.

    In these cases, a line of analysis favors sample dependent
    guarantees.  We explore this framework by studying
    rich model classes that may only admit pointwise consistency
    guarantees,
    yet enough information about the unknown model driving the
    observations needed to gauge estimator accuracy can be inferred
    from the sample at hand.
    In this paper
    we obtain a novel characterization of lossless compression
    problems over a countable alphabet in
    the data-derived framework in terms of what we term
    \emph{deceptive} distributions. We also show that the
    ability to estimate the redundancy of compressing memoryless
    sources is equivalent to learning the underlying single-letter marginal
    in a data-derived fashion. 
    We expect that the methodology
    underlying such characterizations in a data-derived estimation
    framework will be broadly applicable to a wide range of estimation
    problems, enabling a more systematic approach to data-derived
    guarantees.
  \end{abstract}
\section{Introduction and Motivation}


Many of the most challenging problems in the data sciences stem from
one or more of the following characteristics associated with data:
high dimensionality; extreme scale (typically requiring that the data
reside on multiple storage nodes); sparsity;
patterns in the data that manifest at multiple scales; dynamic,
temporal, and heterogeneous structure; complex dependencies between
different parts of the data; and noise/ missing data.  Tasks such as
image recognition, classification, control, and many others, which are
built on such data sources, depend on estimating the relevant
underlying structure in the data. Rich model classes, i.e. rich
collections of probabilistic models, such as the collection of all
probability distributions over a large or countably infinite support,
or the set of long memory, slowly mixing Markov processes are often
required to adequately model the complex characteristics of these data
sources.

  Indeed, 
  in bringing rigorous theory to bear on data science, an important
  question we face is related to model selection.  There is often a
  tension between the need for rich model classes to better represent
  data and our ability to handle these collections from a mathematical
  point of view.
  The richness of a model class is often quantified by metrics such as its
  VC-dimension~\citep{bis06}, Rademacher
  complexity~\citep{Kol01,BBL02,BM02}, or -- what is most relevant 
  in the context of universal compression -- its asymptotic per-symbol
  redundancy~\citep{Sht87,Fit72,kt81,ris84,brb98,ds04,SzW12}. The
  traditional uniform consistency paradigm
  would want 
  an estimation algorithm with a model-agnostic guarantee on its
  performance, depending only on the sample size.

  Many applications, particularly in the big data
  regime, force us to consider model collections that are too complex
  to admit estimators with traditional model-agnostic {\it uniformly}
  consistent guarantees.
  When the model classes we are interested in are too complex to admit
  uniformly consistent estimators,
  the common belief is that the best we
  can do is to have estimators with convergence guarantees dependent
  on not just the sample size but also on the underlying model in the model
  class that governs the statistics of the observations.  These are
  \emph{pointwise consistent} estimators~(\cite[see][]{Dav73} in the context
  of universal compression).
  This is often difficult (and as we will see, sometimes impossible)
  to use predictively as one cannot necessarily verify if the estimator
  has converged till the underlying model, the very quantity being
  estimated, becomes known. 

  To tackle these rich classes, several approaches consider obtaining
  guarantees that hold samplewise, for example, bounds from the
  PAC-Bayes approaches~(\cite{McAll99,Cat07}) for rich classification
  tasks, data-dependent structural risk minimization~(\eg~\cite{BS}) as
  well as its development via the \emph{luckiness}
  framework~(\cite{mdlbook}), or as in~\cite{APS14} for slow mixing
  Markov setups. We adopt the same philosophy---we express any
  estimator accuracy or confidence using empirically observed
  quantities. Our notion of data-derived consistency is also closely
  related to other formulations in compression, statistics and
  learning theory.  In particular, we note hierarchical universal
  compression in~\cite{MF98} and the more general framework of making
  finitely many errors along the lines
  of~\cite{Cov73,DP94,KT94}. We have approached this angle
  under the framework of regularization
  in~\cite{WS21:alt,WS21:aistats}. To get a flavor of the results in
  this line of work, for example,~\citet{Cov73} asks 
  whether one can estimate the rationality of the mean of a Bernoulli
  process in finitely many samples, showing that the answer is
  affirmative if the mean comes from a Baire first category set with
  Lebesgue measure 1 and that also contains every rational number,
  see~\cite{koplowitz1995cover,WS21:aistats} for extensions.
  
  Fundamental to all
  these approaches is to balance the sample complexity of learning
  with the desire for richer model collections (or hypothesis collections as the
  case may be).

  This paper builds a natural information theoretic framework in the
  ambit of this philosophy: however we choose to obtain the 
  data-derived bounds, when can they be made strong enough to answer
  convergence questions with arbitrary pre-specified confidence? Or
  equivalently, when is the data a sufficient statistic for the
  convergence rate of the estimator (or a non-trivial bound on it)? 
  To retain focus in understanding this \emph{data-derived}
  consistency, in this paper we concentrate on universal compression
  to bring out the salient features of this framework. We also make
  connections to a related prediction problem that was analyzed by us
  earlier in~\cite{SA12:jmlr}, and is now seen to fit into this
  broader framework. We note also that universal compression of \iid data
  is equivalent to finding the marginal from samples with data-derived
  guarantees.


  
  We illustrate the salient aspect of the data-derived setup we
  consider with a simple example below.

\bExample\label{eg:ent} \textbf{(Hiding entropy)}

For $\epsilon>0$ and
$M\in\naturals$, where $\naturals$ is the set of natural numbers, and let $p_{\epsilon, M}$ be the probability distribution
that assigns probability $1-\epsilon$ to the natural number $1$ and assigns
probability $\epsilon/M$ to the natural numbers $2$ through $M+1$. Denote the probability distribution
that assigns probability 1 to the natural number $1$ by $p_0$.
Let $\cW$ be the set comprised of the probability distributions $p_{\epsilon, M}$ for
$\epsilon > 0$
and $M\in\naturals$, as well as $p_0$.  

Our task is to estimate the Shannon entropy of a probability distribution in $\cW$
using \iid samples from it. 
However, we do not know which probability distribution in $\cW$ is governing the law of the observed samples.
The natural \emph{plug-in estimator}
assigns to a sample $X_1\upto X_n$ the entropy of its empirical distribution.
Since every probability distribution in $\cW$ has finite support, the plug-in
estimate is consistent almost surely, no matter which underlying
distribution from $\cW$ is generating the observations.
But at what point do we know that the plug-in estimate is close to the
correct answer? Indeed, can we, at any point, get an upper bound for the true entropy
using the plug-in estimate with, say, a confidence probability
$3/4$, regardless of what the true probability distribution in $\cW$ is?

It turns out that it is \emph{impossible} to provide
such guarantees for $\cW$.  
To see why, suppose we have a
sequence of $n$ successive $1$s. This could have come from $p_0$, or,
with high probability, from any probability distribution $p_{\epsilon,M}$ with
$0<\epsilon \ll \frac1n$.
%
What is worse, for any upper bound $\hat{h}$ we may provide, however large,
even if $0<\epsilon\ll \frac1n$, the entropy of $p_{\epsilon,M}$ where
$M \ge 2^{\hat{h}/\epsilon}$
is
$
 h(\epsilon)+\epsilon\log M \ge \hat{h}.  
$
Every such $p_{\epsilon,M}$
gives the sample of $n$ successive $1$s a probability of at least $>3/4$
if $\epsilon$ is sufficiently small,
so our upper bound fails.

This argument applies whether we obtained $\hat{h}$ from the plug-in
estimator or {\em any} other estimator of the entropy. No upper bound that we propose
on the entropy based
on any finite sequence of $1$s can hold with confidence probability
$3/4$ under all probability distributions in $\cW$.  To make matters worse, 
the sequence of all $1$s occurs with
probability $1$ when the underlying model in force is $p_0$. Therefore,
even when we could estimate the entropy consistently, we could never obtain
even a trivial upper bound on the entropy with a confidence probability $\ge 3/4$. 
\eExample


Universal compression posits that we have a model class
of source probability measures,
while we 
are required to 
come up with a universal probability measure that attempts to compress
any source in the model class
as well as possible
without prior knowledge of the
source.  Since the universal probability measure is not exactly matched to any
single source probability measure in the model class
it incurs a redundancy, measured using the Kullback-Leibler (KL) divergence, against any 
source in the model class
when compressing a sequence of observed samples 
whose statistics are governed by this source.
The uniform consistency
setup in this case corresponds to what is commonly known as the \emph{strong}
compression formulation,
where we find universal probability measures whose
per-symbol redundancy incurred against any source in 
the model class
can be
uniformly bounded over the entire 
model class and, in addition,
diminishes to 0 as the sample size grows to infinity.  The pointwise
consistency setup in this case corresponds to what is commonly known as
the \emph{weak} compression formulation and
is one where the universal probability measure incurs
asymptotically zero per-symbol redundancy against each source in the model class, but the
convergence to zero is not necessarily uniform over the 
entire model class.

The \emph{data-derived} weak compression formulation (\dwc) identifies
when, in the weak compression setup, we can also estimate from the
sample the redundancy of the universal probability measure relative
to the underlying source model generating the data.  Broadly speaking,
we aim to find a universal estimator/encoding with a given accuracy as
well as a corresponding stopping rule that allows us to find out at
what point the KL divergence from the true source becomes (and
remains) small from that point on. We also prove that this
characterization is completely equivalent to that of estimating, in a
data-derived fashion, a distribution $q$ over naturals that is within
a specified accuracy from the underlying marginal.\footnote{ We thank
  the anonymous reviewer for suggesting this comparison.}

To characterize the classes of probability distributions on
$\naturals$ that are data-derived weakly compressible, we shall
introduce the notion of what it means for a probability distribution
in the class to be {\it deceptive} relative to the class.
At a high level, a source probability distribution,
viewed as a member of a collection of probability distributions,
is {\it deceptive} if the 
asymptotic per-symbol redundancy of
neighborhoods of 
the source within the model class
is bounded away from $0$, in the limit as the
neighborhood shrinks to $0$.  
Then, in our main finding,
Theorem~\ref{thm:ncssff}, we show that {\it a collection of
  probability measures is data-derived weakly compressible iff no 
  source in the model class
  is
  deceptive}.
As we delve deeper into this formulation, we will see that
data-derived consistency changes how we think of model classes.
It shifts the focus away from the global complexity of the model class to some
form of local complexity of each model within the model class, viewed as a member
of the model class. 


 \ignore{ In this paper, as we attempt to explain when the \dwc property is
  preserved under (finite or countable) unions, we encounter
  hierarchical compression as a natural allied problem. In the
  hierarchical framework discussed in depth in the last section of this
  paper, one considers a nested sequence of model classes. Given a
  sample at hand, we ask if it is possible to compress it universally,
  with a redundancy that corresponds to the smallest component model class the source
  belongs to. This allows for the fact that the union of the nested
  model classes on the whole may be rich with high redundancy, but when a
  source is in a component model class such that it can be compressed well with small
  redundancy as a member of that model class, we do so.}

  \ignore{ However, this elegant formulation has a fatal drawback. As
    observed by the authors and as we point out as well,} 
    \ignore{Of course,
  compressing sources hierarchically does not necessarily imply that we
  find out which component model class the model belongs to---in other words, the
  component model classes may not be distinguishable. This property turns out to be
  the crux to understanding when the \dwc property is preserved under
  finite and countable unions.}
  
  \ignore{In the strong compression case, where this is usually studied, the
  redundancy of compression is upper bounded by essentially the sub-collection
  redundancy asymptotically and a term that depends on the weight to the
  subcollection by a prior---again both terms are unknown at this point.}

  \ignore{As it turns out, distinguishability is key to figuring out what the
  underlying redundancy is in the strong compression setup~(\cite{MF98}).
  The \dwc setup fundamentally tells us when we can tell what the
  redundancy is from samples in any one collection. A data-derived version of
  distinguishability becomes becomes the key to retain the \dwc property
  when we consider unions of \dwc collections.  We also answer the converse
  question---when can we split a given model collection into subcollections such
  that we know the identity of the underlying subcollection the model belongs
  to?}

The paper is organized as follows. In the next section we develop our
data-derived approach. Section~\ref{sec:back} recalls some of the
central prior results on universal compression that we build on in our
work.  Section~\ref{s:characterization} discusses our main result
(Theorem~\ref{thm:ncssff}), which completely characterizes \dwc model
classes of \iid probability distributions on a countable
set. 
Theorem~\ref{thm:equiv} and Appendix~\ref{app:equiv} contain an
equivalent formulation, that of estimating in a data-derived
fashion a distribution $q$ over naturals that is within a specified
KL divergence from the underlying marginal. 
We then illustrate several nuances in our
formulation and results using several examples in Section~\ref{s:ex}.
Sections~\ref{s:ncs} and~\ref{s:sff} are devoted to proving the main
result.
  \ignore{Finally, we consider unions of \dwc model classes in
    Section~\ref{s:hc} and show how the notion of hierarchical
    compression naturally plays a major role in understanding when the
    \dwc property is preserved under finite or countable unions.}  The
  main thread of the discussion is supported by several appendices.
  Appendix~\ref{app:basics} reconciles the traditional definitions of
  strong and weak compressibility with those we work with in this
  paper.  Appendix~\ref{app:redbasics} gathers several basic results
  on entropy and redundancy that we draw upon throughout the
  paper. Appendix~\ref{app:equiv}, as mentioned, proves an operational equivalence
  between our notion of data-derived compressibility and a natural
  definition of {\em learnability} of a class of probability distributions
  (Definition~\ref{dfn:lrn})\footnote{We are grateful to the anonymous reviewer
    for observing and suggesting one direction of this useful connection.}.
  Appendix~\ref{app:dn} contains the details of the proof for the
  claims made regarding one of the examples in Section~\ref{s:ex}.
  Appendix~\ref{app:tail} proves a lemma needed for the proof the
  sufficiency part of the main theorem. The last bit of the proof of
  the necessity part of the main theorem is in Appendix~\ref{s:phione}
  and that of the sufficiency part in Appendix~\ref{s:phibt}. Finally,
  Appendix~\ref{s:fake} corrects an erroneous claim made in passing in
  the concluding remarks in~\cite{SA12:jmlr} (which does not in any
  way affect the rest of that paper), and in addition
  illustrates why, in general, finite unions of \dwc classes, while
  weakly universal, need not be \dwc.
  
\section{Formulation of the Problem}
  \label{s:frm}
  We consider 
  here the lossless compression problem for collections of large alphabet \iid sources.
  The main contribution of this work is to characterize when
  data-derived guarantees for estimation problems can be made
  sufficiently strong. The large alphabet \iid compression problem is
  the vehicle we have used to do this, but this framework leads to
  interesting developments in other problems as well. We compare with
  Example~\ref{eg:ins}, the problem of estimation of percentiles of
  the probability distribution defining the source -- this has been
  studied in depth in~\cite{SA12:jmlr}, and here we show that this
  estimation task also lies in the data-derived framework proposed in
  this document.  Another example is that of entropy estimation, see
  Example~\ref{eg:entest}, and which we have studied
  in~\cite{WS20:isit} from a related, almost-sure hypothesis testing
  framework.

  \subsection{Notations}
  Before embarking on the discussion, we introduce notational
  conventions adopted in the paper.  The symbol $:=$, and occasionally
  $=:$, is used to denote equality by definition.  We write $\log$ for
  logarithms to base $2$ and $\ln$ for logarithms to the natural base.

  \textbf{Strings, sets and types:} The set of natural numbers,
  denoted $\naturals$, is the set $\sets{1,2, \ldots}$, thought of as
  endowed with its usual $\sigma$-algebra comprised of all subsets of
  $\naturals$.  For $n \ge 1$, we use $\naturals^n$ to denote the set
  of strings of length $n$ of natural numbers, with the product
  $\sigma$-algebra. The set of infinite sequences of natural numbers
  is denoted $\naturals^\infty$, and is thought of as endowed with the
  corresponding product $\sigma$-algebra.  We will adopt the
  convention of thinking of a probability measure on $\naturals$ as
  defined by a distribution, which assigns a probability to each
  natural number.  A string of integers
  $(x_1, \ldots, x_n) \in \naturals^n$ will be denoted by $\x$, or by
  $x^n$
  when it seems important to emphasize the specific length of the
  string. The \emph{type} of a string of integers
  $\x := (x_1, \ldots, x_n) \in \naturals^n$ will refer to the pair
  $(n,t)$, where $n$ is the sequence length and $t$ its empirical
  distribution.
  
  $\naturals^*$ denotes the set of strings of naturals of finite
  length, including the empty string. For the purposes of this paper
  it suffices to think of $\naturals^*$ as a set with no additional
  structure.
  Similarly $\{0,1\}^*$ denotes the set of binary strings of finite
  length.  The notation $\{0,1\}^* \backslash \emptyset$ is used for
  the set of binary strings of finite length, excluding the empty
  string.  For $\bb \in \{0,1\}^* \backslash \emptyset$, the length of
  $\bb$ is denoted by $l(\bb)$.
 For $1 \le m \le n$ and strings
  $\y \in \naturals^m$ and $\x \in \naturals^n$, we write
  $\y \preceq \x$ to denote that $\y$ is a prefix of $\x$. We can also
  use this notation when $\y \in \naturals^m$ and
  $x \in \naturals^\infty$.  The length of a finite string
  $\x \in \naturals^n$ is denoted by $|\x|$.

  \textbf{Probability measures and distributions:} Let $\cP$ be a
  collection of probability distributions
  over 
  $\naturals$. Given $\cP$, we let
  $\cP^\infty$ denote the collection of probability
  measures 
  on $\naturals^\infty$ induced
  by \iid assignments from the individual probability distributions in $\cP$. 
 We will use the 
 term \emph{source} to denote either $p \in \cP$ or
$p^\infty \in \cP^\infty$ as appropriate. For notational simplicity and following the
convention in literature, we will also often drop the superscript in $p^\infty$
and use $p$ both for the probability distribution on $\naturals$ and the corresponding \iid
probability measure induced on 
$\naturals^\infty$. Further, for $n \ge 1$ and a string of natural
numbers $\x := (x_1, \ldots, x_n) =: x^n \in \naturals^n$, we will write $p(\x)$
or $p(x^n)$
for $\prod_{i=1}^n p(x_i)$. Here $p$ can be thought of as a simplified notation for 
the product probability measure $p^n$ on $\naturals^n$ corresponding to the probability distribution
$p$ on $\naturals$.

  For a probability measure $q$ on $\naturals^\infty$, given $n \ge 1$
  and a string $\x \in \naturals^n$, we write $q(\x)$ for the probability under $q$ of the set of strings in $\naturals^\infty$ whose prefix of length $n$ is $\x$. In effect, we are treating $\x$ as also denoting an event in $\naturals^\infty$. Note that, for $p \in \cP$, $n \ge 1$, and 
  $\x \in \naturals^n$, this notational convention is consistent 
  with the earlier conventions of writing $p$ for both $p^\infty \in \cP^\infty$
  and for the product probability measure on $\naturals^n$ corresponding to $p$.
  
  It is a standard fact that a probability measure $q$ on $\naturals^\infty$
  is completely specified by $q(\x)$ for all $\x \in \naturals^n$ for all $n \ge 1$, subject to the consistency conditions $q(\x) = \sum_{\y \in \naturals^m ~:~ \x \preceq \y} q(\y)$ for all $1 \le n \le m$ and $\x \in \naturals^n$.
    
  We write $\mathbbm{1}(A)$ to denote the indicator of an event $A$.

  It is convenient to state some of the supporting results in this
  document at a level of generality where the underlying set is a
  countable set, in which case we denote such a set by $\cX$.  Also,
  we will state some results that apply to arbitrary collections of
  probability measures on $\naturals^\infty$, i.e. not necessarily of
  the form $\cP^\infty$ for some collection of probability
  distributions $\cP$ on $\naturals$. In such cases, we denote such a
  collection of probability measures on $\naturals^\infty$ by
  $\Lambda$.
  
  If $q$ and $r$ are arbitrary probability measures on $\naturals^\infty$, then
  \[
  D_n(q||r)
  :=
  E_{q}\log \frac{q(X^n)}{r(X^n)},
  \]
  denotes the KL divergence over length $n$ strings of $q$ with respect to $r$.
  If $p$ and $\tilde{p}$ are probability distributions on $\naturals$, then $D(p||\tilde{p})$ 
  denotes the KL divergence of $p$ with respect to $\tilde{p}$, which is $E_{p}\log \frac{p(X)}{\tilde{p}(X)}$.
  Note that, with our conventions, the expression $D_n(p||\tilde{p})$ is also well-defined, and can be 
  viewed as a shorthand notation for $D_n(p^\infty||\tilde{p}^\infty)$. 
  We thus have 
  $D_n(p||\tilde{p}) =  n D(p||\tilde{p})$ for all $n \in \naturals$,
  since $p^\infty$ and $\tilde{p}^\infty$ are \iid probability measures on $\naturals^\infty$.
  KL divergence is also called {\em relative entropy}.
  
  For probability distributions $p$ and $\tilde{p}$ on $\naturals$, their
  $\ell_1$ distance is
  \[
  ||p-\tilde{p}||_1 := \sum_{i\in\naturals} |p(i)-\tilde{p}(i)|.
  \]

 \subsection{Strong and  Weak Compressibility}

  In the lossless data compression problem for the collection of probability measures
  $\cP^\infty$ on $\naturals^\infty$ corresponding to a collection of probability 
  distributions $\cP$ on $\naturals$, our estimator
  is a probability measure $q$ 
  on $\naturals^\infty$.
  \footnote{
  It is not required that the probability
    measure $q$ 
    be a product measure.
    } 
    The problem formulation can be understood by thinking of the
    {\em loss} $L(p,q,\x)$ incurred by the estimator $q$ against a source $p$,
  given the length $n$ observation $\x \in \naturals^n$, as being the {\em excess codelength},
  \[
  L(p,q,\x):= \log\frac{p(\x)}{q(\x)}.
  \]
 The terminology is justified by thinking of $\log \frac{1}{p(\x)}$ as an indication of the length of the binary string one would want to use to represent $\x$ in an ideal prefix-free scheme for compressing strings of length $n$ from the source $p$ if one knew what $p$ was, and thinking of $\log \frac{1}{q(\x)}$
 as the length of the binary string one would be led to use for representing $\x$ in the prefix-free compression scheme suggested by the estimator $q$. 
 For more on this, see the discussion in Appendix~\ref{app:basics} on how strong and weak compressibility is typically defined in the literature.
 
 With this loss function in mind, we now make the following definitions.
 
 \bDefinition
\label{dfn:strongcomp}
Let $\cP$ be a collection of probability distributions on $\naturals$,
and $\cP^\infty$ the corresponding collection of probability
  measures
  on $\naturals^\infty$ induced
  by \iid assignments from the individual probability distributions in $\cP$. 
  Then $\cP^\infty$, or equivalently $\cP$, is called \emph{strongly
    compressible} if there is a probability measure
  $q$ on 
  $\naturals^\infty$
  satisfying
  \begin{equation}
  \label{eq:str_rdn}
  \limsup_{n\to\infty} 
  \sup_{p\in\cP^\infty}
  \frac1n E_p \log\frac{p(X^n)}{q(X^n)} =0.
  \end{equation}
\eDefinition

 The preceding definition may seem unusual relative to the definition of strong 
 compressibility that is traditionally encountered in the literature on data compression~\citep[see][]{Fit72,Dav73}. In Appendix~\ref{app:basics} we establish that it is identical to the traditional definition.

  
  Discussions of data compression in the literature are often framed in the language of {\em redundancy}. We formalize this notion in the following definition.
  
  \bDefinition
\label{dfn:terms}
  Let $\Lambda$ be any collection of probability measures
  on $\naturals^\infty$.
  The 
  {\em length-$n$ redundancy} of $\Lambda$ is defined to be
  \begin{equation}
  \label{eq:rdn}
  R_n(\Lambda) := \inf_q \sup_{r \in \Lambda}
   E_r \log\frac{r(X^n)}{q(X^n)},
  \end{equation}
  where the outer infimum is taken over all probability measures on $\naturals^\infty$, or equivalently over all probability measures on $\naturals^n$.
  The redundancy in the special case $n=1$ is called
  the \emph{single letter redundancy} of $\Lambda$, and $R_n(\Lambda)/n$ is called the 
  {\em per-symbol length-$n$ redundancy} of $\Lambda$. 
  The \emph{asymptotic per-symbol redundancy} of $\Lambda$ is
  $\limsup_{n \to \infty} R_n(\Lambda)/n$.
  
  More generally, given a probability measure $\hat{q}_n$ on $\naturals^n$ one can define the length-$n$ redundancy of $\Lambda$ with respect to $\hat{q}_n$ to be
   $\sup_{r \in \Lambda} E_r \log\frac{r(X^n)}{\hat{q}_n(X^n)}$ and similarly for the 
   per-symbol length-$n$ redundancy of $\Lambda$ with respect to $\hat{q}_n$.
   Given a probability measure $q$ on $\naturals^\infty$, one can define the asymptotic-per-symbol redundancy of $\Lambda$ with respect to $q$ to be 
   $\limsup_{n \to \infty} \frac1n \sup_{r \in \Lambda} E_r \log\frac{r(X^n)}{q(X^n)}$.
  
   Even more generally, given a probability measure $\hat{q}_n$ on $\naturals^n$ one can define the length-$n$ redundancy of $r \in \Lambda$ with respect to $\hat{q}_n$ to be
   $E_r \log\frac{r(X^n)}{\hat{q}_n(X^n)}$ and define the 
   per-symbol length-$n$ redundancy of $r \in \Lambda$ with respect to $\hat{q}_n$ similarly.
   Given a probability measure $q$ on $\naturals^\infty$, one can define the asymptotic-per-symbol redundancy of $r \in \Lambda$ with respect to $q$ to be 
   $\limsup_{n \to \infty} \frac1n E_r \log\frac{r(X^n)}{q(X^n)}$.
   
   When $\cP$ is a collection of probability distributions on $\naturals$,
and $\cP^\infty$ the corresponding collection of probability
  measures
  on $\naturals^\infty$ induced
  by \iid assignments from the individual probability distributions in $\cP$,
  we will talk about 
  each of the redundancy quantities as properties of $\cP$ when in fact they are
  defined for $\cP^\infty$.
  Similarly, given a probability measure $\hat{q}_n$ on $\naturals^n$ 
  or a probability measure $q$ on $\naturals^\infty$
  we will talk about each of the redundancy quantities for a given $p \in \cP$
  with respect to $\hat{q}_n$ or $q$ (as appropriate) when we mean the corresponding
  quantities for the $p^\infty \in \cP^\infty$ corresponding to $p$.
\eDefinition

  It is worth noting that a collection of probability distributions on $\naturals$
  is strongly compressible
  iff its asymptotic
  per-symbol redundancy is zero.
  For completeness, we give a proof of this claim in Lemma~\ref{lm:asyred} in 
  Appendix~\ref{app:basics}. We also observe that the asymptotic per-symbol redundancy 
  of a collection of probability measures $\Lambda$ on $\naturals^\infty$ can also
  be written as 
  \[
  \limsup_{n \to \infty} R_n(\Lambda)/n = \limsup_{n \to \infty} \frac1n \inf_q \sup_{r \in \Lambda}
   E_r \log\frac{r(X^n)}{q(X^n)} = \inf_q \limsup_{n \to \infty} \frac1n \sup_{r \in \Lambda}
   E_r \log\frac{r(X^n)}{q(X^n)},
  \]
  where the infimum on both sides of the equality is over probability measures $q$ on $\naturals^\infty$.
  Namely, the $\limsup_{n \to \infty}$ can be interchanged with the $\inf_q$. A proof of this
  is given in Lemma~\ref{lm:asyinter} in Appendix~\ref{app:redbasics}.

  We can allow for much richer collections of probability distributions if we work with a weaker
  notion of compressibility.
  
  \bDefinition
\label{dfn:weakcomp}
Let $\cP$ be a collection of probability distributions on $\naturals$,
and $\cP^\infty$ the collection of probability
  measures 
  on $\naturals^\infty$ induced
  by \iid assignments from the individual probability distributions in $\cP$. 
  Then
  $\cP^\infty$, or equivalently $\cP$,
  is called \emph{weakly compressible} if there exists a
  probability measure $q$ over 
  $\naturals^\infty$
  such that, for all $p\in\cP^\infty$ with finite entropy rate, we have
  \begin{equation}
  \label{eq:wk_rdn}
  \limsup_{n\to\infty} 
  \frac1n E_p \log\frac{p(X^n)}{q(X^n)} =0.
  \end{equation}
  \eDefinition
  
  One artifact of the above definition 
  is
  that any collection of probability distributions on $\naturals$
  where every source has
  infinite entropy is vacuously weakly compressible.
  In Appendix~\ref{app:basics} we establish that this definition of weak compressibility is identical to the definition of weak compressibility
  commonly encountered in the literature on data compression, see for example,~\cite{kie78}. 
  Also, in Lemma~\ref{lm:asyweakred} of Appendix~\ref{app:basics} we 
  formally establish the essentially tautological fact that 
  a collection of probability distributions $\cP$ on $\naturals$ is weakly compressible iff
  there exists a
  probability measure $q$ on 
  $\naturals^\infty$ such that 
  every $p \in \cP$ with finite entropy 
  has vanishing asymptotic per-symbol redundancy with respect to $q$.

\subsection{Compression in the Data-Derived Sense}

  Working with 
  collections of probability distributions on $\naturals$ that 
  are weakly compressible gives us a richer class of models than
  working with those that are strongly compressible.
  Weak compressibility of a collection $\cP$ of probability distributions
  on $\naturals$ ensures that there is a probability measure $q$ on 
  $\naturals^\infty$ such that
  $q$ is essentially as good an encoder as the
  underlying $p$ for long enough strings of natural numbers drawn \iid from $p$, where goodness is measured in terms of the number of
  bits used per symbol encoded. This is what it means to say that
  the asymptotic per-symbol redundancy of every $p^\infty \in \cP^\infty$ with 
  respect to $q$ is $0$,
  
  But observe that what one means by ``long enough'' depends on the
  unknown $p$, since convergence to the limit
  in~\eqref{eq:wk_rdn} need not be uniform over $p \in \cP$. The main contribution of our work
  is to come to grips with this issue without having to back off all the way to being able to deal only with
  strongly compressible collections of probability distributions. 
  
  \subsubsection{Stopping Rule} Our ideas are built around the 
  notion of a {\em universal stopping rule}, which we introduce next.
  Recall that a stopping rule is a function of observed strings
  where the decision to \emph{stop} or not at any given time is based
  only on what has been observed thus far. We formalize a stopping rule
  by a function $\tau$ from $\naturals^*$, the set of all finite strings
  of naturals, to the set $\sets{0,1}$,
  \[
  \tau:\naturals^*\to\sets{0,1}.
  \] 
  When $\tau$ assigns value 0 on a finite string $x^n$, possibly the
  empty string, it indicates that the stopping rule is still waiting
  after having observed $x^n$. A string $x^n$, possibly the empty
  string, is assigned 1 if the stopping rule has stopped on any prefix
  of $x^n$.  From a notational point of view, since $\tau$ quantifies
  a stopping rule, we will have for all strings $x^n$ with prefix
  $x^m$ that $\tau(x^n)\ge \tau(x^m)$. To align with the
  common definition of stopping \emph{time} $T$ defined on the
  standard filtration on $\sets{\naturals^n}_{n\ge 1}$, $\tau$ is a binary (0-1)
  process that assigns to $X^m$ a value 1 if $X^m\in \sets{T\le m}$,
  and 0 else.

  The stopping rule $\tau$ is required to be universal for $\cP$. In other words, the stopping rule
  cannot change depending on the unknown probabilistic model $p\in\cP$
  that is generating the observations. 
  In the formulation that we will develop in this paper, 
  given a threshold $\delta > 0$,
  a stopping rule (call it $\tau$ for now)
  will be based on 
  some fixed probability measure $q$ on $\naturals^\infty$,
  and will signify when the sequence length is ``long enough'' that the
  normalized KL divergence
  between the underlying source distribution and the probability measure $q$
  has fallen below $\delta$ and will remain below
  $\delta$ henceforth. 
  We will insist that $\tau$ stops at a
  finite time for all $p\in\cP$, \ie
  \begin{equation}
      \label{eq:taustops}
  p(
  \lim_{n\to \infty} \tau(X^n)=1)=1, \mbox{ for all $p \in \cP$}.
  \end{equation}
  We will include the condition in \eqref{eq:taustops} in the concept of what we mean by a universal stopping rule.

  To understand this requirement better, fix a probability measure $q$ on $\naturals^\infty$, and for $p\in\cP$ let
  \[
  \cN_{p,\delta;q} := \sets{n: \frac1n E_p\log \frac{p(X^n)}{q(X^n)} > \delta}.
  \]
  Thus $\cN_{p,\delta;q}$ is the set of all lengths $n \ge 1$ such that the length-$n$ KL divergence
  of the \iid probability measure $p^\infty$ corresponding to $p$ with respect to the probability measure $q$ 
  is worse than the accuracy required.
  Now consider the set
  \[
  \cN_{\delta;q} := \union_{p\in\cP} \cN_{p,\delta;q}.
  \]
  In the
  trivial case where $\cN_{\delta;q}$ is a finite set, 
  let $N$ denote the largest element in $\cN_{\delta;q}$.
  Then,
  for all $n\ge N$, we have
  \[
  \sup_{p\in\cP} \frac1n E_p\log \frac{p(X^n)}{q(X^n)} \le \delta.
  \]
  Clearly we can choose the stopping rule to be 0 for all sequences with 
  length $n\le N$ and 1 for all sequences with length $> N$, and this is universal.

  \subsubsection{$\delta-$Premature Rules} It is more interesting when
  $\cN_{\delta;q}$ defined above is not a finite set. Even in this
  case, the stopping rule $\tau$ has to stop at a finite time almost
  surely no matter which source is governing the observations.
  Naturally, no matter when $\tau$ stops waiting, the sequence length
  may not be long enough for some sources in $\cP$, so $\tau$ fails on
  such sequences. More formally, for $\delta>0$, $\tau$ fails with
  respect to $q$ or is \emph{$\delta$-premature with respect to $q$}
  for a source $p\in\cP$ and at time $i$ if there is some string $x^i$
  such that
  \begin{equation}
  \label{eq:dprem}
  \tau(x_1^i)=1 \text{ and } \frac1{i} E_p \log\frac{p(X^{i})}{q(X^{i})} >\delta.
  \end{equation}
  For $p\in\cP$, consider the subset of $\naturals^\infty$ defined as
  \begin{equation}
      \label{eq:badset}
  \Sets{
  x_1^\infty \in \naturals^\infty:
  \exists\, i \text{ such that }
  \tau(x_1^i)=1 \text{ and } \frac1{i} \sum_{y^i\in\naturals^i} p(y^i)\log\frac{p(y^{i})}{q(y^{i})} >\delta}.
  \end{equation}
  For $p\in\cP$, 
  the above set 
  is the set of
  strings on which $\tau$ is $\delta-$premature with respect to $q$. 
  While this set depends on which $p \in \cP$
  is driving the observations,
  this set is an event in the product $\sigma$-algebra on $\naturals^\infty$
  whatever the underlying $p \in \cP$.
  To see this, note that it is a countable union of sets of the form 
  $\Sets{
  x \in \naturals^\infty:
  \tau(x^i)=1}$, $i \ge 1$ (which of the components sets lie in the union is determined, for the fixed probability measure $q$ on $\naturals^\infty$, by the underlying source probability distribution $p$).
  
  While
the set in \eqref{eq:badset}
  may not be an empty set, we can at least try to ensure
  that 
  its probability
  under $p$ is 
  small.
  This thought process leads to
  what we mean by a collection of probability distributions on
  $\naturals$ being weakly compressible in the
  data-derived sense,
  formalized below. 
  This is the central concept investigated in this paper.
  

  \bDefinition\label{dfn:dwc} 
  Let $\cP$ be a collection of probability distributions on $\naturals$ and $\cP^\infty$ the associated collection of probability measures on $\naturals^\infty$ got by \iid assignments from the individual distributions in $\cP$.
  We say that $\cP^\infty$, or equivalently $\cP$, is 
  {\em weakly compressible in the data-derived sense} 
  or
  {\em data-derived weakly compressible (\dwc)}
  if there is a 
  probability measure $q$ on $\naturals^\infty$
  such that,
  for any accuracy $\delta>0$ and confidence probability $0 < 1-\eta < 1$, there is a
  universal stopping rule $\tau_{\delta,\eta}$ with the property that,
  no matter what $p^\infty \in\cP^\infty$ is in force, we have
  \begin{eqnarray}
  \label{eq:dwcdef}
  && p(\tau_{\delta,\eta} \text{ is $\delta-$premature with respect to $q$ for $p$})\\
  &&~~~~~~~~ :=
  p(
  \exists\, i \text{ such that }
  \tau_{\delta,\eta}(X^i)=1 \text{ and } \frac1{i} \sum_{y^i\in\naturals^i} p(y^i)\log\frac{p(y^{i})}{q(y^{i})} >\delta) < \eta, \nonumber
  \end{eqnarray}
  where in the second statement the random variables $X_i$ are
  generated \iid $p$.
  \eDefinition

  While the above definition recalls the
  compression/information-theoretic angle of our problem, we also note
  that characterizing \dwc classes will be equivalent to
  characterizing when we can learn the underlying marginals of the
  generating distribution, with a certificate that assures us that the
  estimate is accurate. We state this formally in
  Section~\ref{s:oper}, Definitions~\ref{dfn:lrn} and
  Theorem~\ref{thm:equiv} as the operational interpretation of
  the above definition of \dwc classes.
  
  \bClaim \label{eg:sc}
  \textbf{(Strongly compressible implies \dwc)} 
  Suppose $\cP$ is a collection of probability distributions on $\naturals$ 
  that is strongly compressible,
  namely
  there exists a probability measure $q$ on $\naturals^\infty$ that satisfies~\eqref{eq:str_rdn}.
  It follows then that, for all $\delta>0$, the sets
  \[
  N_{\delta;q} :=\{n : \sup_{p\in\cP^\infty}
  \frac1n E_p \log\frac{p(X^n)}{q(X^n)} >\delta
  \}
  \]
  are finite.
  For any $\eta > 0$, suppose we set
  $
  \tau_{\delta,\eta}(x^i)
  =
  1$ if $i> \max N_{\delta;q}$ and 0 else, we obtain
  for all $p\in\cP^\infty$ that
  $p( \tau_{\delta,\eta} \text{ is $\delta-$premature with respect to $q$} ) =0$.
  Thus every strongly compressible collection of probability distributions on $\naturals$ is \dwc.
  \eClaim
  
  \bClaim \label{eg:dwcimplieswc}
  \textbf{(\dwc implies weakly compressible)} 
  Suppose $\cP$ is a collection of probability distributions on $\naturals$ 
  that is \dwc, as in Definition~\ref{dfn:dwc}. Let $q$ be a probability measure on $\naturals^\infty$ such that, for every accuracy $\delta >0$ and confidence probability $0 < 1- \eta < 1$ there is a universal stopping rule $\tau_{\delta,\eta}$ satisfying
  \eqref{eq:dwcdef} for every $p \in \cP$.
  Fix $p \in \cP$. From \eqref{eq:dwcdef} we conclude that, for all $i \ge 1$, we have
  \[
  p(\tau_{\delta,\eta}(X^i) = 1) \mathbbm{1}\Paren{\frac1{i} \sum_{y^i\in\naturals^i} p(y^i)\log\frac{p(y^{i})}{q(y^{i})} >\delta} < \eta.
  \]
  However, since the stopping rule $\tau_{\delta,\eta}$ is universal, it
  must satisfy \eqref{eq:taustops}, i.e. it stops eventually. Hence we have 
  \[
  \lim_{i \to \infty}p(\tau_{\delta,\eta}(X^i) = 1)= 1.
  \]
  From this, it follows that 
  \[
  \limsup_{i \to \infty} \frac1{i} \sum_{y^i\in\naturals^i} p(y^i)\log\frac{p(y^{i})}{q(y^{i})} \le \delta,
  \]
  (in fact, for this to hold, it suffices to have the condition in \eqref{eq:dwcdef} hold for some $0 <1- \eta < 1$ and not necessarily for all $\eta > 0$, for the given $\delta > 0$).
  Letting $\delta \to 0$, we see that the condition in \eqref{eq:wk_rdn} holds, for the given probability measure $q$ on $\naturals^\infty$, for all $p \in \cP$.
  This means, by definition, that $\cP$ is weakly compressible.
  \eClaim
  
  Claims~\ref{eg:sc} and~\ref{eg:dwcimplieswc} imply that 
  \[
  \mbox{ Strongly compressible} \subseteq \dwc \subseteq \mbox{ weakly compressible}.
  \]
  In Section~\ref{ssec:strict} we will see examples of model classes
  demonstrating that each of these inclusions is strict. Note also
  that the distinctions between the definitions hold when the
  underlying alphabet is infinite---for finite alphabets, all versions
  are equivalent.

  As can be seen from the preceding discussion, our formulation of
  \dwc model classes is aimed at addressing the most interesting case
  from a statistical modeling viewpoint, which is the case where
  $\cP^\infty$ is weakly compressible, but not strongly compressible.
  Typically, we need global constraints on the collection of sources
  that comprise a model class to render the model class strongly
  compressible -- for example, that the square root of the Fisher
  information be integrable over the model class for a class to be strongly
  compressible~(\cite{ris84}).  
  By contrast, as we will see, data-derived weak compressibility does
  not depend on controlling the entire class $\cP^\infty$, but
  requires only that local neighborhoods of each
  $p \in\cP$, viewed as a member of $\cP$, be simple.  Indeed, one of the main
  contributions of this paper is to obtain a condition that is both
  necessary and sufficient for an \iid collection $\cP^\infty$ to be
  \dwc.

  \subsection{Operational Characterization of Data-Derived Compressibility}
  \label{s:oper}

  We provide the following operational perspective for \dwc from a learning
  theoretic perspective.

  Let ${\mathbb P}(\naturals)$ be the set of all probability distributions
  on $\naturals$ and, as before, let $\cP\subset{\mathbb P}(\naturals)$
  be a collection of probability distributions on $\naturals$.
  Let $X_1,X_2,\ldots$ be \iid samples generated by an unknown
  $p\in \cP$ and let $\hat{q}_{X_1\upto X_n}$ be a distribution on
  $\naturals$ that is considered to be an estimate of the underlying
  distribution $p$ obtained using samples $X_1\upto X_n$. Abbreviating
  $\hat{q}_{X^n}$ by $\hat{q}$, the loss incurred by the
  estimate $\hat{q}$ is 
  the single-letter divergence
  $D_1(p||\hat{q})$.

  \bDefinition\label{dfn:lrn} $\cP$ is \emph{learnable} if 
  for all $\eta > 0$ and $\delta > 0$ 
  there is an estimator $\hat{q}:\naturals^*\to{\mathbb P}(\naturals)$ and
  a universal stopping rule $\tau_{\delta,\eta}$ for $\cP$ such that
  for all $p\in \cP$,
  \[
    p( \exists i \text{ s.t. $\tau_{\delta,\eta}(X^i)=1$
      and $D_1(p||\hat{q}_{X^i})>\delta$ } ) < \eta,
  \]
  where $X_1,X_2,\ldots$ above are generated \iid $p$.
  (Here the left hand side of the preceding equation will be abbreviated as 
  $p(\tau_{\delta,\eta} = 1 \mbox{ and } D_1(p||\hat{q})>\delta)$.) 
  \eDefinition
  
  \bTheorem\label{thm:equiv}
   $\cP$ is learnable iff $\cP$ is \dwc,
   \Proof Please see Appendix~\ref{app:equiv}. 
  \eTheorem

  \subsection{Other Examples of Data-Derived Problem Formulations} 
  To clarify that the ideas in our framework have the potential to
  apply much more broadly to estimation problems other than the
  lossless compression problem that we have focused on in this
  document, we highlight in this section data-derived formulations for
  two other estimation problems. The first is a prediction task
  from~\cite{SA12:jmlr}, which we call the \emph{insurance} problem,
  while the second is an entropy estimation task. In later sections,
  we will also make some comparisons between
  the insurance problem 
  and the universal lossless
  compression problem studied here.

  \bExample\label{eg:ins} \textbf{(Insurability)} 
  Suppose we have a collection $\cP^\infty$
  of \iid measures over 
  $\naturals^\infty$.
  Given a finite sample
   $(X_1, \ldots, X_n)$ 
   with \iid marginals 
   from an unknown 
   $p \in \cP$
   we want to estimate a
  finite upper bound on the next symbol $X_{n+1}$ in a data-derived
  sense. 
  If there are $p \in \cP$ with unbounded support then
  for any finite upper
  bound we propose there is a probability 
  under such $p$ 
  that it may not be valid.
  In our data-derived
  formulation, we therefore want to provide an estimated upper bound $\Phi(X_1^n)$,
  and a universal stopping rule $\tau$ that tells us from what point we should
  believe that our estimates $\Phi(X_1^n)$ are at least as big as $X_{n+1}$,
  while allowing for some probability of being wrong.

  Formally, given a confidence probability $0 < 1 - \eta < 1$, we seek to come up with a
  mapping $\Phi:\naturals^*\to\reals$ and a stopping rule
  $\tau$ such that, for all
  $p \in \cP$,
  we have
  \[
    p
    \Paren{
      \,\exists i\in\naturals
      \text{ such that }
      \Phi(X^i) < X_{i+1}
      \text{ and } 
      \tau(X^i)=1} 
    < \eta.
  \]
  If this is possible, we say that the model class 
  $\cP^\infty$ is \emph{insurable}. In prior work, in~\cite{SA12:jmlr}, 
  the collections $\cP^\infty$ that are insurable were completely characterized.
  See Corollary~\ref{corr:tight} and Corollary~\ref{corr:dwcins} for more details and connections with 
  the results developed in this document. \eExample

  \bExample\label{eg:entest} \textbf{(Entropy estimation)} 
   Let $\cP$ be a collection of probability distributions on $\naturals$.
  Given a finite
  sample $(X_1, \ldots, X_n)$ sampled \iid from an unknown $p\in\cP$, we want
  to provide a data-derived finite upper bound $\hat{H}$ on the entropy of $p$. 
  Formally, given a confidence probability $0<  1 - \eta < 1$, we would like to come up with a
  mapping $\hat{H}:\naturals^*\to\reals$ and a universal stopping rule
  $\tau$ such that, for all
  $p \in \cP$, we have
  \[
    p
    \Paren{
      \,\exists i\in\naturals
      \text{ such that }
      \hat{H}< H(p)
      \text{ and } 
      \tau(X^i)=1} 
    < \eta. \eqed
  \]
  While this remains open, we have worked on a related formulation
  in~\cite{WS20:isit} on entropy property testing---namely, given a set
  $A\subset\reals$, to determine whether $H(p)\in A$ or not. We show there that, under mild
  conditions on the underlying distribution, we can
  resolve the property testing problem in finitely many samples iff $A$ and
  $A^c$ are $F_\sigma-$separable, adding to a related line of work
  developed in~\cite{Cov73,DP94,KT94}
  \eExamplep

\section{Background}		\label{sec:back}
  This section highlights some interesting prior results on
  universal compression that will be used in this paper. 
  Readers can skip 
  the proofs in 
  this section if they are
  willing to take the results here at face value when they are referred to. We have collected in this section 
  the more interesting prior results we use. Other, more
  basic, prior results that we also use
  are collected in Appendix~\ref{app:redbasics}.

  \subsection{Weak Compression}
  \label{ssec:wc}
  
  Let $\cP$ be a collection of probability distribution on
  $\naturals$ and $\cP^\infty$ the collection of probability
  measures 
  on $\naturals^\infty$ induced
  by \iid assignments from the individual probability 
  distributions in $\cP$. 
  In Appendix~\ref{app:basics} we have demonstrated that 
  the notion of
  weak compressibility of $\cP^\infty$ in the sense of 
  Kieffer~(\cite{kie78}) is identical to the definition of weak compressibility of $\cP^\infty$ that we have made in 
  Definition~\ref{dfn:weakcomp}.

 
 The following lemma gives a useful characterization of weak compressibility.
 
 \bLemma
  \label{lem:kie}
  Let $\cP$ be a collection of probability distributions on $\naturals$
  and $\cP^\infty$ the associated set of \iid probability measures on $\naturals^\infty$.
  Then $\cP^\infty$ is
  weakly compressible iff there exists a distribution $q$
 on $\naturals$ such that for all $p\in\cP$ with finite
  entropy we have
  \begin{equation}		\label{eq:wcbase}
  \sum_{x\in\naturals} p(x) \log \frac{1}{q(x)} <\infty.
  \end{equation}
  \Proof 
  From~\cite[Theorem 1]{kie78}) we know that $\cP^\infty$ is
  weakly compressible iff there is a
  countable set $\cQ := \{q_1, q_2, \ldots \}$ of probability distributions on 
  $\naturals$
  such that for all $p\in\cP$ with finite entropy there is
  some $q_i\in\cQ$ satisfying
  \[
  \sum_{x\in\naturals} p(x) \log \frac{1}{q_i(x)} <\infty.
  \]
  Therefore, if there is a probability distribution $q$ on $\naturals$
  satisfying \eqref{eq:wcbase} for all $p \in \cP$,
  we can 
  immediately conclude that $\cP^\infty$ is
  weakly compressible. It remains to show the converse.
  
  To do this, suppose that $\cP^\infty$ is
  weakly compressible and let $\cQ$ be a choice of the countable set of 
  probability distributions on 
  $\naturals$
  guaranteed by~\cite[Theorem 1]{kie78}).
  Fix some enumeration of $\cQ$ as
  $\cQ=\sets{q_1,q_2,\ldots}$.

  Consider the probability distribution $q$ on $\naturals$ given by
  \[
  q(n) := \frac{\sum_{i=1}^{|\cQ|}\frac{q_i(n)}{i(i+1)}}{\sum_{j=1}^{|\cQ|}\frac1{j(j+1)}},~~ n \in \naturals,
  \]
  where the upper limit of the summation is understood to be $\infty$
  if $\cQ$ is countably infinite.
  Observe that, for all $i$ and for all $n$, we have
  \[
  q(n) \ge \frac{q_i(n)}{i(i+1)}.
  \]
  Therefore, for all $p\in \cP$ with finite entropy and all $q_i \in \cQ$, we have
  \[
  \sum_{x\in\naturals} p(x) \log \frac{1}{q(x)}
  \le
  \sum_{x\in\naturals} p(x) \log \frac{i(i+1)}{q_i(x)}.
  \]
  Since the right hand side of the preceding equation is finite for at least
  one $q_i \in \cQ$,
  this completes the proof.
  \eLemma

\subsection{Tightness, Percentiles and Relevance to Redundancy}

Let us recall the definition of {\em tightness} of a collection of probability 
distributions on $\naturals$.

\bDefinition 	\label{def:tightness}
A collection $\cP$ of probability distributions on $\naturals$ is
said to be tight if for every $\gamma>0$ there is a natural number $M_\gamma$ such
that
\[
  \sup_{p\in\cP} p(X > M_\gamma) < \gamma.
\]
\eDefinition
Informally, tightness can be expressed as saying that all percentiles of distributions in $\cP$ can be uniformly
bounded over $\cP$. Formally, to define percentiles, we will use the
{\em linearly interpolated cumulative distribution function} of a
probability distribution on $\naturals$, defined as follows.

  \bDefinition For a probability distribution $q$ on $\naturals$, the linearly
  interpolated cumulative distribution $\dotF_q(n)$ for 
  $n\in\naturals \cup \{0\}$ follows the standard definition of the cumulative distribution function, i.e.
  \begin{equation}
    \label{eq:naturals}
    \dotF_q(n) := F_q (n) = \prob(X \le n)
  \end{equation}
  where $X$ is a random variable distributed according to $q$. For
  $n \in \naturals \cup \{0\}$
  and a real
  number
   $ n \le x \le n+1$, however, we define
  \[
    \dotF_q(x) := (n+1-x) \dotF_q(n) + (x-n) \dotF_q(n+1).
  \]
  Note that
  $\dotF_q$ is a nondecreasing function with domain the nonnegative real numbers
  and range either $[0,1]$ or $[0,1)$.
  For $t \in [0,1)$, we define
  $\dotF_q^{-1}(t)$ 
  to be the right continuous inverse of $\dotF_q$, i.e.
  \[
  \dotF_q^{-1}(t) := \sup \{ x \ge 0 : \dotF_q(x) \le t \}.\eqed
  \]
  \eDefinitionp

  \bProposition
  \label{prop:asconv}
  For all distributions $p$ over $\naturals$, if $X_1, X_2, \ldots$
  are generated \iid $p$, and $t_n$ be the empirical distribution of
  $X_1,\cdots, X_n$, then for all $0 < \gamma \le 1$,
  \[
    \dotF_{t_n}^{-1}(1-\gamma) \to \dotF_p^{-1}(1-\gamma) \text{ a.s.}
  \]
  \Proof 
  For any probability distribution $q$ on $\naturals$, any $0 < \gamma \le 1$, and any positive real number $x > 0$
  that is not an integer (so $\lceil x \rceil - \lfloor x \rfloor = 1$), 
  we have 
  \[
  \dotF_{q}^{-1}(1 - \gamma) < x \Longleftrightarrow F_{q}(x) > 1-\gamma \Longleftrightarrow 
  (x - \lfloor x \rfloor) F_q( \lceil x \rceil) + (\lceil x \rceil - x) F_q(\lfloor x \rfloor) > 1 - \gamma.
  \]
  Also, for $M \in \naturals$, we have
  \[
  \dotF_{q}^{-1}(1 - \gamma) < M+1 \Longleftrightarrow F_{q}(M+1) > 1-\gamma.
  \]
  Hence the claim is a consequence of fact that for all integers $M$ we have 
  $F_{t_n}(M) \to F_{p}(M)$ a.s., which in turn follows from the strong law of large numbers.
  \eProposition
  
We now show that tightness of a collection of probability distributions on
$\naturals$ is implied 
by finiteness of
the single letter redundancy of the collection. The 
result we present is a well-known folk theorem,
see for
example~\cite[Lemma 4]{haussler1997general}.
Here we give an
elementary proof of this result.

\ignore{\[ \sum_{x\in\cX} p(x) \left|\log \frac{p(x)}{q(x)}\right| =
    \sum_{x\in\cX} p(x) \log \frac{p(x)}{q(x)} -2\sum_{x\in S}
    \frac{p(x)} \log \frac{p(x)}{q(x)} \le D(p||q) +2\frac{\log e}e.
  \]}

\ignore{\bRemark Note that both~\eqref{eq:str_rdn}
  and~\eqref{eq:wk_rdn} are usually phrased with encoders or
  distributions for length $n$ sequences. However, since we will be
  concerned mainly with the limits, we can use the simpler
  formulations above. See~\cite{S06:thesis} for a formal explanation
  of why these formulations are completely equivalent.  \eRemark}

\bLemma
  \label{lm:bndprc}
  Let $\cP$ be a collection of probability distributions on $\naturals$.
  If the single letter redundancy of $\cP$ is finite, then $\cP$ is tight.  
  \Proof
  We prove the contrapositive here. If $\cP$ is not tight then,
  for some $\epsilon>0$, there is a sequence of probability distributions
  $p_n$ in $\cP$, such that
  \[
    p_n(X > n) \ge \epsilon.
  \]
  For any probability distribution $q$ over $\naturals$ and any positive real number $R$, there is some 
  natural number $M$ such that
  $ q(X > M )< \epsilon/2^R$. 
  Thus we have
  \[
   D(p_M|| q) \ge p_M(X \le M)\log\frac{p_M(X \le M)}{q(X \le M)} + p_M(X > M)\log \frac{p_M(X > M)}{q(X > M)} \ge -\frac1e + \epsilon R.
  \]
  Noting that $R$ can be made arbitrarily large, we conclude that the redundancy
  of $\cP$ is infinite. 
  \eLemma

\subsection{Bounds on Redundancy}
  The following technical lemma is used in Example~\ref{eg:uniform} and in Example~\ref{eg:b}.
  Its roots go back to~\cite{MF98}.

  \bLemma 
  \label{lm:sr}
  Let $\cX$ be a countable set, and 
  $\cP$ be a collection of probability distributions on $\cX$.
  For $i$ ranging over the finite set of indices $\{1, \ldots, M\}$
  or over all indices $i \ge 1$, let
  $S_i\subset \cX$ be a subset of $\cX$, and assume that these sets are
  pairwise disjoint.
  Suppose that for each $i$ there exists $p_i \in\cP$ such
  that
  \[
  p_i(S_i)\ge \delta.
  \]
  Then, for all probability distributions $q$ on $\X$, we have
  \[
  \sup_{p \in\cP} D(p||q) \ge \delta\log (M)
  - 1,
  \]
  if the number of subsets in the collection is finite, equal to $M$, and 
  \[
  \sup_{p \in\cP} D(p||q) = \infty,
  \]
  if the number of subsets in the collection is infinite.
  \Proof This is a simplified formulation of
  the \emph{distinguishability} concept in~\cite{MF98}.  
  To prove the claim, note that for any $m$ at most equal
  to the number of subsets in the collection, we must have $q(S_i) \le 1/m$ for some
  $i$. For such a choice of $i$ we can write
  \begin{eqnarray*}
  D(p_i||q) &=& \sum_{x \in S_i} p_i(x) \log \frac{p_i(x)}{q(x)}
  + \sum_{x \in S_i^c} p_i(x) \log \frac{p_i(x)}{q(x)}\\
  &\stackrel{(a)}{\ge}& p_i(S_i) \log \frac{p_i(S_i)}{q(S_i)}
  + p_i(S_i^c) \log \frac{p_i(S_i^c)}{q(S_i^c)}\\
  &\ge& p_i(S_i) \log \frac{1}{q(S_i)} + p_i(S_i^c) \log \frac{1}{q(S_i^c)} -1\\
  &\ge& \delta \log m - 1,
  \end{eqnarray*}
  where step (a) is from the log sum inequality.
  This completes the proof.
  \eLemma


\section{Characterization of \dwc Model Classes}		\label{s:characterization}

In this section we state our primary result, which is a necessary and sufficient
  condition for a model class comprised of a collection of probability distributions $\cP$ on 
  $\naturals$ to be data-derived weak compressible.

  We will see that what decides whether a model class $\cP$ is \dwc or not
  is a \emph{local} property of the probability distributions in $\cP$, viewed
  as members of $\cP$.
  Namely, the characterization of data-derived weak compressibility
  is based on considering a property of 
  local neighborhoods, as defined in Section~\ref{s:local}, of the individual probability distributions in the
  model class. Distributions having bad local
  neighborhoods are what we call \emph{deceptive} distributions,
  defined and studied 
  in detail
  in Section~\ref{s:deceptive}. The notion of deceptive distributions lies at the heart
  of our characterization, in Theorem~\ref{thm:ncssff}, of which 
  model classes are \dwc.

\subsection{Local Neighborhoods}
  \label{s:local}
  We will see in this section that 
  what makes the local neighborhoods of a probability distribution $p \in \cP$ 
  bad and kills \dwc is that when
  a stopping rule is forced by $p^\infty\in\cP^\infty$ into
  certifying the accuracy of the estimate at some time (which will
  have to be the case, since the stopping rule has to stop with
  probability $1$ under $p$), it will nevertheless be the case that there
  are other probability distributions in $\cP$, potentially arbitrarily close to $p$,
  which induce inadequate performance on the estimator.  \ignore{ As
    mentioned before, since $\cP^\infty$ contains \iid measures, the
    sources therein can be identified without confusion using their
    single letter marginals---we will use the same notation for the
    marginal and the probability measures where there is no ambiguity.
    The collection of the single letter marginals of $\cP^\infty$ will
    be denoted by $\cP$. We will refer to both the measure and the
    single letter distribution as a \emph{model}.}
    We now proceed to make this vague description of the underlying ideas precise.
 \bDefinition
\label{dfn:nbhds}
An $\epsilon-$\emph{neighborhood} of $p\in\cP$ is the set
  $\ngpe$ of all $p'\in\cP$ such that we have
  $||p-p'||_1<\epsilon$, where $||\cdot||_1$ denotes the $\ell_1$ distance. 
  \eDefinition

\subsection{Deceptive Distributions} 
  \label{s:deceptive}
  Data-derived compressibility of a collection $\cP^\infty$ is captured by how
  neighborhoods of measure in $\cP^\infty$ can be compressed. To formalize
  this, we define the notion of deceptive measures that have very complex
  neighborhoods.

  \bDefinition  
    \label{dfn:deceptive}
  $p^\infty\in\cP^\infty$ is said to be \emph{deceptive} if
  the 
  asymptotic per-symbol redundancy 
  of neighborhoods of $p$ is bounded away from 0
  in the limit as the neighborhood shrinks to 0. More precisely, we define
  $p^\infty \in \cP^\infty$, or equivalently
  $p \in \cP$, to be deceptive if
  \begin{equation}
  \label{eqn:dcp}
  \lim_{\epsilon\to 0} \inf_q \limsup_{n\to\infty} \sup_{p'\in\ngpe} \frac1n D_n(p'||q) >0.
  \end{equation}
  In the above, the infimum is over all $q$ that are probability measures on $\naturals^\infty$ 
  (not necessarily obtained by \iid assignments). 
  The verbal description of this condition in terms of
  the asymptotic per-symbol redundancy of the neighborhoods of $p$ is justified by 
  Lemma~\ref{lm:asyinter}, which is proved in Appendix~\ref{app:redbasics}.
  \eDefinition

Our main result is  the following Theorem~\ref{thm:ncssff}.
The necessity part of this theorem is proved
in Section~\ref{s:ncs} and the sufficiency part in Section~\ref{s:sff}.
  \bTheorem
  \label{thm:ncssff}
  Let $\cP$ be a collection of probability distributions on $\naturals$ and $\cP^\infty$ the
  associated collection of probability measures on $\naturals^\infty$ got by \iid assignments.
  Then
  $\cP^\infty$ is \dwc iff no $p\in\cP$ is deceptive.
  \eTheorem

 In the rest of this section we explore the concept of deceptive distributions to 
 flesh out a few properties of such distributions
  and their neighborhoods. This will help to better understand
  Definition~\eqref{eqn:dcp} and will set the stage for understanding the proof of Theorem~\ref{thm:ncssff}.

\subsubsection{A Simpler Characterization of Deceptive Distributions}
  In determining whether a source $p\in\cP$ is deceptive,~\eqref{eqn:dcp} allows
  us to choose $q$ depending on 
  $\epsilon$.
  We now show that this degree of
  freedom is unnecessary.

  \bLemma
  \label{lm:om}
  If $p\in\cP$ is not deceptive, then there is a single probability measure
  $q^*$ on $\naturals^\infty$ such that
  \[
  \lim_{\epsilon\to 0} \limsup_{n\to\infty} \sup_{p'\in\ngpe} \frac1n D_n(p'||q^*) =0.
  \] 
 On the other hand, we have that $p$ is deceptive iff
  \[
  \inf_q\lim_{\epsilon\to 0} \limsup_{n\to\infty} \sup_{p'\in\ngpe} \frac1n D_n(p'||q) >0.
  \]
  where the inf is over all probability measures $q$ on
  $\naturals^\infty$.
  \Proof
  Because $p$ is not deceptive, there exists a sequence 
  $(\delta_m > 0, m \ge 1)$, with $\lim_{m\to\infty}\delta_m\to 0$,
  and a
  sequence of probability measures $(q_m, m \ge 1)$ on $\naturals^\infty$ 
  such that, for all sufficiently large $m \ge 1$, we have
  \[
  \limsup_{n\to\infty} \sup_{p'\in\ngpom} \frac1n D_n(p'||q_m) \le \delta_m.
  \]
  Define the probability measure $q^*$ on $\naturals^\infty$ that,
  for each $n \ge 1$ and $\x \in \naturals^n$, assigns to the string $\x$ the probability
  \[
  q^*(\x) :=\sum_{m\ge1} \frac{q_m(\x)}{m(m+1)}. 
  \]
  For all $m \ge 1$, $n \ge 1$ and $p'\in\ngpom$,
  we have 
  \[
 \frac1n D_n(p'||q^*) \le \frac1n D_n(p'||q_m) + \frac{\log(m(m+1)}{n}.
  \]
  This implies that
  \[
  \limsup_{n\to\infty} \sup_{p'\in\ngpom} \frac1n D_n(p'||q^*) 
  \le
  \delta_m+\lim_{n\to\infty} \frac{\log\Paren{ m(m+1)}}{n}
  =\delta_m,
  \]
  and so
  \[
  \lim_{\epsilon\to 0} \limsup_{n\to\infty} \sup_{p'\in\ngpe} \frac1n D_n(p'||q^*) 
  =
  \lim_{m\to \infty} \limsup_{n\to\infty} \sup_{p'\in\ngpom} \frac1n D_n(p'||q^*)
  \le
  \lim_{m\to\infty} \delta_m
  =0.
  \]
  On the other hand, if $p$ is deceptive, then
  \[
    \inf_q\lim_{\epsilon\to 0} \limsup_{n\to\infty} \sup_{p'\in\ngpe} \frac1n D_n(p'||q)
    \ge
    \lim_{\epsilon\to 0}\inf_q \limsup_{n\to\infty} \sup_{p'\in\ngpe} \frac1n D_n(p'||q)
    >0.
  \]
  The converse follows from the first part of the Lemma.
  \eLemma

\subsubsection{Neighborhoods of Non-Deceptive Distributions are Tight}
  \label{s:tight}
  Recall the definition of \emph{tightness} of a collection
  of probability distributions on $\naturals$ 
  from Definition~\ref{def:tightness}.
  The following corollary is immediate.

  \bCorollary\label{corr:tight} 
  If $p\in\cP$ is not deceptive, then some neighborhood of $p$ is tight.
  \Proof If $p\in\cP$ is not deceptive then,
  for some $\epsilon>0$,
  there exists $n \ge 1$ and a probability measure $q$ on $\naturals^\infty$ such that
  \[
    \sup_{p'\in B(p,\epsilon)} D_n(p'||q) <\infty.
  \]
  From Proposition~\ref{prop:monotone} in Appendix~\ref{app:redbasics}, it follows that
  the single letter redundancy of the neighborhood
  $B(p,\epsilon)$ 
  is finite, which implies that
  $B(p,\epsilon)$ is tight, from Lemma~\ref{lm:bndprc}.  \eCorollary

The above corollary helps to make a connection between
two data-derived formulations -- \dwc, which is considered in this document,
and insurability, from
Example~\ref{eg:ins}. 
We showed in~\cite{SA12:jmlr} that a collection of \iid probability measures $\cP^\infty$ 
on $\naturals^\infty$ is insurable iff some
neighborhood, exactly as defined here, of every $p\in\cP$ is tight. 
We therefore obtain

\bCorollary \label{corr:dwcins} 
Let $\cP$ be a collection of probability distributions on $\naturals$ and 
let $\cP^\infty$ denote the associated collection of \iid probability measures on
$\naturals^\infty$. If $\cP^\infty$ is \dwc, then
$\cP^\infty$ is insurable.  \eCorollary

In both cases, note that the condition relies on some neighborhood within
the model class of
every model being simple. We expect this kind of locality to appear as a feature of
the characterization of which model classes admit data-derived estimators in most
data-derived formulations.

\section{Examples}
  \label{s:ex}

  \begingroup\makeatletter\ifx\SetFigFont\undefined%
      \gdef\SetFigFont#1#2#3#4#5{%
        \reset@font\fontsize{#1}{#2pt}%
        \fontfamily{#3}\fontseries{#4}\fontshape{#5}%
        \selectfont}%
      \fi\endgroup%
  \begin{figure}[!t]
  \begin{center}
    \scalebox{.5}{%
      \begin{picture}(0,0)%
        \includegraphics[bb=0 0 0 0]{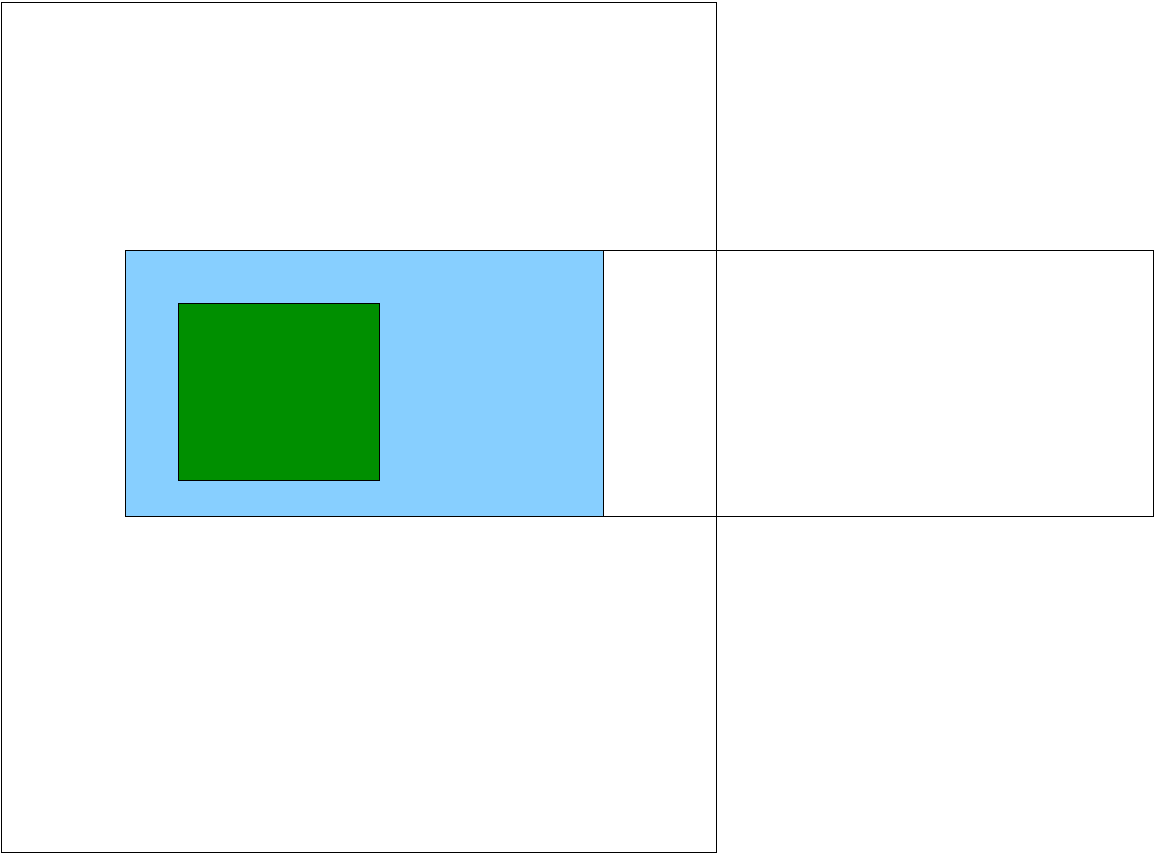}%
      \end{picture}%
      \setlength{\unitlength}{4144sp}%
      \begin{picture}(8811,6504)(3139,-7183)
        \put(6301,-3301){\makebox(0,0)[lb]{\smash{{\SetFigFont{20}{24.0}{\familydefault}{\mddefault}{\updefault}{\color[rgb]{0,0,0}{\Huge ${\cal U}^\infty$}}%
}}}}
\put(6301,-4111){\makebox(0,0)[lb]{\smash{{\SetFigFont{20}{24.0}{\familydefault}{\mddefault}{\updefault}{\color[rgb]{0,0,0}{\Huge ${\cal F}_h^\infty$}}%
}}}}
\put(9721,-3616){\makebox(0,0)[lb]{\smash{{\SetFigFont{14}{16.8}{\familydefault}{\mddefault}{\updefault}{\color[rgb]{0,0,0}{\Huge ${\cal I}^\infty$}}%
}}}}
\put(4996,-3616){\makebox(0,0)[lb]{\smash{{\SetFigFont{12}{14.4}{\familydefault}{\mddefault}{\updefault}{\color[rgb]{0,0,0}{\Huge ${\cal M}_h^\infty$}}%
}}}}
\put(10351,-3031){\makebox(0,0)[lb]{\smash{{\SetFigFont{20}{24.0}{\familydefault}{\mddefault}{\updefault}{\color[rgb]{0,0,0}Insurable}%
}}}}
\put(3826,-1186){\makebox(0,0)[lb]{\smash{{\SetFigFont{20}{24.0}{\familydefault}{\mddefault}{\updefault}{\color[rgb]{0,0,0}Weakly Compressible}%
}}}}
\put(4096,-5641){\makebox(0,0)[lb]{\smash{{\SetFigFont{12}{14.4}{\familydefault}{\mddefault}{\updefault}{\color[rgb]{0,0,0}{\Huge ${\cal N}^\infty$, ${\cal M}^\infty$}}%
}}}}
\put(7876,-3616){\makebox(0,0)[lb]{\smash{{\SetFigFont{14}{16.8}{\familydefault}{\mddefault}{\updefault}{\color[rgb]{0,0,0}{\Huge ${\cal B}^\infty$}}%
}}}}
\put(4726,-2896){\makebox(0,0)[lb]{\smash{{\SetFigFont{20}{24.0}{\familydefault}{\mddefault}{\updefault}{\color[rgb]{0,0,0}\dwc}%
}}}}
\end{picture}%
}
\end{center}
  \caption{Summary of examples: $\cM_h^\infty$ is strongly compressible
    (hence \dwc, insurable and weakly compressible), $\cU^\infty$ and
    $\cF_h^\infty$ are \dwc (hence insurable and weakly compressible),
    $\cB^\infty$ is weakly compressible and insurable but not \dwc,
    $\cN^\infty$ and $\cM^\infty$ are weakly compressible, but not
    insurable nor \dwc, while $\cI^\infty$ is insurable but not weakly
    compressible. Note that Corollary~\ref{corr:dwcins} shows that all
    \dwc collections are insurable, while Claim~\ref{eg:sc} and Claim~\ref{eg:dwcimplieswc} show that strong compressibility implies \dwc and that \dwc implies weak compressibility respectively.}
  \label{fig:comparisons}
  \end{figure}

  We now discuss a series of examples that highlight various aspects of
  our formulation. These examples also help flesh out the notion
  of what it means for a probability distribution to be deceptive.

  \subsection{Strongly Compressible $\subsetneq$ \dwc $\subsetneq$ Weakly Compressible}
  \label{ssec:strict}
  
  We first give examples showing that 
  weakly compressible collections of probability distribution on $\naturals$ 
  are a strictly richer class of models than \dwc collections.
  We also show that there are collections of probability distributions on $\naturals$ that are \dwc
  but are not strongly compressible.

\subsubsection{Weakly Compressible but Not \dwc} 

  We consider two examples in this category.
  

  A monotone probability distribution $p$ on $\naturals$ is one that
  satisfies $p(y) \ge p(y+1)$ for all $y \in \naturals$. 
  Let $\cM$ denote the collection of all monotone probability distributions on $\naturals$ and
  $\cM^{\infty}$ be the corresponding collection of
  \iid probability measures on $\naturals^\infty$.

\bExample
  \label{eg:mnt}
  \textbf{($\cM^{\infty}$ is weakly compressible but not \dwc.)}
  
  To see that $\cM^\infty$ is weakly compressible~(\cite{eli75})
  note that, for all $p\in \cM$ and all $n \in \naturals$, we have
  \[
  p(n)\le \frac1n.
  \]
  It follows that every $p\in\cM$ with finite entropy must satisfy
  \begin{equation}
        \label{eq:monotonefinite}
  \sum_{n\ge 1} p(n) \log n
  \le
  \sum_{n\ge 1} p(n) \log \frac1{p(n)} 
  <
  \infty.
  \end{equation}
  Now consider the probability distribution $q$ on $\naturals$ assigning
  probability $q(n)=\frac{6}{\pi^2 n^2}$ to $n \in \naturals$. From~\eqref{eq:monotonefinite}
  we see that, for all $p\in\cM$ with finite entropy, we have
  \[
  \sum_{n\ge 1} p(n) \log \frac1{q(n)} < \infty.
  \] 
  From Lemma~\ref{lem:kie} we conclude that 
  $\cM^\infty$ is weakly compressible.

  It turns out that all the probability distributions $p\in\cM$ are deceptive. To conclude this, we
  show that no neighborhood around any $p\in\cM$ is tight and then appeal to
  Corollary~\ref{corr:tight}.  
  This would then imply, by Theorem~\ref{thm:ncssff}, that $\cM^\infty$ is not \dwc.  
  In fact, it would have been enough to show that there exists some $p \in \cM$
  such that that no neighborhood of $p$ is tight.
  
  Let
  $\cU$ denote the collection of all uniform distributions over finite
  supports of form $\sets{m,m+1\upto M}$ where $m$ and $M$ are positive
  integers with $m \le M$.  \newcommand{\cme}{\cM(p,\epsilon)} For
  $p \in \cM$ and
  $\epsilon>0$, consider the collection
  \begin{equation}
     \label{eq:monotonethick}
  \cme := \sets{p': p' =(1-\alpha)p +\alpha q \text{ for } q\in\cU\cap\cM
  \text{ and } 0 \le \alpha < \epsilon}.
  \end{equation}
  In~\eqref{eq:monotonethick} $q$ can be any monotone uniform distribution, namely a
  uniform distribution with support $\sets{1\upto M}$ for some $M>0$.
  Clearly $\cme\subset\cM$.
  Note also that $\cme$ is a subset of an $\ell_1-$neighborhood of $p$
  corresponding to $\ell_1-$distance $2\epsilon$. 
  We will show that $\cme$ is not tight for all $p$ and all $\epsilon>0$.
  By the definition of neighborhoods in Definition~\ref{dfn:nbhds},
  it follows that no neighborhood of any $p\in\cM$
  is tight.
  
  For $0 < \alpha < \epsilon$, let
  $0<\delta <\alpha$ and $n \ge 1$. Observe
  that if the support $\{1, \ldots, M\}$ of a uniform distribution $q'\in\cU\cap\cM$
  satisfies $M \ge \frac{n}{1-\frac{\delta}{\alpha}}$, then we have
  \[
  q'\sets{j: j> n} =1-\frac{n}M \ge \frac\delta\alpha.
  \]
  Thus, given any $p \in \cM$,
  we have a distribution $p'=(1-\alpha)p+\alpha q' \in\cme$ that satisfies
  $p'\sets{j: j> n}\ge \delta$. 
  Therefore, $\cme$ is not tight. 
  This completes the argument.
  \eExample

  For our second example, we consider the set $\cN_1^{\infty}$ of all \iid probability measures on $\naturals^\infty$ 
 corresponding to
  the set of all probability distributions $p$ on $\naturals$ such
  that $\Exp_pX<\infty$, denoted $\cN_1$.

  \bExample
  \label{eg:fm} 
  \textbf{($\cN_1^{\infty}$ is weakly compressible but not \dwc.)}
  
  Note that every $p\in\cN_1$ has finite entropy.
  Also, by definition, all $p\in\cN_1$ satisfy $\sum_{i\ge1} i p_i <\infty$. Therefore
  the simplified version of Kieffer's condition for weak compressibility, 
  as stated in Lemma~\ref{lem:kie}, is satisfied by the distribution
  $q(i) :=1/2^i$ ($i\ge 1$). Thus we conclude that $\cN_1$ is weakly compressible. 

  We can show that every $p\in\cN_1$ is deceptive by showing
  that no neighborhood of any $p\in\cN_1$ is tight. The approach is
  similar to that in Example~\ref{eg:mnt}. Given $\epsilon > 0$, consider distributions of the
  form $p'=(1-\alpha)p +\alpha q$, where $q\in\cU$ is a uniform
  distribution over a support of the form
  $\{m, m+1, \ldots, M\}$, 
  and $0 < \alpha < \epsilon$. 
  Since $q$ has finite support, we have $p'\in\cN_1$.  

  As in Example~\ref{eg:mnt} we observe that (i) the $\ell_1$ distance between $p'$ and
  $q$ is 
  strictly less than 
  $2\epsilon$; (ii) 
  for all
  $0 < \delta <\alpha$ and $n \ge 1$, we can pick $q' \in\cU$, with $\cU$
  defined as in Example~\ref{eg:mnt}, whose
  support 
  satisfies
  $M \ge \frac{n}{1-\frac{\delta}{\alpha}}$,
  which then implies that the $(1-\delta)-$percentile
  of $p':= (1-\alpha)p +\alpha q'$ can be made to lie above $n$. 
  Since the above construction works for
  arbitrary $n \ge 1$ and in view of 
  the way in which neighborhoods are defined in Definition~\ref{dfn:nbhds},
  no neighborhood of any $p\in\cN_1$ is tight, which shows that every $p\in\cN_1$ is deceptive 
  and hence, by Theorem~\ref{thm:ncssff}, that $\cN_1$ cannot be \dwc. As in Example~\ref{eg:mnt},
  to apply Theorem~\ref{thm:ncssff} it would have been enough to show that there is at least
  one $p\in\cN_1$ which is deceptive.
  \eExample

\subsubsection{ \dwc but Not Strongly Compressible} 

  The example we consider in this category is $\cU$, which is defined
  in Example~\ref{eg:mnt}.
  Let $\cU^{\infty}$ denote the collection
  of all \iid probability measures on $\naturals^\infty$
  corresponding to $\cU$.

\bExample
  \label{eg:uniform} 
  \textbf{($\cU^\infty$ is not strongly compressible but is \dwc.)}
  
  We first show that $\cU$ has infinite single letter redundancy. To see this, we
  partition $\naturals$ into disjoint subsets $(T_i, i\ge0)$, where $T_i :=\sets{ 2^i\upto
    2^{i+1}-1}$.
For each $T_i$ there is an associated distribution
  $p_i\in\cU$ such that $p_i(T_i)=1$. Since the number of these disjoint sets
  $T_i$ is infinite, we conclude from Lemma~\ref{lm:sr} that 
  the single redundancy of
  $\cU$ is $\infty$. 
  
  From the second part of Proposition~\ref{prop:monotone} we can now conclude that
  the length-$n$ redundancy of $\cU$ is $\infty$ for all $n \ge 1$, so its 
  asymptotic per-symbol redundancy is also $\infty$, which means, by 
  Lemma~\ref{lm:asyred}, that $\cU$ is not strongly compressible.
  

  To see that $\cU$ is \dwc, note that around each probability distribution 
  $p \in \cU$ there is
  an $\ell_1$-neighborhood that contains no other probability distribution in
  $\cU$. Such a neighborhood has length-$n$ redundancy equal to 0 for
  all $n$ because the only possible distribution in the neighborhood is
  $p$. Hence the asymptotic per-symbol redundancy of all sufficient small 
  neighborhoods of each $p \in \cU$ is zero, which means, by definition,
  that each $p \in \cU$ is not deceptive, see Definition~\ref{dfn:deceptive}.
  \eExample

\subsubsection{Strongly Compressible and \dwc }

  For completeness we next give an example of a collection 
  of probability distributions on $\naturals$ which is strongly
  compressible, hence automatically \dwc. 
  \ignore{The example we pick is also
  helpful as a 
  prelude to Section~\ref{s:unions} where we bring out some of the nuances regarding
  unions of \dwc model classes.}

  For $h>0$, we consider the set $\cM_h\subset\cM$ of all
  monotone probability distributions on $\naturals$ 
 where the second moment
  of the self information satisfies the bound
  \[
  E_p \Paren{\log \frac1{p(X)}}^2 \le h.
  \]
  Let $\cM_h^\infty$ denote the set of all \iid probability measures on
  $\naturals^\infty$ corresponding to $\cM_h$.

\bExample
  \label{eg:mnth}
  \textbf{($\cM_h^\infty$ is strongly compressible, hence \dwc.)}

  Note that for any monotone probability distribution $p$ on $\naturals$ and all $i \ge 1$ 
  we have $p(i)\le 1/i$.
  Therefore for any $p\in\cM_h$, if $X$ is a random variable taking values in 
  $\naturals$ with the probability distribution $p$, we have
  \[
  E_p\log^2 (X) \le {E_p\log^2 \frac1{p(X)}}\le h.
  \]
  Therefore, for all $p\in\cM_h$, we have by the Cauchy-Schwartz
  inequality that $E_p \log X \le \sqrt{h}$.
  Now, for the probability distribution $q$ on $\naturals$ given by $q(i)=\frac1{i(i+1)}, i \ge 1$, we have
\[
  \sup_{p\in\cM_h} E_p\Paren{\ceil{\log \frac{1}{q(X)}}}^2
  \le
  \sup_{p\in\cM_h} E_p\Paren{\log (X^2+X) + 1 }^2
  \le
  \sup_{p\in\cM_h} E_p\Paren{2\log X +2 }^2
  \le
  4 (\sqrt{h}+1)^2,
\]
where the last inequality follows because, for all $p\in\cM_h$, we have
\[
  E_p\Paren{2\log(X) +2 }^2
  =
  4 E_p\Paren{\log^2(X) +2 \log X + 1}
  \le
  4 (h + 2\sqrt{h} +1) = 4 (\sqrt{h} +1 )^2.
\]
Therefore (see Appendix~\ref{app:dn} for a proof), 
we can construct a probability 
measure $q^*$ on $\naturals^\infty$ such that
  \[
  \sup_{p\in\cM_h^\infty}
  \frac1n 
  D_n(p||q^*) 
  \le 
  \frac{2h^{\frac{1}{4}}(\sqrt{h} + 1)}{\sqrt{\ln n}}
  +
  \pi\sqrt{\frac2{3n}} \log e.
  \]
 From this it follows that the collection $\cM_h^\infty$ is strongly compressible, and therefore
  \dwc trivially from Claim~\ref{eg:sc}.
  \eExample

Comparing Examples~\ref{eg:mnt} and~\ref{eg:mnth}, we observe, that
  countable unions of \dwc model classes need not be \dwc. In fact,
  as we will see in Example~\ref{eg:b}, even
  finite unions of \dwc model classes need not be \dwc. 
  \ignore{The problem of
  determining when \dwc model classes may be combined to yield another \dwc class 
  is a fascinating one, examined in more detail in Section~\ref{s:hc}.}

\subsection{\dwc Collections} 
Thus far, we have seen two \dwc
classes -- $\cU^\infty$ and $\cM_h^\infty$.  But neither is 
completely satisfying. In the collection $\cU$ above, there was a
neighborhood around each probability measure $p\in\cU$ with no other element of
$\cU$. Thus $\cU$ trivially satisfied the local
condition characterizing \dwc in Theorem~\ref{thm:ncssff}. The $\cM_h$ case falls into
another extreme -- the entire model collection $\cM_h$ is strongly
compressible, and therefore the condition characterizing \dwc in Theorem~\ref{thm:ncssff}
was again satisfied in a trivial way. 

We now therefore 
construct two additional examples
of \dwc model classes that are much more interesting. 
Our first example is of \dwc
model classes $\cF_h$, where neither of the two extreme situations mentioned above
holds. 
Our second example is of a \dwc model class $\cH$ with a source none of
whose neighborhoods are strongly compressible, but where the asymptotic
per-symbol redundancy diminishes to 0 as the neighborhood shrinks to the defining
probability distribution.

\subsubsection{More Interesting \dwc Model Classes} 

  For a probability distribution $p$ on $\naturals$ and a number $M>0$, define the probability measure
  \[
  p^{(M)}(n) :=
  \begin{cases}
  p(n-M) & n \ge M+1\\
  0 & \text{ else. }
  \end{cases}
  \] 
  Namely, $p^{(M)}$ shifts $p$ to the right by $M$. Furthermore, let the
  {\em span} of any probability distribution $p$ on
  $\naturals$ having finite support be defined to be 
  the largest natural number which has non-zero probability under $p$.
  
  For $h > 0$, we consider the model classes
  \[
  \cF_h := \Sets{ (1-\epsilon)p_1+\epsilon p_2^{(\text{\tiny{span}}(p_1)+1)} : p_1\in\cU, p_2\in\cM_h \text{ and } 0 < \epsilon < 1 }.
  \]
  As usual, let $\cF_h^\infty$ denote the set of \iid probability measures on $\naturals^\infty$ 
  associated to $\cF_h$.
Note that the initial uniform component of any $p \in \cF_h$ is uniquely determined.

  \bExample 
  \textbf{$\cF^\infty_h$ is \dwc.}
  

  \Proof 
  Let the {\em base} of any probability distribution over the naturals be the 
  smallest natural number which has non-zero probability.
  Consider any probability distribution
  $p=(1-\epsilon)p_1+\epsilon p_2^{(\text{\tiny{span}}(p_1)+1)} \in
  \cF_h$ with $p_1\in\cU$, $p_2\in\cM_h$, and $0 < \epsilon < 1$.
  Let $m$ denote base($p$) (which clearly equals base($p_1$)), 
  and let $m+M-1$ denote the span($p_1$), where $M \ge 1$. Thus $|\mbox{support}(p_1)| = M$.

  \renewcommand{\ceil}[1]{\left\lceil#1\right\rceil}
Consider any probability distribution
  $u\in\cF_h$,
  written as
  $u=(1-\epsilon')u_1+\epsilon' u_2^{(\text{\tiny{span}}(q_1)+1)}$,
  where $u_1\in\cU$, $u_2\in\cM_h$, and $0< \epsilon' <1$.
  Suppose that $u$ is 
  within $\ell_1$ distance 
  $\frac{(1-\epsilon)^2}{M(M+1)}$ from $p$.
  We show that
  \[
  |\text{span}(u_1)|\le m+\ceil{\frac{M}{1-\epsilon}}.
  \] 

  To see this, suppose to the contrary that 
  we have
  \[
  |\text{span}(u_1)|\ge  m+\ceil{\frac{M}{1-\epsilon}}+1.
  \] 
  If base($u_1$)$\le m$, all elements in the support of $p_1$ are
  assigned probability $\le\frac{1}{\frac{M}{1-\epsilon}+1}$ from
  $u$. If base($u_1$)$>m$, then $u$(base($p_1$))=0. 
  Thus, in either case, we have $u$(base($p_1$)) $\le\frac{1}{\frac{M}{1-\epsilon}+1}$.

  We can now
  lower bound the $\ell_1$ distance between $p$ and $u$ by
  \[
  \frac{(1-\epsilon)}{M}- 
  \frac{1}{\frac{M}{1-\epsilon}+1}
  =
  \frac{(1-\epsilon)^2}{M(M+1- \epsilon)}
  >
  \frac{(1-\epsilon)^2}{M(M+1)}.
  \]
  This contradiction proves the claim.

  Now, for fixed numbers $m'$ and $M'$, consider the collection
  $\cP_{m',M'}\subseteq \cF_h$ of all probability distributions with base $m'$, and
  whose support of the initial uniform component is $M'$.  Recall that $\cM_h$
  was shown to be strongly compressible in
  Example~\ref{eg:mnth}. Observe that the redundancy of $\cP_{m',M'}$
  will be at most the redundancy of $\cM_h$ plus 1. Therefore we must
  also have that $\cP_{m',M'}$ is strongly compressible.

  The set of all probability distributions in the $\ell_1-$neighborhood of $p\in\cF_h$ with
  radius $\frac{(1-\epsilon)^2}{M(M+1)}$ can be decomposed into the
  finite union
  \[
  \bigcup_{\substack{m', M'\\ m'+M'\le \ceil{m+\frac{M}{1-\epsilon}}}} \!\!\!\!\!\!\!\!\!\!\!\!\cP_{m',M'}.
  \]
  Each component of the finite union is strongly compressible. Therefore
  it follows that this neighborhood of $p\in\cF_h$ is strongly
  compressible. Thus no $p\in\cF_h$ is deceptive and the collection
  is \dwc.  \eExample

  \ignore{It suffices to show that there is an $\ell_1$ ball around $p$ of sufficiently 
  small radius, such that for all $\delta > 0$ we can find a uniform bound on the 
  $(1-\delta)$-th percentile of all $q$ in this ball.
  If $m' > m$, then the $\ell_1$ distance between $p$ and $q$ is at least
  $\frac{1 - \epsilon}{M}$. Hence, whenever the 
  $\ell_1$ distance between $p$ and $q$ is strictly less than 
  $\frac{1 - \epsilon}{M}$ we must have $m' \le m$. 
  Thus we may assume that $m' \le m$. 
  Suppose $m' + M' -1 \ge m + \frac{2M}{1-\epsilon}$.
  Then $\frac{M}{M'} \le \frac{1 - \epsilon}{2}$, from which, because
  $\mbox{support}(p_1) \subseteq \mbox{support}(q_1)$, we can conclude that 
  the $\ell_1$ distance between $q$ and $p$ is at least $\frac{1-\epsilon}{2}$.
  Thus we may assume that $m' + M' -1 < m + \frac{2M}{1-\epsilon}$.
  Now, for any $i \ge 0$, we have 
  \[
  q( m' + M' + i) = \epsilon' q_2(i) \le \epsilon' \frac{1}{i+1} \le \frac{1}{i+1}~.
  \]
  Thus for any $K \ge 0$ we have
  \[
  q(X > m + \frac{2M}{1-\epsilon} + K) \le q(X > m' + M' + K) \le \frac{h}{\log (K+1)}~
  \]
  by an argument similar to that in the preceding example, which gives the desired
  conclusion that no $p \in \cF^\infty_h$ is deceptive, and hence that 
  $\cF^\infty_h$ is insurable.}


  \newcommand{\pmkj}{p_{{}_{m,k,j}}}
  \newcommand{\plkj}{p_{{}_{l,k,j}}}
  
  We construct a \dwc collection $\cH$ where one of the probability distributions in $\cH$
  has no non-zero neighborhood that is also strongly compressible.

  We again partition $\naturals$ into $(T_i, i \ge 0)$ as before, where
  $T_i=\sets{2^i\upto 2^{i+1}-1}$ for $i \ge 0$. Let $\cH$ contain the probability 
  distribution
  $p_0$ that assigns probability $\frac1{(i+1)(i+2)}$
  to $2^i$ for all $i \ge 0$. We will 
  construct $\cH$ in such a way that while
  $p_0$ is not going to be
  deceptive in $\cH$, no neighborhood of $p_0$ in $\cH$ will be strongly compressible.

  We construct $\cH$ in several steps. We first fix a sequence
  $(\epsilon_m, m \ge 2)$ 
  such that 
  $0 <\epsilon_m < \frac{1}{2}$ 
  and
  \[
  \lim_{m\to\infty} \epsilon_m =0.
  \]
  Next, for $m\ge 2$, $k\ge m$,
  and 
  $j\in \Sets{2^k+1, \upto 2^k+2^{\lceil{k\epsilon_m}\rceil}}$,
  we define
  the probability 
  distribution
  \[
  \pmkj(r)
  :=
  \begin{cases}
  p_0(r), &\mbox{ if } 1\le r\le 2^{m-1}-1,\\
  \frac1m-\frac1{k+1}, & \mbox{ if } r=2^{m-1}+1,\\
  \frac1{k+1}, & \mbox{ if } r=j, \\
  0 ,& \text{ else.}
  \end{cases}
  \]
  Now, for $m \ge 2$ and $k \ge m$, let
  \[
  \cH_{m,k} := \Sets{\pmkj: 2^k+1\le j \le 
      2^k+2^{\lceil{k\epsilon_m}\rceil}},
  \]
  let 
  \[
  \cH_m :=\union_{k\ge m} \cH_{m,k},
  \]
  and, finally, let
  \[
   \cH := \sets{p_0}\cup \left( \union_{m\ge 2} \cH_m \right).
  \]

  A few observations about our construction.
  For all $m \ge 2$,
  all the probability distributions
  in $\cH_m$ assign probabilities 
  exactly as $p_0$ does
  to every element in $\union_{i=0}^{m-2} T_i$,
  and the rest of their support is disjoint from that of $p_0$. 
  It follows that, for all $m \ge 2$.
  for all ${p\in\cH_m}$, we have
  \[
  ||p-p_0||_1 =\frac2m.
  \]
  Hence, for all $m \ge 2$, the set of probability distributions in $\cH$
  within $\ell_1$
  distance 
  $\le \frac2m$ 
  from $p_0$ is precisely
  $\{p_0\} \cup (\union_{r \ge m} \cH_r)$.  
  Around any probability distribution in $\cH$ other
  than $p_0$, there is a non-zero neighborhood containing no other
  probability distribution that belongs to $\cH$. Therefore, none of the probability distributions in $\cH$ other
  than $p_0$ can possibly be deceptive. Hence, to show that $\cH$ is \dwc, we
  have to prove that $p_0$ is not deceptive.

\bExample
  \label{eg:h}
  \textbf{None of the neighborhoods of $p_0 \in \cH$ is strongly compressible.}
  
  We show that for all $m \ge 2$ the collection of probability distributions 
  $\cH_m$ is not strongly compressible, \ie its asymptotic
  per-symbol redundancy is bounded away from zero. 
  \ignore{To see this, we use
  Lemma~\ref{lm:sr} to show that for all $k$, the length-$(k+1)$ redundancy 
  of $\cH_{m,k}\subset \cH_m$ is bounded away from 0. }

  To see this, 
  for $2^k+1\le j \le 2^k+2^{\ceil{k\epsilon_m}}$, 
  let
  $S_j\subset\naturals^{k+1}$ be the set of all length-$(k+1)$ sequences all of
  whose symbols but one are from $\union_{i=0}^{m-1} T_i$, and there is
  exactly one occurrence of the number $j$ in the sequence. Clearly,
  for distinct $j$, $S_j$ are disjoint. Observe that 
  \[
  \pmkj(S_j)
  =\Paren{1-\frac1{k+1}}^{k}\ge \frac1e.
  \]
  \ignore{where the last inequality follows since $\Paren{1-\frac1{k+1}}^{k}$
  is decreasing as the integer $k$ increases, and has limit $\frac1e$
  as $k\to\infty$.}  
  Therefore, from Lemma~\ref{lm:sr}, we have that
  the length-$(k+1)$ redundancy of $\cH_{m,k}$, 
  which we denote by $R_{k+1}(\cH_{m,k})$, satisfies
  \[
    \frac{R_{k+1} (\cH_{m,k})}{k+1}
    \ge
    \frac1{k+1}\Paren{\frac{\log|\cH_{k,m}|}e -1}
    =
    \frac1{k+1}\Paren{\frac{\ceil{k\epsilon_m}}e -1}.
  \]
  Since for all $k\ge m \ge 2$ we have $\cH_{m,k}\subset \cH_m$, it follows 
  that for $m \ge 2$
  the length-$n$ redundancy of $\cH_m$, for $n\ge m+1$, which we denote by
  $R_n(\cH_m)$, satisfies
  \[
    \frac{R_n(\cH_m)}{n} \ge \frac{R_n(\cH_{m,n-1})}{n}
    \ge \frac1n \Paren{\frac{\ceil{(n-1)\epsilon_m}}e-1}.
  \]
  Hence, the asymptotic per-symbol redundancy of $\cH_m$ satisfies
  \begin{equation}
\label{eq:nc}
    \limsup_{n\to\infty} \frac{R_n(\cH_m)}n \ge \frac{\epsilon_m}e.
  \end{equation}
  Thus $\cH_m$ is not strongly compressible and, in particular, neither is any
  $\ell_1$ neighborhood of $p_0$.
 
  Nevertheless, we can show that 
  $p_0$ is not deceptive. We will verify that, as
  $m\to \infty$, the asymptotic per-symbol redundancy of 
  an $\ell_1$ neighborhood of
  radius $\frac{2(m+1)}{m^2}$ around $p_0$ goes to 0.
  \footnote{The choice of radius $\frac{2(m+1)}{m^2}$ is made
  since it satisfies $\frac{2}{m} < \frac{2(m+1)}{m^2} < \frac{2}{m-1}$
  for $m \ge 2$,
  and we defined $\ell_1$ neighborhoods to be open sets.}

  To do so, observe from Proposition~\ref{prop:rnn} that the
  asymptotic per-symbol redundancy of any collection of probability distributions on $\naturals$ is upper bounded by the
  single-letter redundancy of the collection. Recall that for $m \ge 2$
  the $\ell_1$ neighborhood of radius $\frac{2(m+1)}{m^2}$
  around $p_0$ is the collection 
  $\{p_0\} \cup (\union_{l\ge m} \cH_l)$. 
  We
  will verify that the single-letter redundancy of
  $\{p_0\} \cup (\union_{l \ge m} \cH_l)$ 
  diminishes to 0 as $m\to\infty$, which will
  then imply that $p_0$ is not deceptive,
  using
  Proposition~\ref{prop:rnn}.

  For $m \ge 2$, let $q_m$ be the probability distribution on $\naturals$ 
  defined by 
  \[
    q_m(r) :=
    \begin{cases}
      p_0(r), & \mbox{ if } 1\le r\le 2^{m-1}-1,\\
      \frac1m-\frac1{m+1}, & \mbox{ if } r=2^{m-1}+1,\\
      \frac1{(k+1)(k+2)} \frac{1}{2^{\ceil{k\epsilon_m}}}, & \mbox{ if } r \in 
      \Sets{2^k+1, \upto 2^{\ceil{k\epsilon_m}}}, k\ge m,\\
      0, & \text{ else. }
    \end{cases}
  \]
   
  Let $l \ge m \ge 2$.
  Then, for every $k \ge l$
  and 
  $j\in \Sets{2^k+1, \upto 2^k+2^{\lceil{k\epsilon_l}\rceil}}$,
  note that $\plkj \in \cH_{l,k}$ and $q_l$
  assign the same probabilities as those assigned by $p_0$ to
  every number $\le 2^{l-1}-1$. 
  It follows that
  \begin{align}		\label{eq:old17}
    D(\plkj||q_l) 
    &=
    \plkj(2^{l-1}+1)
      \log\frac
      {\plkj(2^{l-1}+1)}
      {q_l(2^{l-1}+1)}
      +
    \plkj(j)
      \log\frac
      {\plkj(j)}
      {q_l(j)} \nonumber\\
    &\le  
      \frac1l\log (l+1) 
     + \frac{1}{k+1} \log(k+2) + \frac{1}{k+1} \log 2^{\ceil{k \epsilon_l}} \nonumber\\
     &\le 
     \epsilon_l + \frac2l\log (l+1) + \frac{1}{l+1}.
  \end{align}
  
Now, for $m \ge 2$, consider the mixture probability distribution 
   $\barq_m$ on $\naturals$
  given by 
  \[
  \barq_m(r) :=
  \sum_{l\ge m} \frac{m}{l(l+1)} q_l(r).
  \]
  
  Fix $m \ge 2$. We have seen that any probability distribution 
  in $\cH$ in the
  $\ell_1$ neighborhood of radius $\frac{2(m+1)}{m^2}$ around
  $p_0$ must belong to $\{p_0\} \cup (\union_{l \ge m} \cH_l)$. 
  For every $k \ge l\ge m$, 
  and 
  $j\in \Sets{2^k+1, \upto 2^k+2^{\lceil{k\epsilon_l}\rceil}}$,
  we observe that $\plkj \in \cH_{l,k}$ and $\barq_m$
  assign the same probabilities as those assigned by $p_0$ to
  every number $\le 2^{m-1}-1$. Also, $p_0$ and 
  $\barq_m$
  assign the same probabilities as those assigned by $p_0$ to
  every number $\le 2^{m-1}-1$. We will now use this observation to 
  find upper bounds for $D(\pmkj||\barq_m)$ for 
  $k \ge m$ 
  and 
  $j\in \Sets{2^k+1, \upto 2^k+2^{\lceil{k\epsilon_m}\rceil}}$,
  then for $D(\plkj||\barq_m)$ for 
  $k \ge l \ge m+1$
  and 
  $j\in \Sets{2^k+1, \upto 2^k+2^{\lceil{k\epsilon_l}\rceil}}$,
  and finally for $D(p_0||\barq_m)$.
  
  For $k \ge m$ and $j\in \Sets{2^k+1, \upto 2^k+2^{\lceil{k\epsilon_m}\rceil}}$,
  we write 
  \begin{align}		\label{eq:mbound}
  D(\pmkj||\barq_m) 
    &=
    \pmkj(2^{m-1}+1)
      \log\frac
      {\pmkj(2^{m-1}+1)}
      {\barq_m(2^{m-1}+1)}
      +
    \pmkj(j)
      \log\frac
      {\pmkj(j)}
      {q_m(j)} \nonumber\\
      &\le 
      \pmkj(2^{m-1}+1)
      \log\frac
      {(m+1)\pmkj(2^{m-1}+1)}
      {q_m(2^{m-1}+1)}
      +
    \pmkj(j)
      \log\frac
      {(m+1)\pmkj(j)}
      {q_m(j)} \nonumber \\
      &\le 
     \epsilon_m + \frac4m\log (m+1) + \frac{1}{m+1},
  \end{align}
where the last step uses~\eqref{eq:old17} for the choice $l = m$.
 
  For $k \ge l \ge m+1$ and 
  $j\in \Sets{2^k+1, \upto 2^k+2^{\lceil{k\epsilon_l}\rceil}}$,
we write
  \begin{align}		\label{eq:abovembound}
    D(\plkj||\barq_m) 
    &= 
      \sum_{n={m-1}}^{l-2} 
      \plkj(2^n)\log\frac{\plkj(2^n)}{\barq_m(2^n)}
      +
      \sum_{r=2^{l-1}}^{\infty} 
      \plkj(r)\log\frac{\plkj(r)}{\barq_m(r)} \nonumber\\
    &\le 
      \sum_{n={m-1}}^{l-2} 
      \plkj(2^n)\log\frac{\plkj(2^n)}{\frac{m q_{n+2}(2^n)}{(n+2)(n+3)}}
      +
      \sum_{r=2^{l-1}}^{\infty} \plkj(r)\log\frac{\plkj(n)l(l+1)}{mq_l(n)} \nonumber\\
    &\le
      \sum_{n={m-1}}^{l-2} 
      \plkj(2^n)\log\frac{\plkj(2^n)}{\frac{q_{n+2}(2^n)}{(n+2)(n+3)}}
      +
      \sum_{r=2^{l-1}}^{\infty} \plkj(r)\log\frac{\plkj(r)}{q_l(r)} + \frac{\log(\frac{l(l+1)}m)}l \nonumber \\
      &=
      \sum_{n={m-1}}^{l-2} 
      \frac{\log\Paren{(n+2)(n+3)}}{(n+1)(n+2)}
      + \sum_{r=2^{l-1}}^{\infty} \plkj(r)\log\frac{\plkj(r)}{q_l(r)} + \frac{\log(\frac{l(l+1)}m)}l \nonumber \\
    &\ale{(a)}
      \sum_{n={m-1}}^{l-2} 
      \frac{\log\Paren{(n+2)(n+3)}}{(n+1)(n+2)}
      +
      \epsilon_l + 4 \frac{\log (l+1)}l + \frac1{l+1} \nonumber \\
      &\le 
      \sum_{n={m-1}}^{\infty} 
      \frac{\log\Paren{(n+2)(n+3)}}{(n+1)(n+2)}
        +      \epsilon_m +
      \frac{4\log (m+1)}{m} +\frac1{m+1},
  \end{align}
  where $(a)$ uses the bound $\log(l(l+1)/m) \le 2 \log(l+1)$, observes that
  $q_{n+2}(2^n)= p_0(2^n) =\plkj(2^n)$, and uses~\eqref{eq:old17}. 
  
   To bound $D(p_0||\barq_m)$ from above, note that 
   $\barq_m(2^n) = \frac{m}{n+2}p_0(2^n)$ for $n \ge m-1$.
   Therefore we have 
   \begin{align}		 \label{eq:0bound}
   D(p_0||\barq_m)
   &= 
      \sum_{n={m-1}}^{\infty} 
      p_0(2^n)\log\frac{p_0(2^n)}{\barq_m(2^n)} \nonumber\\
   &\le
   \sum_{n={m-1}}^{\infty} \frac{\log(n+1)}{(n+1)(n+2)}.
   \end{align}
   From~\eqref{eq:mbound},~\eqref{eq:abovembound}, and~\eqref{eq:0bound},
   the single letter redundancy of all sources
around $p_0$ within $\ell_1$ distance 
$\frac{2(m+1)}{m^2}$ of $p_0$ satisfies the upper bound
\begin{equation}		\label{eq:overallbound}
  \sup_{p\in \{p_0\} \cup (\union_{l\ge m} \cH_l) } D(p||\barq_m) 
  \le       
  \sum_{n={m-1}}^{\infty} 
  \frac{\log\Paren{(n+2)(n+3)}}{(n+1)(n+2)}
  +      
  \epsilon_m +\frac{4\log (m+2)}{m+1} +\frac1{m+1}.
\end{equation}

Note that
  \[
        \sum_{n=1}^{\infty} 
        \frac{\log\Paren{(n+2)(n+3)}}{(n+1)(n+2)} < \infty.
      \]
Hence, as $m\to \infty$, each of the terms on the right side
of~\eqref{eq:overallbound} converges to
0. Since the single letter redundancy of
$\{p_0\} \cup (\union_{l\ge m} \cH_l)$ diminishes to 0 as $m\to\infty$, from Proposition~\ref{prop:rnn},
the asymptotic per-symbol redundancy of 
$\{p_0\} \cup (\union_{l\ge m} \cH_l)$ also
diminishes to zero as $m\to\infty$. Therefore $p_0$ is not deceptive.

In conclusion, none of the neighborhoods of $p_0$ is strongly
compressible, from~\eqref{eq:nc}, since the asymptotic per-symbol
redundancy of a $\frac{2(m+1)}{m^2}$ size $\ell_1$ neighborhood of 
$p_0$ is lower
bounded by $\epsilon_m/e>0$. Yet, as we showed above, $p_0$ is
not deceptive. As noted above, no other probability distribution in $\cH$ can
possibly be deceptive since it has a neighborhood of nonzero radius around
it containing no other probability distribution from $\cH$. Therefore, $\cH$ is \dwc.
  \eExample
\ignore{  \begin{align*}
    D(\plkj||\barq_m) 
    &= 
      \sum_{r={m-1}}^{l-2} 
      \pmkj(2^r)\log\frac{\pmkj(2^r)}{q(2^r)}
      +
      \sum_{n=2^{l}}^{\infty} 
      \pmkj(n)\log\frac{\pmkj(n)}{q(n)}\\
    &\le 
      \sum_{r={m-1}}^{l-2} 
      \pmkj(2^r)\log\frac{\pmkj(2^r)}{\frac{q_{r+2}(2^r)}{(r+2)(r+3)}}
      +
      \sum_{n=2^{l}}^{\infty} \pmkj(n)\log\frac{\pmkj(n)l(l+1)}{q_l(n)}\\
    &\le
      \sum_{r={m-1}}^{l-2} 
      \frac{\log\Paren{(r+2)(r+3)}}{(r+1)(r+2)}
      +
      \frac{l \epsilon_l +3\log (l+2)}{l+1}\\
      &\le 
      \sum_{r={m-1}}^{\infty} 
      \frac{\log\Paren{(r+2)(r+3)}}{(r+1)(r+2)}
        +      \epsilon_m +
      \frac{3\log (m+2)}{m+1}.
\end{align*}
Therefore, we have that 
\[
  \sup_{p\in \union_{l\ge m} \cH_l} D(p||q) 
  \le       
  \sum_{r={m-1}}^{\infty} 
  \frac{\log\Paren{(r+2)(r+3)}}{(r+1)(r+2)}
  +      
  \epsilon_m +\frac{3\log (m+2)}{m+1}.
\]
as $m\to \infty$, all three terms on the right side above shrink to
0. Therefore, we have that the single letter redundancy of
$\union_{l\ge m} \cH_l$ diminishes to 0 as $m\to\infty$, and thereby
the asymptotic per-symbol redundancy of $\union_{l\ge m} \cH_l$ also
diminishes to zero as $m\to\infty$. Therefore $p_0$ is not deceptive.}

\ignore{
In conclusion, none of the neighborhoods of $p_0$ are strongly
compressible from~\eqref{eq:nc}, since the asymptotic per-symbol
redundancy of a $2/m$ size $\ell_1$ neighborhood of $p_0$ is lower
bounded by $\epsilon_m/e>0$. Yet, as we showed above, $p_0$ is
not deceptive. As noted above, no other probability measure in $\cH$ can
possibly be deceptive since they have a non-zero neighborhood around
them with no other probability measure from $\cH$. Therefore, $\cH$ is \dwc.}

  \ignore{Again using~\cite{HS15:all} (or grammar based codes from~(\cite{HY05})) with routine arguments, we infer that
  \[
  \lim_{m\to\infty}\lim_{n\to\infty} \inf_q\sup_{p\in\union_{k\ge m}\cH_k} \frac1n D_n(p||q) =0,
  \]}

\subsection{Non-\dwc Collections} 
We now construct two examples of
  non-\dwc model classes to illustrate some additional points.
  
  In Example~\ref{eg:b} we define a model class $\cB$ where exactly one source in
  the model class is deceptive.
  This would mean that $\cB$ is not \dwc.
  However, even though $\cB$ is not \dwc, removing the single
  deceptive source renders the rest of the model class \dwc. Put another
  way, adding a single source to a \dwc model class may make the resulting bigger model class not
  \dwc. Since a model class with one source is trivially \dwc, it follows
  that even finite unions of \dwc classes may not be \dwc.

  The second example we give here
  is of an insurable model class $\cI$
  that is not \dwc.
  See Example~\ref{eg:ins} for the definition of insurability of a model class.


  Partition $\naturals$ into $(T_i, i \ge 0)$, where $T_i :=\sets{ 2^i\upto 2^{i+1}-1}$, $i\ge 0$.
  For $0 < \epsilon < 1$, let $n_\epsilon=\ceil{\frac1\epsilon}$. Note that
  $\epsilon$ lies in the range $[\frac1{n_\epsilon}, \frac1{n_\epsilon-1})$.
  For $1\le j\le 2^{n_\epsilon}$, let $p_{\epsilon,j}$ be the probability distribution on
  $\naturals$ that assigns probability $1-\epsilon$ to the natural number 1 (or
  equivalently, to the set $T_0$), and $\epsilon$ to 
  the natural number
  $2^{n_\epsilon}+j-1$. Finally, let $p_0$ be a singleton
  probability distribution assigning probability 1 to the natural number 1. 

  Now, let $\cB$ (mnemonic for binary, since every probability distribution in $\cB$
  has support of cardinality at most $2$) be the collection of probability distributions
  on $\naturals$ defined by
  \[
    \cB := \sets{ p_{\epsilon,j}: 0 < \epsilon < 1,\, 1\le j\le
      2^{n_\epsilon}}\union \sets{p_o}.
  \] 
  As usual, $\cB^\infty$ denotes the set of \iid probability measures on $\naturals^\infty$
  corresponding to $\cB$.

 \bExample 
  \label{eg:b}
  \textbf{($p_0$ is the unique probability distribution in $\cB$ that is deceptive.)}
  
  An $\ell_1$ neighborhood of radius $\delta$ around $p_0$ is comprised of $p_0$
  and the 
  $p_{\epsilon,j}$ for all $0 < \epsilon < \delta/2$, and all
  $1\le j\le 2^{n_\epsilon}$. 
  For all $n \ge 1$ and $j\in \cT_n$,
  let $S_{n,j}$ denote the set of all length $n$ strings of natural numbers with exactly
  one appearance of $j$ and the remaining $n-1$ elements of the string being 1. 
  Then, we have 
  \[
    p_{\frac1n,j}(S_{n,j}) = \Paren{1-\frac1n}^{n-1}\ge \frac1e.
  \]
  For each $n \ge 1$, the sets $S_{n,j}$ are disjoint as $j$ ranges over $\cT_n$.
Further, they are subsets of $\naturals^n$.
  Therefore, Lemma~\ref{lm:sr} implies that the 
  length-$n$ redundancy
  of the collection $\sets{p_{\frac1n,j}: j\in \cT_n}$ is lower
  bounded by
  \[
    \frac{n}{e} - 1.
  \]
  Therefore, for all 
  $n > \frac2\delta$, 
  the length-$n$ redundancy of
  the $\ell_1$ neighborhood of radius $\delta$ is 
  bounded below by
  $\frac{n}{e} -1$.
  This implies that the asymptotic per-symbol redundancy
  of the $\ell_1$ neighborhood of size $\delta$ is 
bounded below by
  $\frac 1e$. From 
  the second part of Lemma~\ref{lm:om}, we conclude that
  $p_0$ is
  deceptive.

  On the other hand, for $0 < \epsilon < 1$, around every other
  probability distribution $p_{\epsilon,j}\in\cB$, there is an $\ell_1$-neighborhood of radius
  $\frac1{n_\epsilon}$
  that contains only probability distributions in $\cB$ that have
  support equal to $\sets{1, 2^{n_\epsilon}+j-1}$. 
  For $n \ge 1$, let $\hat{r}_n$ denote the probability measure on $\naturals^n$
  giving probability 
$\frac{1}{(n+1) {n \choose k}}$ 
  to each of the strings in $\naturals^n$
  comprised of $k$ occurrences of $2^{n_\epsilon}+j-1$
  and $n-k$ occurrences of $1$, $0 \le k \le n$. 
  Let $r_n$ be the probability measure corresponding to $\hat{r}_n$, as in 
  Lemma~\ref{lm:match}.
  Then, for all $p \in \cB$ in this $\ell_1$-neighborhood of $p_{\epsilon,j}\in\cB$, we have for all $n$
  \[
  D_n(p||r_n) \le \log(n+1).
  \]
  Noting that the measure $r$ on $\naturals^\infty$ that assigns probability
  \[
    r(\x) = \sum_{m\ge 1} \frac{r_m(\x)}{m(m+1)} 
  \]
  satisfies
  \[
    \limsup_{n\to \infty}
    \sup_{p:|p- p_{\epsilon,j}| < \frac1{n_\epsilon}}
    \frac1n D_n(p||q)
    \le \lim_{n\to \infty} \frac{\log n}{n} =0,
  \]
  we conclude that for every $p_{\epsilon,j}\in\cB$ there is an
  $\ell_1$-neighborhood of $p_{\epsilon,j}$ that has zero asymptotic
  per-symbol redundancy. Hence, 
  there is a
  neighborhood of $p_{\epsilon,j}$ that has zero asymptotic per-symbol
  redundancy.  We conclude that, while $p_0$ is deceptive, no other
  probability distribution in $\cB$ is deceptive.

  Indeed, this is quite intuitive when we think about what is involved
  operationally in compressing strings of integers whose statistics are 
  \iid and governed by a probability distribution in $\cB$.
  If at any point we see two distinct symbols in such a string, there
  is no ambiguity about what the underlying distribution is from that
  point on, and very little ambiguity in the probabilities of the two
  distinct symbols seen, of which one must be the symbol $1$.  But if
  we see a string of all $1$s we can never be sure (no matter what the
  length of the string) what the underlying source is. One possibility
  is that the source is $p_0$.

  But having seen a string of $1$s of length $m$, there is also a
  reasonable chance that the underlying source could be
  $p_{\epsilon,j}$ for some $\epsilon\ll \frac1m$ and any
  $j\in T_{n_{\epsilon}}$. There are $2^{n_\epsilon}$ such possible
  values $j$ can take in $T_{n_\epsilon}$, so any description of $j$
  requires an additional $n_\epsilon$ bits or $\gg m$ bits.

  However, if we remove $p_0$ from the collection, we have no such
  trouble. We have no obligation to stop on any finite length string of all $1$s,
  no matter how long it is, since the sequence of all $1$s has probability $0$ under
  every source in $\cB$ other than $p_0$.  \eExample


  The last example is a collection $\cI$ of probability measures over
  $\naturals$ that is insurable but not \dwc. In fact $\cI$ is not
  even weakly compressible.
  
  Partition $\naturals$ into the sets
  $(T_i, i\ge0)$ as before, where
  $T_i := \sets{2^i \upto 2^{i+1}-1}$.
  For each $i\ge 1$, pick exactly one element of $T_i$ and
  assign it probability $1/( i(i+1))$.
  We define $\cI$ to be
  the collection of all probability distributions on $\naturals$ that can be formed in 
  this way. 
  $\cI^\infty$ denotes the set of \iid probability
  measures on $\naturals^\infty$ corresponding to $\cI$.
  
  \bExample
  \textbf{($\cI$ is insurable but not weakly compressible, hence not \dwc)}
  
  For all $p\in\cI$ and all $k \ge 1$, we have
  \[
  \sum_{n\ge 2^k} p(n) = \frac1{k}.
  \]
  This means that the entire set $\cI$ is tight.
  By~\cite[Theorem 1]{SA12:jmlr}), we can therefore conclude that $\cI$ is insurable.


  On the other hand, for every
  probability distribution $q$ on $\naturals$, for all $i \ge 1$ there is
  $x_i\in T_i$ such that
  \[
  q(x_i) \le \frac1{2^i}.
  \]
  By the definition of $\cI$, 
  there is a probability distribution $p \in \cI$ 
  that has support
  $\sets{x_i:i\ge1}$. Note that $D(p||q) = \infty$.
  Since every probability distribution in $\cI$ has finite entropy (in fact
  they all have the same entropy),
  from Lemma~\ref{lem:kie} we conclude that $\cI$ is not weakly compressible.
  In particular, $\cI$ is not \dwc.
  \eExample

\section{Necessity Part of Theorem~\ref{thm:ncssff}}
  \label{s:ncs}
 
 In this section we prove the necessity part of Theorem~\ref{thm:ncssff}.
 Namely, we prove that the existence of deceptive distributions kills \dwc.
More precisely, we prove that if
$\cP$ is a collection of probability distributions on $\naturals$ and $\cP^\infty$
  the associated collection of \iid probability measures on $\naturals^\infty$,
  then
  $\cP^\infty$ is \dwc only if no $p\in\cP^\infty$ is deceptive.

  To prove this,
  suppose $p \in \cP$ is deceptive. Then, by the second part of
  Lemma~\ref{lm:om}, for every probability measure $q$ on $\naturals^\infty$
  we can find $\delta > 0$ such that 
  \[
    \lim_{\epsilon'\to 0} \limsup_{n\to\infty}
    \sup_{p'\in\ngpepp} \frac1n D_n(p'||q) > \delta.
  \]
  Pick any $0 < \eta < 1$, and let $\tau$ be a stopping rule. We will demonstrate that
  there is some $\tildep \in \cP$ such that
  \[
  \tildep( \tau\text{ is
    $\delta-$premature with respect to $q$ for $\tildep$})>\eta,
  \]
  where we refer to the discussion around~\eqref{eq:dprem} to
  recall what it means for a stopping rule to be $\delta-$premature for the 
  probability distribution $\tildep \in \cP$, with respect to the probability measure
  $q$ on $\naturals^\infty$.
  
  In order to do this, for all $n \ge 1$ let
  \[
  A_{n} := \sets{x^n\in\naturals^n: \tau(x^n)=1}
  \]
  denote the set of sequences of length $n$ on which $\tau$ has entered. Note 
  that $p(A_n)$ is increasing with $n$ and $\lim_{n \to \infty} p(A_n) = 1$.
  We can therefore pick $n \ge 4/(1- \eta)$ large enough such that
  $p(A_n) \ge (1 + \eta)/2$. 
  
  

 Let 
  \footnote{Please note that in the interest of simplicity, we have
    not attempted to provide the best scaling for $\epsilon$ or the
    tightest possible bounds.}  
   $\epsilon :=\frac1{2n^4}$.
    Applying Lemma~\ref{lm:jn} in Appendix~\ref{app:redbasics} to \iid probability measures over
  length-$n$ strings, we see that for all $\tildep\in\cP$ such
  that $||p-\tildep||_1\le\epsilon$, we have 
\[
   \tildep(A_n) > (1+\eta)/2-\frac2{n}\ge \eta,
 \]
 and for all $m\ge n$, since $A_m$ is an increasing sequence of events with $m$,
  \[
   \tildep(A_m) \ge \tildep(A_n).
 \]
 Since
 $\limsup_{m\to\infty} \sup_{p'\in\ngpepp} \frac1m D_m(p'||q)$
  is nondecreasing in $\epsilon'$ as $\epsilon'$
 increases, we can choose $\tildep\in\ngpe$ such that for some
 $m \ge n$ we have
  \[
  \tildep(A_{m}) > \eta  \text{ and } \frac1m D_m(\tildep|| q)>\delta.
  \]
  This in turn means, for the choice of $\eta$ and $\delta$ above, that
  $$
  \tildep(\tau\text{ is $\delta-$premature with respect to $q$ for $\tildep$})>\eta.
  $$
  This completes the proof of the necessity part of Theorem~\ref{thm:ncssff}.
  

   As a caveat regarding the structure of this proof, we remark that 
   the presence of a deceptive
  distribution $p \in \cP$ does not automatically imply that any other 
  probability distribution in any
  neighborhood of the deceptive distribution $p$ is also deceptive. For
  example, the class $\cB$ in Example~\ref{eg:b} has only $p_0$ deceptive,
  while no other distribution in its neighborhood is.
  \ignore{In particular, we will prove the sufficiency part of Theorem~\ref{thm:ncssff}
  in the next section, i.e. that if $\cP$ is not \dwc then there must be a deceptive 
  distribution in $\cP$.
  However, when $\cP$ is 
  is not \dwc, all we know is that $\cP$
  has at least one deceptive distribution. }
  \ignore{In certain
  cases, all such deceptive measures in the collection have zero Lebesgue
  measure, and removing such deceptive measures would make the collection
  \dwc. Example~\ref{eg:b} in the previous section illustrates such a case.}

\section{Sufficiency Part of Theorem~\ref{thm:ncssff}}
  \label{s:sff}
  
  In this section we prove the sufficiency part of
  Theorem~\ref{thm:ncssff}.  Namely, we prove that if a collection
  $\cP$ of probability distributions on $\naturals$ does not contain
  any deceptive distributions, then $\cP$ is \dwc.  We do this by
  explicitly constructing a probability measure $q^*$
  on $\naturals^\infty$ such that, given any desired confidence
  probability $0 < 1 -\eta < 1$ and accuracy $\delta >0$, there is a
  stopping rule $\tau$ such that, for every $p \in \cP$, under $p$,
  $\tau$ is $\delta-$premature with respect to $q^*$ for $p$, as
  defined in~\eqref{eq:dprem}, with probability at most $\eta$.

  Note that it suffices to prove this for all $\delta$ of the form
  $\frac{1}{m}$ for $m \ge 1$. So will restrict attention to this
  case, set $\delta=\frac1m$ for the rest of the proof, and denote the
  corresponding stopping rule we construct by $\tau_{\eta,m}$.

  \newcommand{\ngped}{B(p,\epsilon_{p,\delta};\cP)}
  \newcommand{\ngpepd}{B(p,\epsilon_{p,\delta}';\cP)}
  
  \newcommand{\ngpem}{B(p,\epsilon_{p,m};\cP)}

We proceed in three steps. Using the fact that $\cP$ does not have
  deceptive distributions, in Section~\ref{s:sff:cover} we
  cover $\cP$ by countably many 
  $\ell_1$
  neighborhoods, each of
  which has asymptotic per-symbol redundancy $<\frac1m$.
  In the second step,
  in Section~\ref{s:sff:q}, we construct a universal measure $q^*$
  of the kind desired
  by
  taking advantage of the 
  countable covering.

  In the third step, in Section~\ref{s:sff:thr}, we use the type of the
  sequence generated to estimate which of the neighborhoods from the
  first step the underlying source may be in. If we get the
  neighborhood right, note that in that neighborhood the 
  asymptotic per symbol
  redundancy
  is bounded by $\frac1m$
  uniformly over all sources in the
  neighborhood. This allows us to 
  get the compression rate down to
  the desired accuracy
  by pretending that the marginal distribution in force is the one
  determining the neighborhood, i.e. its centroid.
  
  Ideally, we would like to be able identify one of the neighborhoods
  from the first step that cover the underlying source (a ``good''
  neighborhood). This requires some care since different neighborhoods
  may have different sizes, and the rate of convergence of the
  empirical statistics to the source statistics is usually pointwise
  and not uniform. But when there are no deceptive distributions,
  given any confidence, a stopping rule can be constructed that can
  certify against prematurely deciding a bad neighborhood to the required
  confidence.
  
  \subsection{Covering $\cP$ by Countably Many Neighborhoods}
  \label{s:sff:cover}
  Using the fact that $\cP$ does not have deceptive distributions, we
  cover $\cP$ by countably many neighborhoods, each of
  which has asymptotic per-symbol redundancy $<\frac1m$.

  Suppose $p\in\cP$ is not deceptive. 
  From Lemma~\ref{lm:om}, there is a probability
  measure $q_p$ on $\naturals^\infty$ such that for all $m \ge 1$
  we can pick $\epsilon_{p,m} > 0$ satisfying
  \begin{equation}		\label{eq:reachm}
    \limsup_{n\to\infty} \sup_{p'\in B(p,\epsilon_{p,m};\cP)} \frac1n D_n(p'||q_{p}) <\frac{1}{m}.
  \end{equation}
  We fix such an $\epsilon_{p,m} > 0$ for each $p \in \cP$ and
  $m \ge 1$. 
  
  \textbf{Reach of $p\in\cP$} For $\delta\ge 1$, let $m=1$ and for
  $0<\delta<1$ let $m=\ceil{1/\delta}$. Therefore $m$ is the natural
  number such that $\frac1m \le \delta < \frac1{m-1}$.  For any
  $\delta>0$, we call $\epsilon_{p,\ceil{1/\delta}}$ the
  $\delta-$\emph{reach} of $p$.  In particular, $\epsilon_{p,m} > 0$
  is the $\frac{1}{m}-$reach of $p$.
  We do not require any regularity of
  $\epsilon_{p,m}$ over $p \in \cP$, in particular $\inf_{p\in\cP}\epsilon_{p,m}$
  can be 0.

   \ignore{Then, by definition, for all
  $\delta > 0$, from Lemma~\ref{lm:om}, there is a probability
  measure $q_{p,\delta}$ on $\naturals^\infty$ and a choice of
  $\epsilon'_{p,\delta} > 0$ such that
  \begin{equation}
    \label{eq:epsilonp}
  \limsup_{n\to\infty} \sup_{p'\in\ngpepd} \frac1n D_n(p'||q_{p,\delta}) <\delta.
  \end{equation}
  We fix such an $\epsilon_{p,\delta}' > 0$, satisfying the additional
  technical requirement that $\epsilon_{p,\delta}' < 16 \log e$. 
  Apart from this requirement, we do not require any regularity over $p \in \cP$
  and $\delta > 0$ 
  of $\epsilon_{p,\delta}'$, i.e. one can just think of these as being 
  strictly positive numbers chosen individually for each $p \in \cP$ and
  $\delta > 0$.
  The reason this is okay will become apparent soon.}

  \ignore{For $\delta\ge 1$, let $m=1$ and for $0<\delta<1$ let
  $m=\ceil{1/\delta}$. Therefore $m$ is the natural
  number such that $\frac1m \le \delta < \frac1{m-1}$. Now let
  \begin{equation}
    \label{eq:epsilonpnew}
    \epsilon_{p,\delta} :=\epsilon_{p,\frac 1m}'.
  \end{equation}
  For any $\delta>0$, we
  call $\epsilon_{p,\delta}$ above the $\delta-$\emph{reach} of $p$.
  }

  \ignore{The intuitive meaning of the reach $\epsilon_{p,\delta}$ of a
  probability distribution $p \in \cP$ is as follows. Suppose the
  statistics of the observations are being determined by some
  probability distribution of $\cP$ within the reach of $p$ that is
  not necessarily $p$. Suppose also that we have decided instead that the
  statistics of the observations are being determined by $p$. Yet,
  by waiting long enough, we will have control over the amount
  of harm, determined by $\delta$, that will be done by misinterpreting
  the source statistics. This rough
  insight will be made more precise in what follows.}

  \ignore{
  The reason this does not matter
  is basically because, for each
  $m \ge 1$, it will suffice to focus on only a countable collection
  of $p \in \cP$.
    We will find the following technical lemma useful in our proofs. The
  lemma essentially proves that we can choose one probability measure
  that is good for all $\delta-$reaches of a non-deceptive probability distribution, rather
  than have the probability measure depend on $\delta$.
  \bLemma
  \label{lm:uq}
  Let $p\in\cP$ be non-deceptive.
  For $\delta > 0$, let $\epsilon_{p,\delta}$ denote the $\delta$-reach of $p$. Then there exists a probability
  measure $q_p$ such that for all $\delta>0$, we have
  \[
    \limsup_{n\to\infty} \sup_{p'\in\ngped} \frac1n D_n(p'||q_{p}) <\delta.
  \]
  \Proof For any $\x\in\naturals^*$, let 
  \[
    q_p(\x) := \sum_{m\ge 1} \frac1{m(m+1)} q_{p, 1/m}(\x).
  \]
  For $\delta>1$, let $m=1$ and for $0<\delta<1$, let
  $m=\ceil{1/\delta}$, and note that $\delta\ge 1/m$.  From the
  definition of the $\delta-$reach in~\eqref{eq:epsilonpnew}, we have
  $\epsilon_{p,\delta}=\epsilon_{p,\frac1m}$.}

  \ignore{Therefore, 
  for all $\delta>0$ 
  we have 
  \[
    \limsup_{n\to\infty} \sup_{p'\in\ngped} \frac1n D_n(p'||q_p) \le \limsup_{n\to\infty} \left(
    \sup_{p'\in\ngped} \frac1nD_n(p'||q_{p,1/m}) + \frac{\log m(m+1)}n  \right) < \frac 1m  \le
    \delta.
  \]
  The lemma follows.
\eLemma}

\ignore{\subsection{Topology of $\cP$ With the $\ell_1$ Metric}\label{ss:lindel}
  To prove that $\cP^\infty$ is \dwc if no measure is deceptive, we
  will need to find a way to cover $\cP$ with countably many sets of the
  form $\ngpep$ above.
  Unfortunately, $\dist(p,q)$ is not a metric, so
  it is not immediately clear how to go about doing this. On the other
  hand, since Lemma~\ref{lm:dist} proves that
  \[
  \dist(p',p)\le |p-p'|_1/\ln 2,
  \] 
  where $|p-p'|_1$ denotes the $\ell_1$ distance between the single letter
  marginals of $p$ and $p'$,
  we can bootstrap off an understanding of the topology induced on $\cP$
  by the $\ell_1$ metric.}

  \ignore{The topology induced on $\cP$ by the $\ell_1$ metric is Lindel\"of,
  i.e. any covering of $\cP$ with open sets in the $\ell_1$ topology has
  a countable subcover (see~\cite[Defn. 6.4]{Dugundji}) for definitions
  and properties of Lindel\"of topological spaces). See~\cite{SA12:jmlr}
  for the proof of why $\cP$ is Lindel\"of.
  }

  \ignore{We can show that $\cP$ with the $\ell_1$ topology is Lindel\"of by
  appealing to the fact that the set of all probability measures on
  $\naturals$ with the $\ell_1$ topology, is second countable, i.e.
  that it has a countable basis. The set of all probability measures on
  $\naturals$ along with $\ell_1$ topology has a countable basis because
  it has a countable norm-dense set (consider the set of all probability
  measures on $\naturals$ with finite support and with all
  probabilities being rational). Now, $\cP$, as a topological subspace of
  a second countable topological space is also second
  countable~\cite[Theorem 6.2(2)]{Dugundji}). Finally, every second
  countable topological space is Lindel\"of \cite[Thm. 6.3]{Dugundji},
  hence $\cP$ is Lindel\"of.}



  \ignore{Our primary conceptual contribution is to connect the topology
  of the model class with the notion of data-derived consistency. The
  proof below will reflect this explicitly. In effect, we show to how
  to identify a source to within one of its \emph{good} neighborhoods
  (its reach), while controlling for any erroneous identification of
  the reach. This step also constructs a stopping rule $\tau$, which stops
  waiting once the neighborhood is confidently identified. 
  Given a confidence probability $1 > 1 -\eta > 0$ and accuracy $\delta$, the stopping rule
  $\tau$ ensures that for all $p^\infty\in\cP^\infty$ we have
  \[
  p\Paren{\tau \text{ is $\delta-$premature with respect to $q$ for $p$ }}<\eta.
  \]
  The second part is simpler---exploiting the fact that we are in a good
  neighborhood.}

  \ignore{In fact, to ensure we are within the reach of the generating
  probability measure $p\in\cP$, we will define its \emph{zone} $\Qpd$ below as
  a set of probability distributions on $\naturals$ around it. The probability distributions
  in $\Qpd$ are not necessarily in $\cP$.}
  
\textbf{Zone of $p\in\cP$} 
Given $m \ge 1$, the \emph{zone} $\Qpm$ of
a probability distribution $p \in \cP$ is defined to be the set of
probability distributions $u$ on $\naturals$ given by
  \begin{equation}
    \label{eq:zone}
  \Qpm \ed \Sets{ u: ||p-u||_1 < \frac{\epsilon_{p,m}}2},
\end{equation}
where, $\epsilon_{p,m}$ is the $\frac{1}{m}$-reach of $p$.  
Note that
the probability distributions in $\Qpm$ are not necessarily in
$\cP$. 

  \ignore{For $p\in\cP$, we have $\epsilonpd$ to be the reach of $p$. Then,
  \begin{equation}
    \label{eq:zone}
  \Qpd \ed \Sets{ u: |p-u|_1 < \frac{{\epsilonpd}^2(\ln 2)^2}{16}},
\end{equation}
  where $p$ is understood to be the one dimensional marginal (\ie a
  probability distribution on $\naturals$), and as mentioned above, $u$ above is
  any probability distribution on $\naturals$.}

\textbf{Countable cover of $\cP$} The zone
$\Qpm$ satisfies $\Qpm \cap \cP \subseteq \ngpem$. 
\ignore{Furthermore, no $\Qpd$ is empty since no $p$ is deceptive. For
  large enough $n$, the set of sequences of length $n$ with empirical
  distribution in $\Qpd$ will ensure that the stopping rule $\tau$ to
  be proposed enters with probability 1 when $p$ is in force. The
  reason we shrink $\Qpd$ beyond the reach is to accommodate the form
  of the pseudo-triangle inequality in Lemma~\ref{lm:dist}.}
Trivially $p\in \Qpm \cap \cP$.  Therefore we have
  we have
  \[
  \cP=\union_{p\in\cP} (\Qpm \cap \cP).
  \] 
  Further, since $\Qpm$ is open in the $\ell_1$ topology, each of
  the intersections $\Qpm \cap \cP$ is relatively open in the $\ell_1$ topology
  on $\cP$.
  Since $\cP$ under the $\ell_1$ topology is second
  countable, it is also Lindel\"of (see~\cite[Sec. 6.1]{SA12:jmlr} for
  a proof), \ie there is a countable set
  $\tilde\cP_m \subseteq\cP$, such that $\cP$ is covered by the
  collection of relatively open sets
  $( Q_{\tilde p,m}\cap \cP, \tilde{p} \in\tilde\cP_m)$, i.e. we have
  \begin{equation}
  \label{eq:dec}
  \cP=\union_{{\tilde p}\in\tilde\cP_m} ( Q_{\tilde p,m}\cap \cP ).
  \end{equation}
  For any fixed $m \ge 1$,
  we will make a choice of such a $\tilde\cP_m$ and refer to it as the \emph{quantization} of
  $\cP$ and to elements of $\tilde\cP_m$ as  the \emph{centroids} of
  the quantization, borrowing from commonly used literature.
  \ignore{Letting $m$ be a number such that
  $\frac1m \le \delta\le \frac1{m-1}$, for all $\delta$ in the range
  $\frac1m \le \delta\le \frac1{m-1}$, the reaches and therefore the
  zones $\Qpd$ do not change. Therefore, for all
  $\frac1m \le \delta\le \frac1{m-1}$, we retain the same quantization
  in~\eqref{eq:dec}---henceforth, with this in mind, we will denote
  these quantizations as $\tilde\cP_m$.}
  We index
  the countable set of centroids, $\tilde\cP_m$ by
  $\iota_m:{\tilde \cP}_m\to \naturals$. 
  \ignore{Essentially~\eqref{eq:dec} assures
  us that no matter what the generating probability measure is, it belongs
  within the zone (and hence reach) of one of the centroids in
  $\tilde\cP_m$. Consequently, our task will now be to ensure that
  we associate with the correct centroid in $\tilde\cP_m$.}

  \subsection{Construction of the Probability Measure $q^*$ on $\naturals^\infty$}
  \label{s:sff:q}
  We now construct a probability measure $q^*$ on $\naturals^\infty$ and,
  for each $0 < \eta < 1$ and $m \ge 1$, a stopping rule $\tau_{\eta,m}$,
  such that the pair  $q^*$ and $\tau_{\eta,m}$ will together satisfy the 
  required guarantee that for every $p \in \cP$,
  the probability that the
  stopping rule $\tau_{\eta,m}$ is $\frac{1}{m}-$premature
  with respect to $q^*$ for $p$ is at most $\eta$. This section details
  the construction of $q^*$, while Section~\ref{s:sff:thr} details the
  construction of $\tau_{\eta,m}$. Note that while the stopping rule
  $\tau_{\eta,m}$ depends on the confidence $\eta$ and accuracy threshold
  $\frac1m$, the measure $q^*$ is universal over all choices of the
  confidence and accuracy.

  \newcommand{\ngtpem}{B(\tilde p, \epsilon_{\tilde p, \frac1m}; \cP)}
  \newcommand{\ngtped}{B(\tilde p, \epsilon_{\tilde p, \delta}; \cP)} 
  \ignore{We construct the measure $q^*$ in two steps. First we construct a measure
  $q_m$ for all $m\ge1$, which we average over to obtain $q^*$.
n
  Fix $m\ge 1$ and let $\frac1{m}\le \delta <\frac1{m-1}$.}

First, fix $\eta$ and $m$. We construct the universal $q^*$ using the
partition in~\eqref{eq:dec}, which holds regardless of what the
confidence $\eta$ is.  Therefore, our construction of the measure is
not dependent on the confidence $\eta$, and is universal over the
choice of $\eta$ automatically. For each $\tildep\in\tilde\cP_m$
  there is a probability measure $q_{\tilde{p}}$ on $\naturals^\infty$
  satisfying~\eqref{eq:reachm} for $\tilde{p}$, with $\epsilon_{\tilde{p},m}$
  denoting the $\frac{1}{m}-$reach of $\tilde{p}$.
  \ignore{such that for all $\frac1{m} \le \delta < \frac1{m-1}$,
  \[
    \limsup_{n\to\infty}
    \sup_{p'\in\ngtped}
    \frac1n
    D_n(p'||q_{\tilde{p}})
    =
    \limsup_{n\to\infty}
    \sup_{p'\in\ngtpem}
    \frac1n
    D_n(p'||q_{\tilde{p}})
    \le \frac1m\le \delta,  \]
  where the first equality follows from~\eqref{eq:epsilonpnew}.}
  Let
  \[
    \tilde{Q}_m := \sets{ q_{\tilde p} : \tilde p\in\tilde \cP_m}
  \]
  denote the collection of these probability measures as $\tildep$
  ranges over $\tilde\cP_m$. Note that $\tilde{Q}_m$ is countable and
  is a collection of not necessarily \iid probability measures on
  $\naturals^\infty$.  For $\tilde{q} \in \tilde{Q}_m$, set the index
  $\iota_{m}(\tildeq)$ to be equal the index assigned to the
  corresponding centroid $\tildep$ in the enumeration of
  $\tilde\cP_m$. Then define a probability measure $q_{m}$ on  
  $\naturals^\infty$ by extending the following assignment for each
  $n \ge 1$ and each $\x \in \naturals^n$,   \[
    q_{m}(\x) :=\sum_{\tilde{q}\in\tilde{Q}_m}
    \frac{\tilde{q}(\x)}{\iota_m(\tildeq)(\iota_m(\tildeq)+1)}.
  \]
  Similarly, to remove dependence on $m$, let $q^*$ be the probability
  measure on $\naturals^\infty$
  extending the following assignment 
  for each $n\ge1$ 
  and each
  $\x \in \naturals^n$,
  \[
    q^*(\x) :=\sum_{m\ge 1} \frac{q_{m}(\x)}{m(m+1)}.
  \]
  Now, for all $\tildep\in\tilde\cP_m$, we have
  \begin{align}
    \nonumber
    \limsup_{n\to\infty} \sup_{p'\in\ngpem} \frac1n D_n(p'||q^*)
    &=
      \limsup_{n\to\infty} \sup_{p'\in\ngpem} \frac1n D_n(p'||q_m)\\
    \nonumber &=
                \limsup_{n\to\infty} \sup_{p'\in\ngpem} \frac1n D_n(p'||q_{\tilde{p}}) \\
  \label{eq:dwu}  
              &< \frac1m.
  \end{align}

  \ignore{In an abuse of notation, let $2^\naturals$ be the set of all
    finite subsets of $\naturals$. Note that $2^\naturals$ is
    countable. To see this, represent every number by its Elias code,
    and every finite subset of naturals by concatenating the
    representations of its components. Each subset of $\naturals$ hence
    maps to a unique rational number. Since rational numbers are
    countable, so is $\naturals^*$.}

  \subsection{Description of the Stopping Rule $\tau_{\eta,m}$}
  \label{s:sff:thr}
  We turn next to construct a stopping rule $\tau_{\eta,m}$ having the
  property that, for all $p\in\cP$, we have
  \[
  p\Paren{\tau_{\eta,m} \text{ is $\frac1m-$premature with respect to $q^*$ for $p$}}<\eta.
  \]
  \ignore{From this point on, we are focusing on a fixed value of the
    confidence $\eta$ and accuracy $\delta$ (therefore, the
    quantization $\cP_\delta$). Therefore, we drop the subscript
    $\delta$ for readability, keeping in mind that the $\delta$
    subscript is understood to be there.  As before, $m$ is a number
    satisfying $1/m \le \delta < 1/(m-1)$.}  The essential task in
  this section is to use the type of the empirically
  observed sequence, say $(n,t)$, to identify one of the
  centroids in the covering~\eqref{eq:dec} that contains the
  underlying source, say $p$, within its $\frac1m-$reach.

\subsubsection{Summary}
  This is a variant of a hypothesis testing problem over a countable
  number of hypotheses---where we either choose one of the hypothesis
  (guaranteeing that at the point of choice our estimate will be
  accurate with the required confidence) \emph{or} defer the decision
  to a point where we have more data.

  To summarize the theoretical approach, 
  initialize $\tau_{\eta,m}$ to be $0$ on the empty string.
  Given a string $X_1^T$, 
  if for any $i < T$ we have 
  $\tau_{\eta,m}(X_1^i)=1$ then set $\tau_{\eta,m}(X_1^T)=1$.
  (Here $X_1^0$ denotes the empty string.)
  Else, we consider the following tests, one for each
  centroid ${\tilde p}\in \tilde \cP_{m}$: test if the empirical
  distribution $t \in Q_{{\tilde p}, m}$, and, if so, additionally
  test for Equations~\eqref{eq:bnkrpt} and~\eqref{eq:bnkrpttwo}
  below. If any centroid passes all the tests, we choose the first
  centroid (according to the enumeration of $\tilde \cP_{m}$ chosen in
  Section~\ref{s:sff:cover}) among them, 
  and we also determine $\tau_{\eta,m}(X_1^n)$ for all $n \ge 1$, as explained in 
  the detailed description of the scheme below.
  If none do, we defer the
  decision to a future point.

  Testing whether a centroid contains the observed type in its zone is
  clearly 
  a natural thing to do.
  Since the
  empirical distribution converges to the underlying probability
  distribution at a rate that is only pointwise and cannot in general
  be uniformly bounded over $\cP$, it could happen on any finite
  sequence (say, with type $(n,t)$) that certain centroids close to
  the empirical distribution $t$ may not contain the generating
  distribution $p$ within their $\frac1m-$reach.

  Resolving which centroids are misleading and which are not cannot
  always be done to arbitrary confidence using finite sequences.
  However if $\cP$ has no deceptive distributions, imposing the
  additional tests in~\eqref{eq:bnkrpt} and~\eqref{eq:bnkrpttwo}
  enables us to attest that the probability the type generated by a
  source $p$ can be captured by \emph{any} centroid in $\tilde \cP_{m}$
  which does not have $p$ in its reach is $<\eta$.

  At this point, we prove that with the desired confidence,
  we have identified a centroid $\hat p$ that contains the generating
  source $p$ within its $\frac1m-$reach. Therefore we can now identify,
  based on the uniform convergence 
  of per symbol redundancy
  within the
  $\frac1m-$reach of $\hat p$, when the per-symbol redundancy drops
  $\le \frac1m$ and stays below the threshold.
 
  \ignore{ stopping rule is a formalization of our
    ability to decide, to the given confidence, if the empirically
    observed distribution $t$ is ``typical'' to at least one of the
    quantized regions (and if so, which ones) or not.  underlying
    (unobserved) source $p$.  But if no source in $\cP$ is deceptive,
    we show that the hypothesis tests in~\eqref{eq:bnkrpt}
    and~\eqref{eq:bnkrpttwo} will be accurate when
    $\tau_{\eta,m} =1$.}

  \ignore{
    The idea is that
  we want sequences generated by the (unknown) $p \in \cP$ to be captured by 
  one of the
  centroids of the quantization $\tilde\cP_m$ that have $p$ in their
  $\frac{1}{m}-$reach. To do this, we examine the sequence generated, the stopping
  rule stops waiting when we can establish with the desired confidence probability
  that the sequence has been \emph{captured} by a centroid with $p$
  within its reach.}

  \subsubsection{Detailed Construction} Fix $0 < \eta < 1$ and $m \ge 1$. 
  Let $p \in \cP$ be the probability distribution in force, 
  which is unknown.
  Consider a length-$n$ sequence $x^n$ on which we have
  not yet decided that $\tau_{\eta,m}(x^r)=1$ for any $1 \le r < n$. 
  Let $x^n$ have type $(n,t)$ where $t$ is the empirical distribution.
  The set of centroids in $\tilde\cP_m$ that can potentially
  \emph{capture} the type is defined to be
  \[
  \tilde \cP_{m,t} := \sets{ {\tilde p} \in{\tilde \cP_m}: t\in Q_{{\tilde p}, m} }.
  \]
  Not every centroid in $\tilde \cP_{m,t}$ is necessarily benign.
  Some of the centroids in $\tilde \cP_{m,t}$ may not have the
  generating probability measure $p$ within their $\frac1m$-reach.
  Therefore, when $\tilde \cP_{m,t} \ne \emptyset$, we refine
  $\tilde \cP_{m,t}$ further to a set of safe centroids 
  $\hat \cP_{m,t} \subset \tilde \cP_{m,t}$ in a way that will allow
  us to use Lemma~\ref{lm:yeung} to bound the probability of wrong
  capture.

  To counter 
  the possibility that the convergence of empirical
  distribution is not necessarily uniform over $\cP$, we
  use a modified convergence result in Lemma~\ref{lm:yeung}. This is a
  distribution free bound that bounds the probability that the
  empirical distribution is far from the
  underlying $p$, but only for empirical distributions
  that are ``top heavy'' (namely, those with at
  least a certain probability mass within the first $k$ symbols).
  To do so, for every ${\tilde p} \in \tilde \cP_{m}$,
  with $\frac1m-$reach $\epsilon_{{\tilde p},m}$, let
  \[
  \gamma_{{}_{{\tilde p},m}}
  :=
  \frac{\epsilon_{{\tilde p},m}}2
  \]
  The quantity above plays the role of $\gamma$ when using 
  Lemma~\ref{lm:yeung}. 
  Note that $\frac{\epsilon_{{p},m}}2$ also played a role in defining the zone 
  $Q_{{p}, m}$ (for given $m \ge 1$) 
  of an arbitrary probability distribution (not just a centroid) $p \in \cP$.
  \eqref{eq:zone}.

  To understand the core of our sufficiency proof, consider
  what happens when the underlying $p$ happens
  to be outside the $\frac1m-$reach of some $p' \in \tilde \cP_{m,t}$.
  Since $p$ is far from $p'$ (out of its $\frac1m-$reach), but $p'$
  is close to the empirical distribution, $t$, of the observed sequence,
  the triangle inequality will lower bound the distance of $t$ from the
  underlying $p$ by  $\gamma_{{}_{{\tilde p},m}}$.

  The centroids in $\tilde \cP_{m,t}$ that get placed
  into the safe set $\hat \cP_{m,t}$ are those that satisfy~\eqref{eq:bnkrpt}
  and~\eqref{eq:bnkrpttwo} in addition. In what follows, the quantity 
  $\log C(p',m)$
  of a centroid $p'\in \tilde \cP_{m,t}$ plays the role of the 
  ``effective size" of the support
  size of $p'$, corresponding to the number $k$ of Lemma~\ref{lm:yeung}.
  Given ${\tilde p} \in \tilde \cP_{m}$, we define
  $C({\tilde p},m)$ via
  \begin{equation}		\label{eq:effectivesize}
  C({\tilde p},m) := 2^{3\Paren{{\sup_{r\in B({\tilde p},\epsilon_{{\tilde p},m};\cP)}} \dotF_r^{-1}(1-\gamma_{{}_{{\tilde p},m}}/6)}},
  \end{equation}
  and we note that $C({\tilde p},m)$ is finite from the tightness result in
  Lemma~\ref{lm:bndprc}. This is because we have
  \[
  \limsup_{n\to\infty} \sup_{r\in B({\tilde p},\epsilon_{{\tilde p},m};\cP)} \frac1n D_n(r||q^*) < \frac1m,
  \]
  from~\eqref{eq:dwu}, which implies that for sufficiently large $n$
  the single letter redundancy of the family of $n$-fold product measures
  on $\naturals^n$ corresponding to the probability distributions in 
  $B({\tilde p},\epsilon_{{\tilde p},m};\cP)$ is finite, which, by Lemma~\ref{lm:bndprc},
  implies that this family of $n$-fold product measures on $\naturals^n$
  is tight, which implies that 
  the family of probability distributions
  $B({\tilde p},\epsilon_{{\tilde p},m};\cP)$ is tight.
  
  With $C(p',m)$ for $p'\in \tilde \cP_{m,t}$ defined as in~\eqref{eq:effectivesize}, the conditions we
  require on $p' \in \tilde \cP_{m,t}$ in order to place it in 
  $\hat \cP_{m,t}$ are
  \ignore{\begin{equation}
  \label{eq:bnkrpt}
  \exp\Paren{-n\gamma_{{}_{{\tilde p},m}}^2/18}
  \le 
  \frac{\eta }{2C(p',m) \iota(p')^2 n(n+1)},
  \end{equation}}
  \begin{equation}
  \label{eq:bnkrpt}
  \exp\Paren{-n\gamma_{{}_{p',m}}^2/18}
  \le 
  \frac{\eta }{2C(p',m) \iota(p')^2 n(n+1)},
  \end{equation}
  and 
  \ignore{\begin{equation}
  \label{eq:bnkrpttwo}
  2 \dotF_t^{-1}(1-\gamma_{{}_{{\tilde p},m}}/6) \le \log C(p',m).
  \end{equation}}
  \begin{equation}
  \label{eq:bnkrpttwo}
  2 \dotF_t^{-1}(1-\gamma_{{}_{p',m}}/6) \le \log C(p',m).
  \end{equation}
  These criteria will be then translated into a bound on the probability of wrong
  capture.
\ignore{  Appendix~\ref{s:phione} 
  $\tilde{p} \in \tilde \cP_m$ and some type $(n,t)$, one could ask if the 
  conditions analogous to~\eqref{eq:bnkrpt} and~\ref{eq:bnkrpttwo} hold or not
  for the pair ${\tilde p}$ and type 
  (i.e. with $p' \in \tilde \cP_{m,t}$ replaced by an arbitrary
  $\tilde{p} \in \tilde \cP_m$);
  this observation will become important in 
  Appendix~\ref{s:phione}.}
  It is also worth remarking that the proof of sufficiency of the necessary and
  sufficient condition for the insurability of a model class
  in~\cite[Thm. 1]{SA12:jmlr}) also
  uses a similar criterion to bound the probability of wrong capture.
  \ignore{Comparison with Lemma~\ref{lm:yeung} will give a hint as to
    why the equations above look this way.}

  We are now in a position to specify the stopping rule
  $\tau_{\eta,m}$.  Consider a sequence of natural numbers, $x^n$,
  having type $(n,t)$. Assume that we have not yet specified
  $\tau_{\eta,m}$ for any prefix $x^l$ of the sequence $x^n$ for
  $1 \le l \le n$.
  
  If 
  $\hat \cP_{m,t} =\emptyset$,
  \ignore{
  \color{red} (commented out ``there is '') \color{black}
  }
  we move on to all the possible single letter extensions of the sequence $x^n$.
  \ignore{
  \color{red} (commented out ``without deciding... '')\color{black}
  }
  
  If $\hat \cP_{m,t} \ne\emptyset$, let
  $\hat p$ denote the probability distribution in $\hat \cP_{m,t}$ with the smallest index. All
  suffixes of $x^n$ are then said to be \emph{trapped} by
  $\hat p$, which means that they are assigned to $\hat p \in \hat \cP_{m,t}$.
  From~\eqref{eq:dwu}, we have
  \[
  \limsup_{n\to\infty} \sup_{r \in B({\hat p},\epsilon_{{\hat p},m};\cP)} \frac1n D_n(r||q^*) 
  <
  \frac1m.
  \]
  This means that the set
  \begin{equation}
    \label{eq:delta}
  N_{{}_{\hat p}} :=\sets{n: \sup_{r \in B({\hat p},\epsilon_{{\hat p},m};\cP)} \frac1n D_n(r||q^*) \ge \frac1m}
  \end{equation}
  is finite. For any suffix $x^N$ of $x^n$, when $N> \max N_{{}_{\hat p}}$,
  we set $\tau_{\eta,m}(x^N)=1$, 0 else.
  
  Finally for each finite string $x^n$ for which the value of $\tau_{\eta,m}(x^n)$ has not
  yet been decided, we set this value to be $0$. It can be checked that 
  $\tau_{\eta,m}$ so defined is a stopping rule. This is because if 
  $\tau_{\eta,m}(x^n) = 0$
  for any sequence $x^n \in \naturals^n$, then we also have 
  $\tau_{\eta,m}(x^m) = 0$ for $1 \le m \le n$, i.e. for all its prefixes.

  \subsubsection{$\tau_{\eta,m}$ Enters With Probability 1} 
  This is proved in Appendix~\ref{s:phione}, using an argument similar to that
  used in the sufficiency proof in~\cite{SA12:jmlr}.

  \subsubsection{$\tau_{\eta,m}$ is $\frac1m-$Premature is $<\eta$}
  Consider any $p\in\cP$.  Among sequences of natural numbers on which $\tau_{\eta,m}$ has entered,
  we will distinguish between those that are in \emph{good} traps and
  those in \emph{bad} traps. If a sequence $x^n$ is trapped by 
  $\hat p \in \tilde \cP_m$ such
  that $p\in B({\hat p},\epsilon_{{\hat p},m};\cP)$, we call
  $\hat p$ is a good trap for that sequence. Conversely, if
  $p\notin B({\hat p},\epsilon_{{\hat p},m};\cP)$, $\hat p$ is called
  a bad trap for that sequence.

  \textit{(Good traps)}
  Suppose a length-$n$ sequence $x^n$ is in a good trap. Namely, it is
  trapped by a probability distribution
  $\hat p \in \tilde \cP_m$ such
  that $p\in B({\hat p},\epsilon_{{\hat p},m};\cP)$. 
  Then, if $\tau_{\eta,m}(x^n) = 1$ it must be the case that 
  $\frac{1}{n}D(p||q^*) < \frac{1}{m}$. Thus such sequences cannot
  contribute to the probability under $p$ of $\tau_{\eta,m}$ being
  $\frac1m-$premature with respect to $q^*$ for $p$.
  
  \ignore{In this case,
  going by the construction of our stopping rule, we therefore have 
  \[
    p( X^\infty:
    \tau \text{ is $\delta-$premature with respect to $q^*$ for $p$}
    \mid
    X^\infty\text{ suffix of $x^n$ in good trap} )=0.
  \]}

  \textit{(Bad traps) } We can show that the probability with which sequences
  generated by $p$ fall into bad traps is strictly less than 
  $\eta$ using an argument,
  which is essentially identical
  to the one used in~\cite{SA12:jmlr}.
  This argument is reproduced in Appendix~\ref{s:phibt} for the sake of
  completeness. Pessimistically, we assume that $\tau_{\eta,m}$ is 
  $\frac1m-$premature
  with respect to $q^*$ for $p$
  on every sequence that falls into a bad trap.

This completes the proof of the sufficiency part of Theorem~\ref{thm:ncssff}.

\acks{We thank the anonymous reviewer for several suggestions that helped
streamline the proofs and improve the readability of the document.
We also thank the reviewer for suggesting 
one direction of the connection to learnability (as noted in two footnotes
and in Appendix~\ref{app:equiv}). \color{black}

This work was in part supported by the NSF Science \& Technology
Center for Science of Information Grant number CCF-0939370. In
addition, Santhanam was also supported by NSF Grants CCF-1065632 and
CCF-1619452; Anantharam was also supported by the ARO MURI grant
W911NF- 08-1-0233, “Tools for the Analysis and Design of Complex
Multi-Scale Networks”, Marvell Semiconductor Inc., the U.C. Discovery
program, the William and Flora Hewlett Foundation supported Center for
Long Term Cybersecurity at Berkeley, and the NSF grants CNS-0910702,
ECCS-1343998, CNS--1527846, CCF--1618145 and CCF-1901004; Szpankowski
was also supported by NSF Grants CCF-2006440, and CCF-2007238.}


\appendix

\section{Alternate Definitions of Strong and Weak Compressibility}
\label{app:basics}

We first establish the following elementary result.

\bLemma
\label{lm:match}
For $n\ge 1$, let $\hat q_n$ be a probability measure on $\naturals^n$.
Then there is a probability measure $q_n$ 
on $\naturals^\infty$
such that, for all $\x \in \naturals^n$, we have
$q_n(\x)=\hat q_n(\x)$. 
\Proof
We define $q_n$ by specifying $q_n(\y)$ for all $\y \in \naturals^m$
for all $m \ge 1$.
If $1 \le m \le n$ and $\y \in \naturals^m$, 
let 
\[
  q_n(\y) := \sum_{\x'\in\naturals^n: \y \preceq \x'} \hat{q}_n(\x').
\]
For $m \ge n$ and $\y \in \naturals^m$, if
$\y$ is $\x$ followed by a string
of $1$s, for some $\x \in \naturals^n$, let 
\[
  q_n(\y) := \hat{q}_n(\x),
\]
else let $q_n(\y) :=0$.  
It can be checked that $q_n$, defined in this way, satisfies the consistency conditions
$q_n(\z) = \sum_{\y \in \naturals^m ~:~ \z \preceq \y} q_n(\y)$ for all $1 \le l \le m$ and $\z \in \naturals^l$.
Hence $q_n$ defines a
probability measure on $\naturals^\infty$.
It can also be checked that 
$q_n$ satisfies the requirement in the statement of the lemma.
\eLemma

Using Lemma~\ref{lm:match}, we now get the 
following result, which will help establish
the equivalence of our definitions 
of strong and weak compressibility with those common in literature. 
\bLemma
\label{lm:eq}
Let $\Lambda$ be any collection of probability measures 
on $\naturals^\infty$ (not necessarily
\iid). Suppose there exists a sequence of probability measures $\hat q_n$ on $\naturals^n$
such that
\[
\limsup_{n\to\infty} \sup_{r \in \Lambda} \frac1n E_r \log \frac{r(X^n)}{\hat q_n(X^n)} =0.
\]
Then there is a probability measure $q$ on $\naturals^\infty$ such that
\[
\limsup_{n\to\infty} \sup_{r \in \Lambda} \frac1n E_r \log \frac{r(X^n)}{q(X^n)} =0.
\]

\Proof
For each $n \ge 1$, let the probability measure $q_n$ on $\naturals^\infty$ 
be constructed to match the probability measure $\hat q_n$ on
$\naturals^n$, 
as in Lemma~\ref{lm:match}. Define the probability measure $q$ 
on $\naturals^\infty$ that, for each $n \ge 1$ and
$\x \in \naturals^n$, 
assigns to $\x$ the probability
\[
q(\x) := \sum_{i=1}^\infty \frac{q_i(\x)}{i(i+1)}.
\]
For all $n \ge 1$
we therefore have
\begin{eqnarray*}
\sup_{r \in \Lambda} \frac1n E_r \log \frac{r(X^n)}{q(X^n)} 
&\le&
\sup_{r \in \Lambda} \frac1n E_r \log \frac{r(X^n)}{q_n(X^n)} 
+ \frac{\log ( n(n+1))}{n}\\
&=& 
\sup_{r \in \Lambda} 
\frac1n E_r \log \frac{r(X^n)}{\hat{q}_n(X^n)} 
+ \frac{\log (n(n+1))}{n}.
\end{eqnarray*}
Hence
\[
\limsup_{n\to\infty} 
\sup_{r \in \Lambda} \frac1n E_r \log \frac{r(X^n)}{q(X^n)} 
=0.
\]
\eLemma

Let $\cP$ be a collection of probability distributions on 
  $\naturals$ and $\cP^\infty$ the collection of probability
  measures 
  on $\naturals^\infty$ induced
  by \iid assignments from the individual probability distributions in $\cP$. 
  In most prior work~(\cite{Fit72,Dav73,kie78}) the collection
  $\cP$ is called strongly compressible if there is 
  a
  sequence of probability measures $\hat q_n$ on $\naturals^n$
  such that
  \[
\limsup_{n\to\infty} \sup_{p \in \cP^\infty} \frac1n E_p \log \frac{p(X^n)}{\hat q_n(X^n)} =0.
\]
 Lemma~\ref{lm:eq} immediately establishes that this definition is equivalent to the definition of strong compressibility that we have made in Definition~\ref{dfn:strongcomp}. 
 
 The most commonly used definition of weak compressibility in 
 prior work is due to Kieffer~(\cite{kie78}), and is framed
 in the language of 
 length functions of compression schemes. 
  Let $\Lambda$ be any collection of stationary ergodic probability measures 
on $\naturals^\infty$ (not necessarily
\iid). A compression scheme is a sequence of mappings 
$\phi_n ~:~ \naturals^n \to \{0,1\}^* \backslash \emptyset$
whose image satisfies the prefix condition, i.e. for any two distinct elements in the domain the image of the first is not a prefix of the image of the second. The collection $\Lambda$
is called weakly compressible if there is a compression scheme 
$(\phi_n, n \ge 1)$ such that, for all $r \in \Lambda$, we have
\[
\lim_{n \to \infty} \frac{1}{n} E_r l(\phi_n(X^n)) = H(r),
\]
where $H(r)$ denotes the entropy rate of $r$. 

Let $\cP$ be a collection of probability distributions on 
  $\naturals$ and $\cP^\infty$ the corresponding collection of \iid probability
  measures 
  on $\naturals^\infty$. 
  Note that $\cP^\infty$ is a collection of 
  stationary ergodic probability measures. We now show that the 
  definition of weak compressibility of $\cP^\infty$ in the sense of 
  Kieffer~(\cite{kie78}) is identical to the definition of weak compressibility of $\cP^\infty$ that we have made in 
  Definition~\ref{dfn:weakcomp}. 
  
  Suppose first that $\cP^\infty$ is weakly compressible in the sense of Definition~\ref{dfn:weakcomp}. If every probability distribution in
$\cP$ has infinite entropy, consider an arbitrary compression scheme 
  $(\phi_n, n \ge 1)$, for instance by defining $\phi_n(x^n)$ by 
  concatenating symbol by symbol the representation of $i \in \naturals$
  by a bit string of length $\ceil{\log \frac{1}{(i+1)(i+2)}}$ coming from 
  a prefix code for $\naturals$ corresponding to the probability distribution
  assigning probability $\frac{1}{(i+1)(i+2)}$ to $i \in \naturals$.
Then we have
\begin{equation}		\label{eq:Gibbs}
\frac{1}{n} E_p l(\phi_n(X^n))  \stackrel{(a)}{\ge} \frac{1}{n} E_p \log\frac{1}{p(X^n)} = \infty,
\end{equation}
and so 
\[
\lim_{n \to \infty} \frac{1}{n} E_p l(\phi_n(X^n)) = H(p),
\]
for all $p \in \cP$. 
Here $(a)$  in~\eqref{eq:Gibbs} 
can be seen by picking a probability measure $q_n$ on 
$\naturals^n$ that satisfies $l(\phi_n(X^n)) \ge \log \frac{1}{q_n(x^n))}$
and observing that $E_p \log \frac{p(X^n)}{q_n(X^n)} \ge 0$.
 If there are probability distributions in $\cP$ with finite entropy,
 let $q$ be a probability measure on $\naturals^\infty$ verifying the requirements in Definition~\ref{dfn:weakcomp}. For $n \ge 1$, let 
  $\hat{q}_n$ denote the probability measure on $\naturals^n$
  resulting from restricting $q$ to $\naturals^n$. We can then define a compression scheme $(\phi_n, n \ge 1)$
  such that $l(\phi_n(\x)) = \lceil \log \frac{1}{\hat{q}_n(\x)} \rceil$ 
  for all $\x \in \naturals^n$ for all $n \ge 1$. Hence, for every $p \in \cP$, we have
  \[
  \frac{1}{n} E_p l(\phi_n(X^n)) = \frac{1}{n} E_p \lceil \log \frac{1}{\hat{q}_n(X^n)} \rceil = \frac{1}{n} E_p \lceil \log \frac{1}{q(X^n)} \rceil.
  \]
  Suppose $H(p) = \infty$. 
  By the same argument as that used in~\eqref{eq:Gibbs}
  we conclude that $\frac{1}{n} E_p l(\phi_n(X^n)) = \infty$ for all $n \ge 1$ and so, for all such $p$, we have
\[
\lim_{n \to \infty} \frac{1}{n} E_p l(\phi_n(X^n)) = H(p).
\]
On the other hand, if $H(p) < \infty$ we have
\begin{eqnarray*}
\frac{1}{n} E_p l(\phi_n(X^n)) &\le& 
 \frac{1}{n} E_p \log \frac{1}{q(X^n)} + \frac{1}{n}\\
 &=& \frac{1}{n} E_p \log \frac{p(X^n)}{q(X^n)} + H(p) + \frac{1}{n},
\end{eqnarray*}
and so, letting $n \to \infty$, we see that 
\[
\lim_{n \to \infty} \frac{1}{n} E_p l(\phi_n(X^n)) = H(p)
\]
also holds for such $p$. 
We have established that $\cP^\infty$ is also weakly compressible in the sense of Kieffer~(\cite{kie78}), irrespective of whether $\cP$ is comprised entirely of
probability distributions with infinite entropy or also contains probability 
distributions with finite entropy.

For the converse, suppose that $\cP^\infty$ is weakly compressible in the sense of Kieffer~(\cite{kie78}). For each $n \ge 1$ we can find a probability measure $\hat{q}_n$ on $\naturals^n$ such that 
$\hat{q_n}(\x) \ge 2^{- l(\phi_n(\x))}$ for all $\x \in \naturals^n$, where $(\phi_n, n \ge 1)$
is a compression scheme verifying the weak compressibility of 
$\cP^\infty$ in the sense of Kieffer~(\cite{kie78}).
For each $n \ge 1$ we define the probability measure 
$q_n$ on $\naturals^\infty$ in terms of $\hat{q}_n$ as in 
Lemma~\ref{lm:match}, and we define the probability measure $q$ on 
$\naturals^\infty$ which, 
for each $n \ge 1$ and
$\x \in \naturals^n$, 
assigns to $\x$ the probability
\[
q(\x) := \sum_{i=1}^\infty \frac{q_i(\x)}{i(i+1)}.
\]
For each $p \in \cP$ with finite entropy, we have 
\begin{eqnarray*}
\frac{1}{n} E_p \log \frac{p(X^n)}{q(X^n)} &\le& 
\frac{1}{n} E_p \log \frac{p(X^n)}{q_n(X^n)} + \frac{\log n(n+1)}{n}\\
&=& \frac{1}{n} E_p \log \frac{p(X^n)}{\hat{q}_n(X^n)} + \frac{\log n(n+1)}{n}\\
&\le& - H(p) + \frac{1}{n} E_p l(\phi_n(X^n)) + \frac{\log n(n+1)}{n},
\end{eqnarray*}
and so, from $\lim_{n \to \infty} \frac{1}{n} E_p l(\phi_n(X^n)) = H(p)$, we conclude that $\limsup_{n \to \infty} \frac{1}{n} E_p \log \frac{p(X^n)}{q(X^n)} = 0$. This proves that 
$\cP^\infty$ is weakly compressible in the sense of Definition~\ref{dfn:weakcomp}.

To close this section, we give proofs of two statements that allow us to think
about strong compressibility and weak compressibility respectively in terms of
vanishing asymptotic per-symbol redundancy.

\bLemma
\label{lm:asyred}
Let $\cP$ be a collection of probability distribution on
  $\naturals$ and $\cP^\infty$ the collection of probability
  measures 
  on $\naturals^\infty$ induced
  by \iid assignments from the individual probability 
  distributions in $\cP$. Then $\cP^\infty$ is strongly compressible iff it has zero asymptotic per-symbol redundancy.
  
  \Proof
  
If $\cP^\infty$ is strongly compressible, then taking the probability measure $q$ on $\naturals^\infty$ which verifies the strong compressibility condition in~\eqref{eq:str_rdn} from
Definition~\ref{dfn:strongcomp} as the $q$ in~\eqref{eq:rdn} from Definition~\ref{dfn:terms}
for each $n \ge 1$ immediately implies that 
$\cP^\infty$ has zero asymptotic per-symbol redundancy.

Conversely, suppose $\cP^\infty$ has zero asymptotic per-symbol redundancy. Given $\epsilon >0$, for each $n \ge 1$ let $q_n$ be a probability measure on $\naturals^\infty$ for which
$\sup_{p \in \cP^\infty}
   E_p \log\frac{p(X^n)}{q_n(X^n)} \le R_n + \epsilon$, and define the probability measure $q$ on $\naturals^\infty$ by
   \[
q(\x) := \sum_{i=1}^\infty \frac{q_i(\x)}{i(i+1)}.
\]
Then we have
\[
\frac1n \sup_{p \in \cP^\infty}
   E_p \log\frac{p(X^n)}{q(X^n)} \le 
   \frac1n \sup_{p \in \cP^\infty}
   E_p \log\frac{r(X^n)}{q_n(X^n)} + \frac{\log (n(n+1))}{n},
\]
and so 
\[
\limsup_{n \to \infty} \frac1n \sup_{p \in \cP^\infty}
   E_p \log\frac{p(X^n)}{q(X^n)} \le \epsilon.
   \]
   Letting $\epsilon \to 0$ shows that $\cP^\infty$ is strongly compressible.
\eLemma

\bLemma
\label{lm:asyweakred}
Let $\cP$ be a collection of probability distribution on
  $\naturals$ and $\cP^\infty$ the collection of probability
  measures 
  on $\naturals^\infty$ induced
  by \iid sampling from the individual probability 
  distributions in $\cP$. Then $\cP^\infty$ is weakly compressible iff
  there is a probability measure $q$ on $\naturals^\infty$ such that for every $p \in \cP$ with finite entropy
  the corresponding $p^\infty \in \cP^\infty$ has zero asymptotic per-symbol redundancy with respect to $q$.
 
 \Proof
 
The claim is vacuously true if all the probability distributions in $\cP$ have infinite entropy. 
If there are distributions in $\cP$ with finite entropy and
$\cP^\infty$ is weakly compressible, then consider the probability measure $q$ on $\naturals^\infty$ which verifies the weak compressibility condition in~\eqref{eq:wk_rdn} from
Definition~\ref{dfn:weakcomp}. 
By definition, with respect to this $q$, every $p \in \cP$ with finite entropy is such that the corresponding 
$p^\infty \in \cP^\infty$ has zero asymptotic per-symbol redundancy with respect to $q$.
Conversely, if there are distributions in $\cP$ with finite entropy and there is a probability measure $q$ on $\naturals^\infty$
such that for every $p \in \cP$ the corresponding
$p^\infty \in \cP^\infty$ has zero asymptotic per-symbol redundancy with respect to $q$ then, by definition, this 
$q$ satisfies the condition in~\eqref{eq:wk_rdn} from
Definition~\ref{dfn:weakcomp} for all $p \in \cP$ with finite entropy. This establishes that $\cP^\infty$ is weakly compressible.
 \eLemma
 
\section{Basic Properties of Relative Entropy and Redundancy}
\label{app:redbasics}

In this appendix we gather some basic results on the KL divergence
and redundacy, which are used at various points in the document.

\bProposition
\label{prop:abs}
Let $p$ and $q$ be two probability distributions on a countable set $\cX$.
Then
\[
  \sum_{x\in\cX} p(x) \left|\log \frac{p(x)}{q(x)}\right| \le D(p||q)
  +2\frac{\log e}e.
\]
\Proof 
Let
$S\subset\cX$ be the set of all elements $x\in\cX$ such that
$p(x) \le q(x)$. Note that $q(S) >0$.
We have
\begin{eqnarray*}
D(p||q) - \sum_{x\in\cX} p(x) \left|\log \frac{p(x)}{q(x)}\right| &=& 2\sum_{x\in S} p(x) \log \frac{p(x)}{q(x)}\\
&\stackrel{(a)}{\ge}& 2p(S)\log
  \frac{p(S)}{q(S)}\\
  &\ge& 2p(S)\log {p(S)}\\
  &\ge& -2\frac{\log e}e,
\end{eqnarray*}
where step (a) is from the log sum inequality.
The proposition follows.
\eProposition


\bProposition
  \label{prop:monotone}
  For all probability measures $r$ and $q$ on $\naturals^\infty$ and all
  $1 \le m \le n$, we have 
  \[
  D_m(r||q)\le D_n(r||q).
  \]
  In particular, for any collection of probability distributions $\cP$ on 
  $\naturals$, if $\cP^\infty$ denotes the associated collection of \iid probability measures
  on $\naturals^\infty$, we will
  have
  \[
  R_m(\cP) := \inf_q \sup_{p\in\cP} E_p \log \frac{p(X^m)}{q(X^m)} 
  \le
  \inf_q \sup_{p\in\cP} E_p \log \frac{p(X^n)}{q(X^n)} = R_n(\cP),
  \]
  where the outer infimum on both sides is taken over all probability measures $q$ on
  $\naturals^\infty$ and so $R_m(\cP)$ and $R_n(\cP)$ are the length-$m$ redundancy and the 
  length-$n$ redundancy of $\cP$, respectively.
  \Proof
  The first part of the claim follows from convexity, because, for all $y^m \in \naturals^m$,
  we have 
  \[
  r(y^m) = \sum_{x^n ~:~ y^m \preceq x^n} r(x^n) \mbox{ and } q(y^m) = \sum_{x^n ~:~ y^m \preceq x^n} q(x^n).
  \]
  For the second part of the claim, for any $\epsilon > 0$ pick a probability measure $q'$ on
  $\naturals^\infty$ such that 
  \[
  \sup_{p\in\cP} E_p \log \frac{p(X^n)}{q'(X^n)} < R_n(\cP) + \epsilon.
  \]
  It then follows from the first part of the claim that 
  \[
  R_m(\cP) \le \sup_{p\in\cP} E_p \log \frac{p(X^m)}{q'(X^m)} < R_n(\cP) + \epsilon.
  \]
  We let $\epsilon \to 0$ to complete the proof.
  \eProposition

\bProposition
  \label{prop:rnn}
  Let $\cP$ be a collection of probability distributions on $\naturals$ and
  $\cP^\infty$ the corresponding collection of probability measures on
  $\naturals^\infty$ got by \iid sampling from the individual 
  probability distributions in $\cP$. For $n \ge 1$, let $R_n$ denote the
  length-$n$ redundancy of $\cP^\infty$, 
  as defined in~\eqref{eq:rdn}.
  Then, for all $n\ge 1$, the per-symbol length-$n$ redundancy of $\cP^\infty$
  satisfies $R_n/n \le R_1$.
  \Proof
  Let $\epsilon>0$. Let $\tilde{p}$ be a
  probability distribution on $\naturals$ such that the single letter
  redundancy of $\cP^\infty$ with respect to $\tilde{p}$ is strictly less than $R_1+\epsilon$. 
  With the usual abuse of notation, let $\tilde{p}$ also denote the \iid probability measure 
  on $\naturals^\infty$ corresponding to $\tilde{p}$. 
  Then,
  for all $p \in\cP$, we have
  \[
  \frac{1}{n} E_p \log \frac{p(X^n)}{\tilde{p}(X^n)} = E_p \log \frac{p(X)}{\tilde{p}(X)} < (R_1+\epsilon).
  \]
  By letting $\epsilon \to 0$, the proposition follows.
  \eProposition
\color{black}
  \bCorollary Let $\cP$ be any collection of distributions over
  $\naturals$ and let $\cP^\infty$ the set of probability measures
  obtained by \iid sampling from distributions in $\cP$. Then
  $\limsup_{n\to\infty} \frac1n R_n(\cP) <\infty$ iff $R_1<\infty$.

  \Proof Immediate from the Propositions~\ref{prop:monotone}
  and~\ref{prop:rnn} since for all $n$,
  \[ \frac1n R_1(\cP) \le \frac1n R_n(\cP)\le R_1(\cP).\eqed\]
    \eCorollaryp

  \color{black}
  \bLemma
  \label{lm:asyinter}
  Let $\Lambda$ be
  a collection of probability measures on $\naturals^\infty$.
  Then we have
  \begin{equation}
      \label{eq:asyinter}
  \limsup_{n \to \infty} \frac1n \inf_q \sup_{r \in \Lambda}
   E_r \log\frac{r(X^n)}{q(X^n)} = \inf_q \limsup_{n \to \infty} \frac1n \sup_{r \in \Lambda}
   E_r \log\frac{r(X^n)}{q(X^n)},
  \end{equation}
  where the infimum is taken over all probability measures $q$ on $\naturals^\infty$.
  Namely, the $\limsup_{n \to \infty}$ can be interchanged with the $\inf_q$ in the definition of
  the asymptotic per-symbol redundancy of $\Lambda$.
  \Proof
  Fix $\epsilon > 0$. For $n \ge 1$, let $q_n$ be a probability measure on $\naturals^\infty$ 
  such that 
  \[
  \frac1n \sup_{r \in \Lambda}
   E_r \log\frac{r(X^n)}{q_n(X^n)} < \frac1n R_n + \epsilon.
  \]
  Define the probability measure $\barq$ 
on $\naturals^\infty$ that, for each $n \ge 1$ and
$\x \in \naturals^n$, 
assigns to $\x$ the probability
\[
\barq(\x) := \sum_{i=1}^\infty \frac{q_i(\x)}{i(i+1)},
\]
where, as usual, $q_i(\x)$ is the probability under $q_i$ of the 
event in $\naturals^\infty$ comprised of the sequences having the prefix $\x$.
We then have 
\[
\frac1n \sup_{r \in \Lambda}
   E_r \log\frac{r(X^n)}{\barq(X^n)}
   \le \frac1n \sup_{r \in \Lambda}
   E_r \log\frac{r(X^n)}{q_n(X^n)} + \frac{\log(n(n+1)}{n} < \frac1n R_n + \epsilon
   + \frac{\log(n(n+1))}{n}.
\]
Thus
\[
\inf_q \limsup_{n \to \infty} \frac1n  \sup_{r \in \Lambda}
   E_r \log\frac{r(X^n)}{q(X^n)}
   \le
   \limsup_{n \to \infty} \frac1n \sup_{r \in \Lambda}
   E_r \log\frac{r(X^n)}{\barq(X^n)} \le \limsup_{n \to \infty} \frac1n R_n + \epsilon.
   \]
   Letting $\epsilon \to 0$, we see that the term on the right hand side of~\eqref{eq:asyinter} is no bigger than 
   the term on its
   left hand side. Showing the inequality in the other direction is straightforward, since
   \[
   \frac1n \inf_q \sup_{r \in \Lambda}
   E_r \log\frac{r(X^n)}{q(X^n)} \le \frac1n \sup_{r \in \Lambda}
   E_r \log\frac{r(X^n)}{q(X^n)},
   \]
   for each probability measure $q$ on $\naturals^\infty$. This completes the proof.
  \eLemma

The following lemma will be needed in Appendix~\ref{app:equiv}.

\color{black}
\bLemma
  \label{lm:finiteunion}
  Let $\cP_1, \ldots, \cP_L$ be classes of probability distributions on $\naturals$. 
  Let $\cP := \cup_{l=1}^L \cP_l$ denote their union. Then, for each $n \ge 1$ we have 
  \[
  \max_l R_n(\cP_l) \ge R_n(\cP) - \log L.
  \]
  \Proof 
  For any $\epsilon > 0$, for each $1 \le l \le L$, let $q_l$ be a probability measure on $\naturals^\infty$ such that
  \[
  \sup_{p \in \cP_l} E_p \log \frac{p(X^n)}{q_l(X^n)} \le R_n(\cP_l) + \epsilon.
  \]
  Let $q := \frac{1}{L} \sum_{l=1}^L q_l$. Then we have
  \begin{align*}
  R_n(\cP) &= \inf_{{\tilde q}} \sup_{p \in \cP} E_p \log \frac{p(X^n)}{\tilde{q}(X^n)}\\
  &\le \sup_{p \in \cP} E_p \log \frac{p(X^n)}{q(X^n)}\\
  &= \max_l \sup_{p \in \cP_l} E_p \log \frac{p(X^n)}{q(X^n)}\\
  &= \left( \max_l \sup_{p \in \cP} E_p \log \frac{p(X^n)}{q_1(X^n) + \ldots + q_L(X^n)} \right) + \log L\\
  &\le \left( \max_l \sup_{p \in \cP} E_p \log \frac{p(X^n)}{q_l(X^n)} \right) + \log L\\
  &\le \max_l R_n (\cP_l) + \epsilon + \log L,
  \end{align*}
  where the infimum in the first line is over probability measures $\tilde{q}$ on $\naturals^\infty$.
  Letting $\epsilon \to 0$ completes the proof.
\eLemma
\color{black}

  The following variation of the result from~\cite{SA12:jmlr}
  will be needed
  to prove the necessity part of Theorem~\ref{thm:ncssff}.

\bLemma
  \label{lm:jn}
\color{black}
Fix $\epsilon>0$.
  Let $p$ and $q$ be probability distributions on $\naturals$
  with $||p-q||_1\le \epsilon$. Fix $n \in \naturals$ with $2n^2\epsilon\le 1$.
  Consider the probability measures on $\naturals^n$ obtained by \iid
  sampling from $p$ and $q$ respectively, which we continue to 
  denote by $p$ and $q$ respectively, following our convention.
  
  Suppose $A_n \subset \naturals^n$ is subset for which
  $p(A_n)\ge 1-\alpha$, for some
  $\alpha>0$. Then we have
  \[
  q(A_n) > 1- \alpha-2n^3\epsilon-\frac1n. \eqed
  \]
 \Proof
  Let 
  \[
  \cB_1 := \Sets{ i\in\naturals : q(i) \le p(i)\Paren{1-\frac1{n^2}} },
  \mbox{ and } 
  \cB_2 := \Sets{ i\in\naturals : p(i) \le q(i)\Paren{1-\frac1{n^2}} }.
  \]
  We are given $||p-q||_1\le \epsilon$, and in addition, we have
  \[
  ||p-q||_1
  \ge
  \sum_{x\in\cB_1} (p(x) - q(x) )
  \ge
  \frac{p(\cB_1)}{n^2}
  \ge
  \frac{q(\cB_1)}{n^2},
  \]
  and similarly
  \[
  ||p-q||_1
  \ge
  \sum_{x\in\cB_2} (q(x) - p(x) )
  \ge
  \frac{q(\cB_2)}{n^2}
  \ge
  \frac{p(\cB_2)}{n^2}.
  \]
  From the preceding inequalities, it follows that 
  \begin{equation}
  \label{eq:tmp}
  p(\cB_1\cup \cB_2) \le 2n^2\epsilon
  \text{ and }
  q(\cB_1\cup \cB_2) \le 2n^2\epsilon.
  \end{equation}
  
  Let $S :=\naturals-(\cB_1\cup\cB_2)$. For all $x\in S$ we have
  \begin{equation}
  \label{eq:tmptwo}
  q(x) \ge p(x) \Paren{1-\frac1{n^2}}.
  \end{equation}
  In addition, from~\eqref{eq:tmp} we have
  \[
    p(S)\ge 1-{2n^2}\epsilon.
  \]
  \ignore{From Lemma~\ref{lm:app},
  \[
  q(S) \ge 1-2\epsilon-2N^2\sqrt{4\epsilon\ln2}.
  \]}
  Let $S_n \subset \naturals^n$ denote the set of all length-$n$
  strings of symbols from $S$. Clearly since $2n^2\epsilon<1$
  \[
  p(S_n) \ge (1-{2n^2}\epsilon)^n > 1- 2n^3\epsilon.
  \]
  Thus we have
  \[
  p(A_n\cap S_n) > 1-2n^3\epsilon-\alpha.
  \]
  From~\eqref{eq:tmptwo}, for all $x^n \in S_n$, we have
  \[
  q(x^n) \ge p(x^n)\Paren{1-\frac1{n^2}}^n > p(x^n)\Paren{1-\frac1n}.
  \]
  Therefore, 
  \[
  q(A_n) \ge q(A_n\cap S_n) > (1-2n^3\epsilon-\alpha)\Paren{1-\frac1n}
  > 1-\alpha-2n^3\epsilon-\frac1n.\eqed
  \]
  \eLemmap
\color{black}

  \section{Operational Formulation of the Problem}
  \label{app:equiv}
\color{black}
  Recall that ${\mathbb P}(\naturals)$ denotes the set of probability distributions
  on $\naturals$ and $\cP\subset{\mathbb P}(\naturals)$
  a collection of probability distributions on $\naturals$.
  We prove Theorem~\ref{thm:equiv} in this section, i.e. that $\cP$ is learnable (see
  Definition~\ref{dfn:lrn}) iff it is \dwc.

  \subsection{Learnable $\Rightarrow$ \dwc}
  To prove that if $\cP$ is learnable then it is \dwc, we use the
  equivalence of \dwc and the existence of deceptive distributions which was proved in
  Theorem~\ref{thm:ncssff}. Specifically, we show that if $\cP$ is learnable,
  then there cannot be any deceptive distributions in $\cP$.

  Suppose, to the contrary, that $\cP$ is learnable but that
  $p \in \cP$ is deceptive. Then, 
  by the definition of what it means to be deceptive, see Definition~\ref{dfn:deceptive},
  we can find $\delta > 0$ such that
  \begin{equation}      \label{eq:weakform}
    \inf_q \limsup_{n\to\infty}
    \sup_{p'\in\ngpepp} \frac1n D_n(p'||q) > \delta,
  \end{equation}
  \color{black} for all $\epsilon' > 0$ 
  \color{black}
  and hence, by Lemma~\ref{lm:asyinter} in
  Appendix~\ref{app:redbasics}, we have
  \begin{equation}      \label{eq:strongform}
     \limsup_{n\to\infty} \inf_q
    \sup_{p'\in\ngpepp} \frac1n D_n(p'||q) > \delta,
  \end{equation}
  \color{black}
  for all $\epsilon' > 0$.
  \color{black}
  In 
  both \eqref{eq:weakform} and \eqref{eq:strongform}
  the infimum is over all probability measures $q$ on $\naturals^\infty$.
  
  Since $\cP$ is assumed to be learnable, from Definition~\ref{dfn:lrn} there must certainly be some 
  $\eta > 0$, a stopping rule $\tau$, and 
  $\hat{q}:\naturals^*\to {\mathbb P}({\mathbb N})$ 
  such that for all $\tilde{p} \in \cP$ we have
  \[
  \tilde{p}(\tau = 1 \mbox{ and } D_1(\tilde{p}||\hat{q})>\delta) < \eta.
  \]

  For all $n \ge 1$ let
  \[
  A_{n} := \sets{x^n\in\naturals^n: \tau(x^n)=1}
  \]
  denote the set of sequences of length $n$ on which $\tau$ has entered. Note 
  that $p(A_n)$ is increasing with $n$ and $\lim_{n \to \infty} p(A_n) = 1$.
 We can therefore pick $n \ge 4/(1- \eta)$ large enough such that
and a finite set $S_n\subset A_n$ such that
  $p(S_n)  \ge (1 + \eta)/2$.

 Let 
 $\tilde\epsilon :=\frac1{2n^4}$.  Applying Lemma~\ref{lm:jn} in
 Appendix~\ref{app:redbasics} to \iid probability distributions over
 length-$n$ strings, we see that for all $\tildep\in\cP$
 such that $||p-\tildep||_1\le\tilde\epsilon$, we have
  \[
  \tildep(S_n) > (1+\eta)/2-\frac2{n} \ge \eta.
  \]
  \color{black}
  
  From \eqref{eq:weakform} and \eqref{eq:strongform} respectively it then follows that \color{black} for all $0 < \epsilon<\tilde\epsilon$
  \color{black}
  we have 
  \begin{equation}      \label{eq:weakformnew}
   \inf_q \limsup_{m\to\infty} \sup_{p'\in\ngpe} \frac1m D_m(p'||q) >\delta,
 \end{equation}
 and 
 \begin{equation}       \label{eq:strongformnew}
   \limsup_{m\to\infty} \inf_q \sup_{p'\in\ngpe} \frac1m D_m(p'||q) >\delta,
 \end{equation}
 and of course, $p'(S_n) \ge \eta$ for all $p'\in\ngpe$.
  
  Fix some
  \color{black}
  $0 < \epsilon<\tilde\epsilon$.
  \color{black}
  Since $S_n$ is finite by choice,
  for each $p'\in\ngpe$ we can choose $\y(p')\in S_n$ such that
  \[
    \y(p') = \arg\min_{\y \in S_n} D(p'||q_{\y}),
  \]
  where $q_{\y} = \hat{q}(\cdot|\y)$. Let
  \[
    \cB_{\y} = \sets{p' \in \ngpe : \y(p') = \y }.
  \]
Therefore,
  \begin{equation}      \label{eq:finiteunion}
    \ngpe = \union_{\y\in S_n} \cB_\y
  \end{equation}
  where the union above is finite. 
  
  
  From \eqref{eq:strongformnew} we have that the asymptotic per symbol redundancy of 
  $\ngpe$ is strictly bigger than $\delta$. Since the union in 
  \eqref{eq:finiteunion} is finite,
  from Lemma~\ref{lm:finiteunion} in Appendix~\ref{app:redbasics} 
  it follows that there is some $\y' \in S_n$
  such that the asymptotic per symbol redundancy of $\cB_{\y'}$ is strictly bigger than $\delta$.
  Hence, from Proposition~\ref{prop:rnn}, we have that
  the single letter redundancy of $\cB_{\y'}$ is strictly bigger than $\delta$, 
  which in turn implies that $\sup_{p'\in\cB_{\y'}} D(p'|| q_{\y'}) > \delta$. Thus we conclude that
  there is some $p'\in\cB_{\y'}$ such that $D(p'|| q_{\y'}) > \delta$.

  Therefore, if $p'$ were in force then  with probability $\ge\eta$, we would have
  \[
    D_1(p||\hat{q}) \ge D_1(p||q_{\y'}) >\delta,
  \]
  which violates the assumption that $\cP$ is learnable.

 
  

  This completes the proof of the necessity part of Theorem~\ref{thm:equiv}.
  \color{black}

\color{black}
\prasadtrue
\ifprasad
\subsection{\dwc $\Rightarrow$ Learnable}
  We thank the anonymous reviewer for observing this direction of the connection.

  Suppose for all $\delta'>0, \eta'>0$, we have a stopping rule
  $\tau_{\delta',\eta'}$ and a universal measure $q^*$, such that
  $\tau_{\delta',\eta'}$ certifies with confidence $1-\eta'$ when the
  per-symbol redundancy of $q^*$ falls (and remains) below $\delta'$.

  Then for any given $\delta>0$ and $\eta>0$ we construct a new stopping
  rule $\sigma_{\delta,\eta}$ 
  \ignore{and an estimate
  $\hat{q}$ over $\naturals$}
  and an estimator
  $\hat{q}:\naturals^*\to{\mathbb P}(\naturals)$ 
  that satisfies for all $p\in\cP$,
  \color{black}
  \begin{equation}
    \label{eq:claim}
    p\Paren{ \sigma_{\delta,\eta}=1 \text{ and } D(p||\hat{q})>\delta } < \eta.
  \end{equation}
  \color{black}
  According to Definition~\ref{dfn:lrn}, this will establish that
  $\cP$ is learnable.

  To see this, let $\delta'=\delta\eta/2$ and $\eta'=\eta/2$. Let
  \ignore{\[
    T = \min_{t\ge 1} \tau_{\delta',\eta'}(X_1^t)=1
  \]
  be the stopping time, }
  \[
    T := \min \{ t\ge 1 : \tau_{\delta',\eta'}(X_1^t)=1 \},
  \]
  and note that $X_{T+1},\cdots,X_{2T-1}$ 
  \ignore{are the
  $T-1$ samples after $\tau$ stops.}
  are the
  $T-1$ subsequent samples.
  Set
  \[
    \sigma_{\delta,\eta}(X^{2T-1})=1,
  \]
  (regardless of what $X_{T+1}^{2T-1}$ are) and output the estimate
  $\hat{q}_*\in{\mathbb P}(\naturals)$, 
  where for all $x\in\naturals$
  \ignore{\[
    \hat{q}(x) = \frac1T\sum_{i=0}^{T-1} q^*(x | X^{T+i}_{T+1}).
  \]}
   \[
    \hat{q}_*(x) = \frac1T\sum_{i=0}^{T-1} q^*(x | X^{T+i}_{T+1}).
  \]
  In the above, $X_{T+1}^T$ is understood to be the empty
  string. Note that 
  \color{black}
  $\hat{q}_*$ 
  \color{black}
  does not use the \emph{observations}
  $X_1\upto X_T$ and, given $T$, 
  $\hat{q}_*$ 
  is conditionally independent
  of $X^T$. Rather, 
  $\hat{q}_*$ 
  applies the marginal distributions of $q^*$
  over $\naturals^i$, $i\le T$, to the observations
  $X_{T+1}\upto X_{2T-1}$. 
  To complete the definition of $\hat{q}$ as a function from $\naturals^*$ to ${\mathbb P}(\naturals)$,
  we define it arbitrarily for finite sequences of naturals on which $\sigma_{\delta,\eta}$ equals $0$
  and on those for which $\sigma_{\delta,\eta}$ equals $1$ we define it to be 
  $\hat{q}_*$.
  We claim 
  that the stopping time $\sigma_{\delta,\eta}$ and 
  estimator
  $\hat{q}:\naturals^*\to{\mathbb P}(\naturals)$ 
  as defined above satisfy~\eqref{eq:claim}. 

  To prove the claim, fix $p\in\cP$. 
  Note that
  \ignore{\[
    D_1(p||\hat{q})= D_1\Paren{p||\frac1T \sum_{i=0}^{T-1} q^*(\cdot| X^{T+i}_{T+1})} \le  \frac1T \sum_{i=0}^{T-1} D_1\Paren{p|| q^*(\cdot| X^{T+i}_{T+1})}.
  \]}
   \[
    D_1(p||\hat{q}_*)= D_1\Paren{p||\frac1T \sum_{i=0}^{T-1} q^*(\cdot| X^{T+i}_{T+1})} \le  \frac1T \sum_{i=0}^{T-1} D_1\Paren{p|| q^*(\cdot| X^{T+i}_{T+1})}.
  \]
  Further,
  we have for any $X^T$ that
  \begin{align}
    \nonumber
   {\mathbb E}\left[ \frac1{T} \sum_{i=0}^{T-1} D_1\Paren{p|| q^*(\cdot| X^{T+i}_{T+1})} \mid X^T \right]
    &=   \sum_{i=0}^{T-1} {\mathbb E}\left[ \frac1{T} D_1\Paren{p|| q^*(\cdot| X^{T+i}_{T+1})}\mid T \right]\\
    \label{eq:ce}
    &= \frac{1}{T} D_{T}(p||q^*),
  \end{align}
  where the first equality holds because (i) $p$ is \iid, and (ii) the
  single letter distributions within any of the KL divergences only
  depend on the length $T$, and not on the values of $X_1\upto
  X_T$. In the last expression, $\frac{1}{T} D_{T}(p||q^*)$ denotes $\frac1m D_m(p||q)$ evaluated at $T$.
  
  Observe that since $\cP$ is \dwc, and $q^*$ is a weak universal
  measure, there exists $N_p$ such that
  $\frac1m D_m(p||q^*) \le \delta'$ for all $m\ge N_p$.  
  The conditional expectation in~\eqref{eq:ce} is a random variable that only
  depends on $T$, and whenever $T\ge N_p$, we have
  \[
    {\mathbb E}\left[ \frac1{T} \sum_{i=0}^{T-1} D_1\Paren{p|| q^*(\cdot| X^{T+i}_{T+1})} \mid X^T \right] =  \frac{1}{T} D_{T}(p||q^*) \le \delta'.
  \]
  When $T \ge N_p$ therefore, Markov's inequality implies that with
  probability $\ge1- \delta'/\delta$, conditioned on $X^T$,
  \ignore{\[
    D(p||\hat{q})\le\frac1T \sum_{i=0}^{T-1} D\Paren{p|| q^*(\cdot| X^{T+i}_{T+1})} \le \delta.
  \]}
  we have
  \[
    D_1(p||\hat{q}_*)\le\frac1T \sum_{i=0}^{T-1} D\Paren{p|| q^*(\cdot| X^{T+i}_{T+1})} \le \delta.
  \]
  Since $\cP$ is \dwc, 
  if we take $N_p$ to be the smallest such integer
  we know $p(T> N_p)\ge 1-\eta'$. Hence, with
  probability under $p$
  $\ge (1-\eta')(1- \frac{\delta'}\delta)\ge
  1-\eta'-\frac{\delta'}{\delta} = 1-\eta$,
  \ignore{\[
    D(p||\hat{q}) \le \delta.
  \]}
  we have
  \[
    D_1(p||\hat{q}_*) \le \delta.
  \]

\else
\color{red}
  \subsection{\dwc $\Rightarrow$ Learnable}
  
  Suppose $\cP$ is \dwc.
  Then, by Definition~\ref{dfn:dwc}, there is a probability measure $q$ on $\naturals^\infty$
  such that for any given $\delta>0$ and $\eta>0$ there is a stopping rule $\tau_{\delta,\eta}$ 
  such that \eqref{eq:dwcdef} holds. 
  Then, for the given $\delta>0$ and $\eta>0$, we construct a new stopping
  rule $\sigma_{\delta,\eta}$ and an estimator
  $\hat{q}:\naturals^*\to{\mathbb P}(\naturals)$ 
  such that, for all $p \in \cP$. we have
  \begin{equation}
    \label{eq:claim}
    p\Paren{ \sigma_{\delta,\eta}=1 \text{ and } D_1(p||\hat{q})>\delta } < \eta.
  \end{equation}
  According to Definition~\ref{dfn:lrn}, this will establish that
  $\cP$ is learnable.

  
  Let $T$ denote the first time that is at least $1$ and at which $\tau_{\delta,\eta}$ is equal to $1$.
  To construct $\sigma_{\delta,\eta}$ and
  $\hat{q}:\naturals^*\to{\mathbb P}(\naturals)$ we take an
  additional $T-1$ samples, $X_{T+1},\cdots,X_{2T-1}$, and at this point, set
  $\sigma_{\delta,\eta}=1$. (Note that if $T=1$ we do not take any additional samples.)
  We construct the estimate
  \[
    \hat{q}_T(x) := \frac1T\sum_{i=0}^{T-1} q(x | X^{T+i}_{T+1}),
  \]
  which is a probability distribution on $\naturals$, where $X_{T+1}^T$ is understood to be the empty string. Note that
  $\hat{q}_T$ does not use the observations $X_1\upto X_T$ except insofar as they determine $T$ - what it does is to
  apply the marginal distributions of $q$ over $\naturals^i$,
  $i\le T$, to the observations $X_{T+1}\upto X_{2T-1}$. 
  Thus, for any fixed $p \in \cP$, $\hat{q}_T$ is conditionally independent of $X_1^T$ given $T$, which means that
  one can think of $\hat{q}_T$ as depending only on $T$ (since $p \in \cP$ is fixed), and the subscript is meant
  to indicate this. To complete the definition of $\hat{q}$ as a function from $\naturals^*$ to ${\mathbb P}(\naturals)$,
  we define it arbitrarily for finite sequences of naturals on which $\sigma_{\delta,\eta}$ equals $0$
  and on those for which $\sigma_{\delta,\eta}$ equals $1$ we define it to be $\hat{q}_T$.
  

  Note that
  \begin{equation}      \label{eq:chain}
    D_1(p||\hat{q}_T)= E[ D_1\Paren{p||\frac1T \sum_{i=0}^{T-1} q(\cdot| X^{T+i}_{T+1})}|T] \le  E[\frac1T \sum_{i=0}^{T-1} D_1\Paren{p|| q(\cdot| X^{T+i}_{T+1})}|T] = \frac{1}{T} D_{T}(p||q),
  \end{equation}
  the last expression here being $\frac1m D_m(p||q)$ evaluated at $T$.
  Since $\cP$ is \dwc, with $q$ being a probability
  measure on $\naturals^\infty$ that verifies this, there exists $N_p$ such that
  $\frac1m D_m(p||q) \le \delta$ for all $m\ge N_p$,
  and if $N_p$ is chosen to be the first such time then we must have $p(T \ge N_p) \ge 1 - \eta$.
  But then, since $\sigma_{\delta,\eta}$ enters only after $T$, it follows from 
  \eqref{eq:chain} that we have $D(p||\hat{q}_T) \le \delta$ with probability at least $1- \eta$, which completes the proof.
  
\fi
 
\color{black}

\section{Length-$n$ Per-Symbol Redundancy of $\cM_h$}
  \label{app:dn}
  We construct a probability measure $q^*$ on $\naturals^\infty$ such that 
  for $\cM_h$ we have
  \[
  \sup_{p\in\cM^n_h} \frac1n D_n(p||q) 
  \le 
  \frac{2h^{\frac{1}{4}}(\sqrt{h} + 1)}{\sqrt{\ln n}}
  +
  \pi\sqrt{\frac2{3n}} \log e.
  \]
  This implies that the per-symbol length-$n$ redundancy of $\cM_h$ diminishes to $0$ 
as $n \to \infty$.
  Hence $\cM_h$ is strongly compressible.


  Consider the probability distribution $q$ on $\naturals$ defined by
  $q(i)=1/i(i+1), i \ge 1$.
  As observed in Example~\ref{eg:mnth}, we have 
  \begin{equation}		\label{eq:4bound}
  \sup_{p\in\cM_h} E_p\Paren{\ceil{\log\frac1{q(X)}}}^2 < 4 (\sqrt{h}+1)^2.
  \end{equation}
  We consider a scheme that encodes patterns~(\cite{OSZ03}) of symbols 
  (i.e. natural numbers in our case) first, followed by
  an encoding using $\ceil{\log\frac1{q(x)}}$ bits to describe every symbol $x$
  that appeared in the string, in the order in which they arrived.
  To clarify, recall that the pattern of a sequence of symbols from $\naturals$
  replaces each symbol by $k \in \naturals$ if the symbol was the $k$-th
  new symbol to appear in the sequence. For example, the pattern of the 
  sequence of natural numbers $(2,3,17,4,3,3,1,2,4)$ is $(1,2,3,4,2,2,5,1,4)$.
  If in addition to the pattern of a finite sequence of natural numbers, in which there
  are $l$ distinct symbols,
  one knows which symbol was the $k$-th symbol
  to appear for each $1 \le k \le l$, one learns the sequence of symbols.

  The expected (not normalized by $n$)
  additional number of bits to encode the pattern of a sequence of symbols of length $n$ from any $p \in \cM_h$ is at most $\pi\sqrt{\frac23n}\log e
  $, using the results in~\cite{OSZ03}, 
  while the expected number of bits
  to describe the symbols of length-$n$ strings using a prefix code based
  on the probability distribution $q$ on $\naturals$ is at most
  \[
  \sum_{i\in\naturals}\Paren{1-(1-p(i))^n} \ceil{\log \frac1{q(i)}}.
  \]
  Note that the distinct symbols appearing the the string will need to be 
  specified in the order in which they arrived.
  Let $M_n$ denote the number of distinct symbols that appear in a
  sequence of length $n$.
  Then the expected number of extra bits the scheme uses for length-$n$ strings
  is (without normalizing by $n$) at most $\pi\sqrt{\frac23n}\log e$ plus at most
  \begin{align*}
  &
  \sum_{i\in\naturals}\Paren{1-(1-p(i))^n} \ceil{\log \frac1{q(i)}}\\
  &\ale{(a)}
  \sqrt{
  \sum_{i\in\naturals}\Paren{1-(1-p_i)^n}
  \sum_{j\in\naturals}\Paren{1-(1-p_j)^n}\Paren{\ceil{\log \frac1{q(j)}}}^2
  }\\
  &\le
  \sqrt{
  \sum_{i\in\naturals}\Paren{1-(1-p_i)^n}
  \sum_{j\in\naturals}\Paren{n p_j }\Paren{\ceil{\log \frac1{q(j)}}}^2
  }\\
  &\ale{(b)}
  \sqrt{
    4 (\E M_n)n (\sqrt{h}+1)^2}\\
  &\ale{(c)}
  \frac{2nh^{1/4}(\sqrt{h}+1)}{\sqrt{\ln n}}.
  \end{align*}
  Here $(a)$ follows from the Cauchy-Schwartz inequality, while
  $(b)$ follows from~\eqref{eq:4bound} and the definition of $M_n$.
  As for $(c)$, a result similar to $(c)$ can be found in~\cite{OSVZ04}, but
  we justify $(c)$ below for completeness. We observe
  that for all $i \in \naturals$ we have
  \begin{align*}
    1-(1-p_i)^n &= p_i\sum_{j=0}^{n-1} (1-p_i)^j\\
                 &\le p_i\Paren{\sum_{j=0}^{n-1} (1-p_i)^j}
                   \frac{\sum_{k=1}^{n}\frac1k}{\ln n}\\
                &\ale{(a)} \frac{np_i}{\ln n}\sum_{j=0}^{n-1} \frac{(1-p_i)^j}j\\
                 &\le \frac{np_i\log\frac1{p_i}}{\ln n}.
  \end{align*}
  Combining the above with the fact that the entropy of any
  $p\in\cM_h$ is at most $\sqrt{h}$, which was shown
  in Example~\ref{eg:mnth}, 
  proves $(c)$ in the previous set of equations.  In the above set of equations,
  inequality $(a)$ follows from Minkowski's inequality
  which says that if $x_i$ and $y_i$ ($0\le i\le n-1$) are both
  decreasing positive sequences, then
  $n\sum x_i y_i\ge \sum x_j\sum y_k$. Minkowski's inequality is
  easily proved by noting
  $\sum x_j\sum y_k= \sum_{m}\sum x_i y_{(i+m)\mod n}$ and that
  $\sum x_i y_i \ge \sum x_i y_{(i+m)\mod n}$ for all $0\le m\le n-1$.

  The claim about the per-symbol length-$n$ redundancy of $\cM_h$ follows after normalization by $n$.

\section{Typicality of Top Heavy Empirical Distributions }
  \label{app:tail}
  
In this section we prove a useful result quantifying how close the 
empirical distribution of a sample drawn \iid from a probability distribution 
$p$ on 
$\naturals$ is to $p$, when the alphabet of symbols showing
up in the sample is not too spread out. 
There is a lemma that looks somewhat similar in~\cite{HY10}. The
difference of the result in 
Lemma~\ref{lm:yeung} from that in~\cite{HY10} is that the 
right side of the inequality in~\eqref{eq:yeung}
does \emph{not} depend on $p$.
The result of  Lemma~\ref{lm:yeung} will be used in the 
sufficiency proof in Appendix~\ref{s:phibt}
and this property is crucial for its use.

  \bLemma
  \label{lm:yeung}
  Let $p$ be any probability distribution on $\naturals$. Let $\gamma>0$
  and let $k \ge 2$ be an integer. Let $X_1^n$ be a sequence
  generated \iid with marginals $p$ and let $t(X^n)$ be the empirical
  distribution of $X_1^n$.  Then
  \begin{equation}		\label{eq:yeung}
  p\Paren{ |t(X^n)-p|_1 >\gamma\text{ and }2 \dotF_t^{-1}(1-\gamma/6) \le k} 
  \le 
  (2^{k}-2) \exp\Paren{-\frac{n\gamma^2}{18}}.
  \end{equation}
  
\ignore{\bRemark
  This bounds the probability that the empirical distribution is far from $p$, but with the
  additional constraint that the empirical distribution is effectively contained over the
  alphabet $\sets{1\upto k}$.
  \eRemark}

\Proof
\color{black}
  For any probability distribution $p'$ on $\naturals$ with finite
  support of size $L$ we have the following well-known result (\eg,~\cite[Proposition 1]{WOSVW05})
  \begin{equation}\label{eq:base}
    p'( |t_{X^n}-p'|_1 \ge \alpha) \le (2^L-2) \exp\Paren{-\frac{n \alpha^2}{2}},
  \end{equation}
  where $t_{X^n}$ is the empirical distribution of $X^n$ generated \iid with marginal 
  distribution $p'$. The above is easily seen by recalling
  \[
   |t_{X^n}-p'|_1 = 2\sup_{E \subset [L]} |t(E)-p'(E)| = 2\sup_{\substack{E \subset [L]\\ |E|\le \floor{L/2}}} |t(E)-p'(E)|,
  \]
  and that for any $E\subset [L]$, from Hoeffding's inequality,
  \[
    p\Paren{ |t_{X^n}(E) - p'(E)| \ge \frac\alpha2} \le 2\exp\Paren{-\frac{n\alpha^2}{2}}.
  \]
  A union bound over all non-empty subsets of size $\le \floor{L/2}$ yields~\eqref{eq:base}.
  
\color{black}  
  Consider the probability distributions $p'$ and $t'$ on $A$ obtained
  from $p$ and $t$ respectively
  via the mapping from $\naturals$
  to $A := \sets{1\upto k-1}\union \sets{-1}$ which maps $i$ to $i$ for
  $0 \le i \le k-1$ and maps all the other natural numbers to $-1$.
  Thus, we have
  \[
    p'(i)=
    \begin{cases}
      p(i), & \mbox{ if $1\le i \le k-1$},\\
      \sum_{j=k}^\infty p(j), & \mbox{ if $i = -1$}.
    \end{cases}
  \]
  Further, 
  sequences of natural numbers generated \iid with marginal distribution $p$
  and with empirical distribution $t$
  are mapped to sequences from $A$ that are \iid with probability 
  distribution $p'$ and have empirical distribution $t'$.

  Applying~\eqref{eq:base} to $p'$, we have
  \begin{equation}		\label{eq:pprime}
    p'(|p'-t'|_1 > \gamma/3)
    \le
    (2^{k}-2)
    \exp\Paren{-\frac{n\gamma^2}{18}}.
  \end{equation}
We first argue that all sequences generated by $p$ with empirical
distributions $t$ satisfying
\[
|p-t|_1 > \gamma
\text{ and }
2 \dotF_t^{-1}(1-\gamma/6) \le k
\]
are mapped into sequences generated by $p'$ with empirical $t'$ satisfying
\[
\label{eq:toshow}
|p'-t'|_1 > \gamma/3
\text{ and }
t'(-1)\le \gamma/3.
\]

This follows from writing
\begin{align*}
|p-t|_1 &-\sum_{i=1}^{k-1} |p(i)-t(i)|\\ 
&\le\sum_{j=k}^{\infty} (p(j)-t(j)) +2\sum_{j=k}^{\infty} t(j)\\
&\le|p'(-1)-t'(-1)|+ \gamma/3,
\end{align*}
where the last inequality above follows from 
the fact that $2 \dotF_t^{-1}(1-\gamma/6) \le k$ implies 
$F_t(k-1) \ge 1 - \gamma/6$, i.e. $\sum_{j=k}^{\infty} t(j) \le \gamma/6$.
Hence we have
\[
|p'-t'|_1 = \sum_{i=1}^{k-1} |p(i)-t(i)| + |p'(-1)-t'(-1)|
\ge |p-t|_1 -\gamma/3 >  \gamma/3,
\]
because $|p-t|_1 > \gamma$.

Thus, from~\eqref{eq:pprime}, we will have
\begin{align*}
&p( |t(X^n)-p|_1 > \gamma \text{ and } 2 \dotF_t^{-1}(1-\gamma/6) \le k) \\
&\le
p'(|t'-p'|_1 > \gamma/3 \text{ and } t'(-1)\le \gamma/3) \\
&\le
(2^{k}-2) \exp\Paren{-\frac{n\gamma^2}{18}}.
\end{align*}
\ignore{where the first inequality follows by simple logical implications and
by observing that $p(\naturals- \sets{1\upto k(q)-1})=p'(-1)$. The lemma
then follows.}
This completes the proof of the lemma.
\eLemma

\section{$\tau$ Enters With Probability 1}
  \label{s:phione}
  We reproduce the argument from~\cite{SA12:jmlr} here for completeness.

  Every probability distribution $p\in\cP$
  is contained in at least one of the elements of the cover 
  \color{black}
  $(Q_{\tilde p,m} \cap \cP, {\tilde p} \in {\tilde \cP}_m)$,
  where $Q_{\tilde p,m}$ denotes the zone 
  \color{black}
  of ${\tilde p} \in {\tilde \cP}_m$.
  Recall the enumeration of $\tilde\cP_m$. Let $p'$ be
  be centroid with the smallest index among all centroids
  in ${\tilde \cP_m}$ whose zones contain $p$.
 \ignore{Let $Q$ be the zone of $p'$.
  There is thus some $\gamma > 0$ such that the neighborhood around $p$,
  \[
  I(p,\gamma) := \sets{t: |p-t|_1 <\gamma}
  \]
  where the elements $t$ are probability distributions on $\naturals$,
  satisfies $I(p,\gamma)\subseteq Q$.  Note in particular that $p$ is in
  the reach of $p'$.}
    With probability 1, sequences generated by $p$
  will eventually have their empirical distribution within
  $\Qpprimem$.
  (see~\cite{Chu61} 
  for a proof).

  Next note that for all $n$ sufficiently large the analog of~\eqref{eq:bnkrpt},
\color{black}
(which makes sense for all $\tilde{p} \in \tilde \cP_m$) 
\color{black}
  will hold.
  This follows since the right hand side of~\eqref{eq:bnkrpt} diminishes to
  zero polynomially with $n$ while the left hand side diminishes
  to zero exponentially fast in $n$.

  Next,
  the analog of~\eqref{eq:bnkrpttwo} 
  will also hold eventually with 
  probability 1, since, if $t$ denotes 
  the \color{black} empirical distribution \color{black} of a sequence 
  of length $n$
  generated by $p$,
  \color{black} then from Proposition~\ref{prop:asconv} \color{black}
  \begin{equation}		\label{eq:cdfconvg}
  \dotF_t^{-1}(1-\gamma_{{}_{{p'},m}}/6)\to \dotF_p^{-1}(1-\gamma_{{}_{{p'},m}}/6)
  \end{equation}
  with probability 1 as $n \to \infty$, where we note that the quantity 
  on the left hand side of~\eqref{eq:cdfconvg} is actually a random
  variable and $t$ determines $n$. Furthermore, we will also have after finitely
  many samples that
  \begin{align*}
    \color{black}
    2 \dotF_t^{-1}(1-\gamma_{{}_{{p'},m}}/6)
    &< \color{black}3 \dotF_p^{-1}(1-\gamma_{{}_{{p'},m}}/6)\\
  &\le 3 \Paren{{\sup_{r\in B(p',\epsilon_{p',m};\cP)}} 
  \dotF_r^{-1}(1-\gamma_{{}_{{p'},m}}/6)}\\
  &=
  \log C(p',m),
  \end{align*}
  where the \color{black} second \color{black} inequality follows since $p$ is in the $\frac1m-$reach of $p'$. \color{black}Note that the 3 in the inequalities above can be replaced by any
  number strictly $>2$ or by an additive constant. \color{black}

  Therefore, both~\eqref{eq:bnkrpt} and~\eqref{eq:bnkrpttwo} will 
  eventually hold with
  probability 1. Furthermore, long enough sequences generated by $p$
  fall into the zone of $p'$ with probability 1. This implies in turn
  that $\tau_{\eta,m}$ enters with probability 1. Note that it is entirely
  possible that 
  \color{black}
  some other centroid
  \color{black}
  traps strings before they can be
  trapped by $p'$, but that does not take away from the fact that 
  $\tau_{\eta,m}$
  will enter with probability 1.

\section{Probability of Falling Into Bad Traps}
  \label{s:phibt}
  \ignore{We need two additional technical lemmas for this proof. The proofs of both
  are in~\cite{SA12:jmlr}.
  \bLemma
  \label{lm:dpq} 
  Let $\epsilon_0 > 0$. If
  \[
  |p_0-q|_1\le \frac{\epsilon_0^2(\ln 2)^2}{16}~,
  \]
  then
  for all $p\in\cP$ with $\dist(p,p_0)\ge\epsilon_0$, we have
  \[
  \dist(p,q)\ge\frac{\epsilon_0^2\ln 2}{16}.\eqed
  \]
  \eLemmap}

\color{black}
  Let $t$ be any length-$n$ empirical
    distribution trapped by $\hat{p}$, which we recall has $\frac1m$-reach 
    $\epsilon_{{\hat p},m}$, such
    that $p\notin B({\hat p},\epsilon_{{\hat p},m};\cP)$.  
   Then we have
  \[
  ||\hat{p}-p||_1\ge \epsilon_{{\hat p},m},
  \]
  because $p\notin B({\hat p},\epsilon_{{\hat p},m};\cP)$, and
  we have 
  \[
  ||\hat{p}-t||_1 
  <
  \frac{\epsilon_{\hat{p},m}}{2},
  \]
  because $t$ 
  has to be in the zone $Q_{\hat{p},m}$ in order to 
  be captured by ${\hat p}$.
  Using the triangle inequality for $\ell_1$ norms,
   we get
  \[
  ||p - t|| \ge\frac{\epsilon_{\hat{p},m}}2 :=\gamma_{{}_{{\hat p},m}}
  \] 
  \color{black}
  This means that for every $p\in\cP$, the probability that 
  length-$n$ sequences with empirical distribution $t$ are trapped
  by a bad $\hat{p}$ can be bounded from above as
  \begin{align*}
  &\le
  p\biggl( 
  | t-p|_1 \ge \gamma_{{}_{{\hat p},m}}
  \text{ and } 
  \ignore{\biggr.\\
  &\qquad\qquad\qquad }
  2 \dotF_t^{-1} (
  1-\frac{\gamma_{{}_{{\hat p},m}}}6
  )
  \le \log C(\hat{p},m) 
  \biggr)\\
  &\ale{(a)}
  (C(\hat{p},m)-2) \exp\Paren{-\frac{n \gamma_{{}_{{\hat p},m}}^2}{18}}\\
  &\ale{(b)}
  \frac{\eta (C(\hat{p},m)-2)}
  {2 C(\hat{p},m) \iota(\hat{p})^2 n(n+1)} \\
  &\le \frac{\eta}{2\iota(\hat{p})^2 n(n+1)},
  \end{align*}
  where the inequality $(a)$ follows from Lemma~\ref{lm:yeung}
  and $(b)$ from~\eqref{eq:bnkrpt}. 
  Therefore, the probability of sequences
  falling into bad traps is bounded above by
  \[
  \le
  \sum_{n\ge1}\sum_{\tilde{p}\in {\tilde \cP}}
  \frac{\eta}{2\iota(\tilde{p})^2 n(n+1)} 
  \le \frac{\pi^2}{12} \eta < \eta,
  \]
  since
  $\sum_{\tilde{p}\in {\tilde \cP}}
  \frac1{\iota(\hat{p})^2} = \frac{\pi^2}{6} \mbox{  and  }
  \sum_{n\ge1}
  \frac1{n(n+1)}
  =
  1.
  $
  
\section{A Fake Proof}
  \label{s:fake}
 In this section we give a fake proof of the following
  mistaken claim: \emph{if $\cP_1$ and $\cP_2$ are \dwc, then
    $\cP_1\union\cP_2$ is also \dwc}. We then explain why it is wrong.  In the concluding remarks in~\cite{SA12:jmlr}
    it was stated, in passing, that if $\cP_1$ and $\cP_2$ are insurable 
    then $\cP_1\union\cP_2$ is also insurable. This statement if false,
    for the reasons explained in this section. This does not affect any of the
    results in~\cite{SA12:jmlr}.
    
    The argument proceeds as
  follows. Since $\cP_i$ is \dwc for each $i = 1, 2$, 
  there is a probability measure $q_i$ on $\naturals^\infty$ for each $i = 1, 2$ such that
   for every $m \ge 1$,
   $0 < 1 - \eta < 1$ and $i = 1, 2$ there is a universal stopping rule $\tau_{\eta,m}^{(i)}$ such that, for all $p\in\cP_i$, we have
   \[
   p\Paren{
   \exists n \mbox{ such that } \frac1n  D_n\Paren{p||q_i}>\frac1m
   \text{ and }
 \tau_{\eta,m}^{(i)}(X^n)=1}\\
   < \eta.
   \]
   Let $q := (q_1+q_2)/2$ 
   and, for accuracy $\frac1m > 0$ and confidence $0 < 1 -\eta < 1$, define
   \begin{equation}
       \label{eq:fakeprooftau}
    \tau_{\eta,m}(\x)
    :=
     \mathbbm{1}(\tau^{(1)}_{\eta,2m}(\x) =1)
     \mathbbm{1}(\tau_{\eta,2m}^{(2)}(\x) =1)
     \mathbbm{1}(|\x| > 2m).
   \end{equation}

   Now, suppose $p\in\cP_1\union \cP_2$. 
   Without loss of generality, assume that
   $p\in\cP_1$. Now,
  if $n> 2m$ and we have
  \[
   \frac1n D_n\Paren{p||\frac{q_1+q_2}2}>\frac1m,
   \]
   then we have
   \[
  \frac1n D_n\Paren{p||q_1}>\frac1m-\frac1n>\frac{1}{2m}
   \]
  Further, from \eqref{eq:fakeprooftau}, if $\tau_{\eta,m}(\x)=1$, then we have $\tau^{(1)}_{\eta,2m}(\x)=1$
  as well. Therefore
   \begin{align*}
   p&\Paren{
   \exists n \text{ such that }
  \frac1n D_n\Paren{p||\frac{q_1+q_2}2}>\frac1m
   \text{ and }
   \tau_{\eta,m}(X^n)=1}\\
   &\le
  p\Paren{
   \exists n \text{ such that $n>2m$, } 
   \frac1n D_n\Paren{p||q_1}>\frac{1}{2m}
   \text{ and }
   \tau_{\eta,2m}^{(1)}(X_1^n)=1}
   < \eta,
   \end{align*}
   where we have used \eqref{eq:fakeprooftau} to see that the event whose probability is being evaluated on the left hand side of the preceding equation cannot occur unless $n > 2m$.
   Since the above holds for all $p\in\cP_1$ and we 
   can use a similar argument for all $p \in \cP_2$, we are ``done''.

   The flaw in the above ``proof'' is that $\tau_{\eta,m}$, as defined in \eqref{eq:fakeprooftau}, does not necessarily eventually equal $1$ almost surely for
   all sources in $\cP_1\union\cP_2$, which would mean that it is not a universal stopping rule for the model class $\cP_1\union\cP_2$. To see why this issue might arise, note that $\tau_{\eta,2m}^{(i)}$ is known to eventually equal $1$ 
   almost surely only for sources in $\cP_i$.
   Thus, if it happens to be the case that there is
   some event
   $A \subsetneq \naturals^\infty$ and
   $p_1\in\cP_1$ with $p_1(A)>0$ for which we have $p_2(A)= 0$ for every source
  $p_2\in\cP_2$, then
  $\tau_{\eta,2m}^{(2)}$ might never stop waiting
   on the sequences in $A$. This doesn't stop $\cP_2$ from being \dwc. But
   when we introduce sources from $\cP_1$, in particular $p_1$, we find
  that $\tau_{\eta,m}$, as defined in \eqref{eq:fakeprooftau}, 
  will never stop waiting under $p_1$.
   The stopping rule $\tau_{\eta,m}$ would then not be a universal stopping
   rule for the model class $\cP_1 \cup \cP_2$.
  \bibliography{univcod}

  \end{document}